\documentclass[12pt, a4paper, oneside]{article}
\usepackage{setspace}
\usepackage[stable]{footmisc}
\long\def\symbolfootnote[#1]#2{\begingroup %
\def\thefootnote{\fnsymbol{footnote}}\footnote[#1]{#2}\endgroup}
\usepackage{tabularx}
\usepackage{parskip}
\usepackage{float}

\usepackage[singlelinecheck=true]{caption}
\usepackage{natbib}
\usepackage[ansinew]{inputenc}
\usepackage[T1]{fontenc}                        
\usepackage{textcomp}                           

\usepackage{longtable}
\usepackage{colortbl}
\usepackage{booktabs}
\usepackage{tabularx}
\newcolumntype{L}[1]{>{\raggedright\arraybackslash}p{#1}} 
\newcolumntype{C}[1]{>{\centering\arraybackslash}p{#1}} 
\newcolumntype{R}[1]{>{\raggedleft\arraybackslash}p{#1}} 

\usepackage{graphicx}
\usepackage{epic}
\usepackage{overpic}
\graphicspath{{graphics/}}
\usepackage{import}
\usepackage{tikz}
\usetikzlibrary{arrows,calc}
\usepackage{booktabs}       
\usepackage{dcolumn}        
\usepackage[labelfont=bf]{caption}
\usepackage{latexsym}
\usepackage{amsmath}
\usepackage{amsfonts}
\usepackage{amssymb}
\usepackage{url}
\usepackage{amsbsy}
\usepackage{verbatim}
\usepackage{chngpage}
\usepackage{rotating}
\usepackage{lscape}
\usepackage{fancyhdr}
\usepackage{lmodern}
\usepackage{tabulary}
\usepackage{bm}
\usepackage{multirow}
\usepackage{rotating}
\usepackage[left]{eurosym}
\usepackage[title]{appendix}
\usepackage{geometry}
\renewcommand{\baselinestretch}{1.1}
\usepackage{array,multirow}
\usepackage{hhline}
\usepackage{adjustbox}
\usepackage{rotating}


\usepackage{adjustbox}
\usepackage{enumerate}
\usepackage{wrapfig}
\usepackage{setspace}
\usepackage{multirow}
\usepackage[super]{nth}
\usepackage[colorlinks=true,allcolors=blue]{hyperref}
\usepackage{booktabs}
\usepackage{siunitx}
\usepackage{threeparttable}
\usepackage{mathpazo}
\usepackage{pdflscape}
\usepackage{needspace}
\usepackage{dsfont}
\usepackage{subcaption}
\usepackage{bm}
\usepackage{setspace}
\usepackage{comment}
\usepackage{siunitx}
\usepackage{dcolumn}
\newcolumntype{d}[1]{D{.}{.}{#1}}
\usepackage{ragged2e}

\begin{document}  
\addtolength{\footskip}{5cm}

\begin{titlepage}

\title{\vspace{-2cm}\Huge Can Digital Aid Deliver During Humanitarian Crises? \thanks{Callen: LSE, CERP and CESifo. Email: \href{mailto:m.j.callen@lse.ac.uk}{m.j.callen@lse.ac.uk}. Fajardo-Steinh{\"a}user: LSE. Email: \href{mailto:m.fajardo-steinhauser@lse.ac.uk}{m.fajardo-steinhauser@lse.ac.uk}. Findley: University of Texas Austin. Email: \href{mailto:mikefindley@utexas.edu}{mikefindley@utexas.edu}. Ghani: Washington University in St. Louis. Email: \href{mailto:tghani@wustl.edu}{tghani@wustl.edu}. We thank Ilaha Eli Omar, Hila-Nawa Alam and the staff at Uplift Afghanistan Fund, Mukhtar Sabri and the staff at CDDO, Sanzar Kakar, Nigel Pont and the staff at HesabPay, and Kathleen McGowan for collaboration in program design and implementation. Cedric Antunes and Michael Cowan provided excellent research assistance. Shahim Kabuli provided excellent field management. Noam Angrist, Tim Besley, Robin Burgess, Michael Denly, Arif Husain, Daniel Nielson, Rohini Pande, Lamar Pierce, Soledad Prillaman, Simone Schaner, Jacob Shapiro, Abu Siddique, Garima Siwach, Rebecca Wolfe, Noam Yuchtman and seminar participants at the ACES summer school, the ADB and Yale conference, Brookings, CEPR/TIME/TCD at Trinity College, HBS Tech for All Lab, Innovations for Poverty Action/Northwestern, the Institute for Fiscal Studies, King's College, John's Hopkins School of Advanced International Studies, The University of Maryland, Washington University in St. Louis, Wharton Management in Emerging Markets Conference, and the World Bank shared insightful feedback. Research funding was provided by the UK Foreign, Commonwealth \& Development Office, awarded through the J-PAL Crime and Violence Initiative. The authors declare no competing interests. The research protocol was approved by the London School of Economics' Institutional Review Board (study number 89546).}
\vspace{-5mm}}
\author{\large Michael Callen
\\ Miguel Fajardo-Steinh{\"a}user 
\\ Michael G. Findley
\\ Tarek Ghani}
\date{\vspace{-0.3cm}\today}
\maketitle
\vspace{-5mm}
\begin{abstract}
Can digital payments help reduce extreme hunger? Humanitarian needs are at their highest since 1945, aid budgets are falling behind, and hunger is concentrating in fragile states where repression and aid diversion present major obstacles. In such contexts, partnering directly with governments is often neither feasible nor desirable, making private digital payment platforms a potentially useful means of delivering assistance. We experimentally evaluated digital payments to extremely poor, female-headed households in Afghanistan, as part of a partnership between community, nonprofit, and private organizations. The payments led to substantial improvements in food security and mental well-being. Despite beneficiaries' limited tech literacy, 99.75\% used the payments, and stringent checks revealed no evidence of diversion. Before seeing our results, policymakers and experts are uncertain and skeptical about digital aid, consistent with the lack of prior evidence on digital payments for humanitarian response. Delivery costs are under 7 cents per dollar, which is 10 cents per dollar less than the World Food Programme's global figure for cash-based transfers. These savings can help reduce hunger without additional resources, demonstrating how hybrid partnerships utilizing digital payment platforms can help address grand challenges in difficult contexts.

\bigskip
\end{abstract}
\end{titlepage}
\pagebreak \newpage

\pagenumbering{Roman}
\setcounter{page}{1} \addtolength{\footskip}{-5cm}
\pagestyle{fancy}
\renewcommand{\headrulewidth}{0pt}
\fancyhead{}
\fancyfoot{}
\fancyfoot[C]{\thepage}
\fancypagestyle{plain}{%
\fancyhead{}
\fancyfoot{}
\fancyhf{} 
\fancyfoot[C]{\thepage} 
}
\newcommand{\citedouble}[4]{(\citealp{#1}, #2; \citealp{#3}, #4)}
\newcommand{\citetriple}[6]{(\citealp{#1}, #2; \citealp{#3}, #4; \citealp{#5}, #6)}

\defcitealias{oecd/2022}{OECD 2022}
\defcitealias{fcdo/2023}{FCDO 2023}
\defcitealias{townsend2021future}{Towsend et al. 2021}
\defcitealias{ghani2022predictable}{Ghani and Gordon 2022}
\defcitealias{grfc2023}{FSIN 2023}
\defcitealias{wfp2023b}{WFP 2023b}


\pagenumbering{arabic}
\setcounter{page}{1} 
\renewcommand{\baselinestretch}{1}
\interfootnotelinepenalty=0

\section{Introduction}

Food insecurity is a systemic global challenge where policy obstacles are multiplying \citep{george2016understanding}. Despite progress over the last half-century, global hunger levels have set new records in each of the last three years and most households experiencing food security crises now live in fragile and conflict-affected states like Afghanistan, Democratic Republic of Congo, and Yemen \citepalias{fcdo/2023, grfc2023, townsend2021future}.\footnote{In 2022, 11 countries - Afghanistan, the Central African Republic, Chad, Democratic Republic of the Congo, Haiti, Nigeria, South Sudan, Sudan, Somalia, Syria, and Yemen - accounted for over 50 percent of the quarter billion facing acute hunger and potential starvation (those with a food insecurity score of 3 or higher among the roughly 1.350B people analyzed) \citep{grfc2023}.} In such settings, oppressive state and non-state actors often seek to control resource flows and can restrict humanitarian access to the most vulnerable populations, raising donor concerns that aid will be diverted from intended beneficiaries \citep{Kurtzer2019, cliffe2023aid}. In such settings, humanitarians face a dilemma: either deliver aid and risk supporting hostile actors and exacerbating conflict \citep{nunn2014us} or suspend operations in the face of urgent needs. Moreover, global aid budgets are failing to keep pace with growing needs, creating pressure to reduce the costs of delivering aid while ensuring it reaches the intended beneficiaries directly.\footnote{Humanitarian food security budgets fell 40\% from USD 85 per person in 2018 to USD 51 per person in 2021 \citep{grfc2023}.}

Can digital payment platforms enable more effective humanitarian crisis response? Humanitarians increasingly transfer physical cash to vulnerable individuals, but that modality is expensive and logistically challenging, subject to diversion by hostile actors that donors do not want to support, and requires in-person contact that is increasingly denied to marginalized populations.\footnote{For example, the share of the World Food Program's (WFP) assistance delivered as cash-based transfers (rather than in-kind aid contributions) rose from 2\% in 2010 to 34\% in 2021 \citep{mcdonough2022}, a move supported by research evidence \citep{jeong2022cash}.} Relative to cash, digital payment systems create value by leveraging scalable technology and integrating private firms to facilitate transactions \citep[cf.][]{dodgson2015managing, wormald2021david,suri2023mobile}.\footnote{The United Nations Better than Cash Alliance defines a digital payment as `the transfer of value from one payment account to another using a digital device or channel. This definition may include payments made with bank transfers, mobile money, QR codes, and payment instruments such as credit, debit, and prepaid cards.' In this study, we examine a specific digital payments platform in Afghanistan, HesabPay, that is available as an app on iOS and Android, and is also accessible to feature phone users via USSD or combined with a QR code card.} Delivering aid digitally offers the potential to reduce coordination costs and delays, increase transparency for donors and preserve privacy of beneficiaries, and utilize local supply chains without relying directly on local authorities \citep{byrd2023,waal2023}. While digital payment platforms are widely utilized by development programs for poverty alleviation \citep{muralidharan2016building, suri2016long}, adoption by humanitarians has been slowed by market and institutional frictions in fragile states \citep{williamson2000new, dorobantu2017nonmarket, wfp2021} -- particularly given limited evidence on the efficacy of digital aid to assist vulnerable, hard-to-reach populations \citep{AkerEtAl2016,PazarbasiogluEtAl2020,Gentilini2022}.\footnote{\citet{gd2024} catalogs the vast literature on cash transfers, including important contributions by \citet{haushofer2016short,bedoya2019no, banerjee2020effects,MuralidharanEtAl2021, mcintosh2022using}. While prior cash transfer studies focus mainly on sustained poverty reduction (i.e. SDG 1), we are concerned with basic human survival (i.e. SDG 2); online Appendix~\ref{RelatedWork} discusses prior cash transfer studies focused on humanitarian goals.}

In this study, we conduct a randomized evaluation to study the impacts of a digital aid program in Afghanistan. Specifically, we compare a treatment group receiving digital aid payments against a comparison group which does not receive aid until later. The program was a hybrid partnership involving community, nonprofit, and private organizations. Locally elected community councils identified extremely poor, tech-illiterate, female-headed households in three cities: Kabul, Herat, and Mazar-i-Sharif. Working with an Afghan digital payments platform and a nationwide Afghan NGO, a U.S.-based nonprofit organization transferred digital value vouchers to beneficiaries' mobile phones that could be spent at participating local private merchants.\footnote{These value vouchers have a denominated local currency value redeemable at participating merchants for any available goods. \citet{wfp2023d} treats value vouchers as an official form of a cash-based transfer, unlike commodity vouchers, which are both tied to specific merchants and goods.} The intervention transferred 45 USD payments every two weeks (the equivalent of 3.75 months worth of average monthly household income in the sample) for two months.\footnote{The Afghanistan Cash and Voucher Working Group (CVWG) estimates a Basic Food Basket for a family of seven costs approximately 96 USD per month \citep{Bete2022}, roughly equivalent to two biweekly direct aid transfers of 4000 AFA and a monthly survey incentive of 350 AFA. This basket was composed of 89 kg wheat flour, 21 kg domestic rice, 7 liters vegetable oil, 9 kg pulses, and 1 kg salt at prevailing exchange rates in August 2022.} The experiment included 2,409 beneficiary households, randomly assigned to treatment and control groups. Given our study's focus on relieving human suffering, we make use of appropriate methods to measure key outcomes and to avoid false positives, including rigid adherence to a Pre-Analysis Plan (PAP).\footnote{The Pre-Analysis Plan is registered \href{https://www.socialscienceregistry.org/versions/160809/docs/version/document}{\textit{here}}.} Crucially, the program leveraged three enabling conditions: widespread access to mobile phones, the digital platform's robust merchant network, and sufficient market availability of food products to meet demand. 

Our experiment yields three results on digital aid's efficacy. First, despite low levels of education and literacy, 99.75\% of our sample successfully used their digital payments, and about 80\% would not be willing to pay a 2.5\% fee to have a physical ``cash out'' option. Second, digital payments improved all prespecified measures of nutritional well-being, with an index of these measures increasing by 0.5$\sigma$ (SE = 0.032; $p$ $\leq$ 0.0001) and all prespecified measures of mental and financial health, with an index of these measures improving by 1.5$\sigma$ (SE = 0.042; $p$ $\leq$ 0.0001). Last, rigorous tests reveal no evidence of diversion either from beneficiaries or from merchants accepting digital payments. Our conservative estimates of the cost of delivery -- including all aspects of recruitment and facilitation -- are 6.7 cents per dollar, or less than 40\% of the \citet{wfp2023b} global benchmark figure of 17 cents per dollar for cash-based transfers.

Our results exceeded the expectations of experts in this field as elicited prior to the release of the study. To benchmark beliefs, we provided a precise description of the intervention to 55 international analysts, practitioners and policymakers working on global development issues (including many engaged specifically on hunger in Afghanistan) and 36 academics, many with expertise related to digital payment systems. Asked to predict key outcomes of the study, respondents were skeptical and uncertain. On average, they predicted only 43\% of women in our study would be able to use digital payments (SD = 25.67); in practice, 99.75\% of women purchased goods at least once. Likewise, they predicted, on average, that local authorities would attempt to tax roughly 40\% of beneficiaries (SD=27.28); in practice, less than 2\% of the treatment group reported any diversion attempts, with no statistically significant difference in the control group.

Our work builds on two extensive bodies of social science research on cash transfers -- one documenting the usefulness of digital transfers to achieve development outcomes \citep{BastagliEtAl2019,gd2024} and the second focused mainly on the benefits of distributing physical cash to vulnerable populations during humanitarian crises \citep{jeong2022cash}. While the former literature in development contexts often takes the existence of robust digital payment platforms for granted, the latter literature in humanitarian crises rarely engages the private sector as a delivery partner. In Online Appendix~\ref{RelatedWork}, we systematically review the intersection of these two literatures and find limited evidence regarding digital payments to hard-to-reach populations in fragile states during humanitarian crises -- the closest digital aid papers examine comparatively stable settings including refugee camps and natural disaster response \citep[cf.][]{deHoopEtAl2019, MercyCorps2022}. Our primary contribution is to provide clear proof-of-concept that digital transfers can cost-effectively address humanitarian needs for highly vulnerable, hard-to-reach groups while avoiding diversion. In exploring the efficacy of digitization for humanitarian response, we echo the emphasis of \citet{george2021digital} on both technological innovation and ``developing business models that infuse innovations with new purpose.'' Indeed, the digital payments platform utilized in this study was developed prior to the Taliban's 2021 return to power with the original intent of serving a growing Afghan middle class, and only afterwards shifted its focus to assist in aid delivery to the most vulnerable citizens.

Our work also contributes to a nascent strategy-related literature on grand challenges affecting large populations in difficult contexts \citep[cf.][]{ballesteros2017masters, mcgahan2023there, fangwa2024governance}. Scholars are increasingly focused on organizational approaches to grand challenges by analyzing the comparative strengths of public and private actors in addressing social issues \citep{luo2019private, george2024partnering}. The hybrid arrangement we study in this program -- involving community, nonprofit, and private organizations -- differs from most previously documented responses to global challenges in that government was not an active partner given donor and humanitarian concerns about engaging the Taliban authorities. This program relied on community councils created by the previous government, which in turn notified  local Taliban authorities to ensure the safety of operations. Limiting the Taliban's opportunities to divert aid was a central motivation for digital aid delivery, and we make considerable methodological efforts to measure diversion rigorously. Given increasing awareness that many of the Sustainable Development Goals will not be met in 2030 absent major innovations in fragile states \citepalias{ghani2022predictable, oecd/2022, fcdo/2023}, our findings indicate one hopeful pathway -- technological and organizational -- for the private sector to help address growing global humanitarian needs. 


\section{Research design} 

We develop, implement, and evaluate an approach to delivering digital transfers to a vulnerable, hard-to-reach population during Afghanistan's humanitarian crisis. We briefly describe the research design in this section, with the supplementary materials providing more details. Specifically, Online Appendix~\ref{main_appendix} describes the context, implementation, estimation strategies, deviations from the PAP and benchmarking of the estimated treatment effects. In addition, Online Appendix \ref{ss:costanalysis} discusses the cost-effectiveness and cost-efficiency estimation and Online Appendix \ref{ss:ethics} discusses ethical considerations. 

\subsection{Theory of change}

In principle, digital payments should enable the transfer of value to vulnerable beneficiaries who can then purchase food and thus alleviate their household's food insecurity. However, several enabling conditions must first be met: vulnerable households must have access to mobile phones, enough merchants must accept digital payments to allow convenient and competitive shopping, and the market must have enough goods to meet demand. Even if these conditions are met, other obstacles might still limit the effectiveness of digital aid: illiterate Afghan women may have difficulty utilizing digital payments, local merchants may find digital transactions overly cumbersome, and the Taliban authorities may decide to interfere. 

Following an extensive piloting process, this digital aid program was designed with the goal of ensuring a successful implementation despite the challenges posed by the context and population. Beneficiaries were selected through locally-elected Community Development Councils (CDCs), who informed local Taliban authorities of their activities but denied them influence over participant selection. Tech illiteracy was addressed through a careful onboarding process described below, which also ensured that merchants were well-prepared to meet the needs of this population. Finally, research procedures were put in place to ensure high quality data and valid inference procedures to minimize the risk of false positives.

\subsection{Partners}

The program was a hybrid arrangement designed to minimize interference by Taliban authorities and address common issues facing digital transfers -- including technical issues like interoperability and smartphone ownership and social issues like tech literacy and trust. The research team composed of academics designed all research-related components of the project. Locally-elected Community Development Councils (CDCs) identified female-headed households as program beneficiaries through community-level, participatory meetings and facilitated their engagement in the program. Uplift Afghanistan, a U.S.-based nonprofit, received grant funding from a private foundation and managed transfers to beneficiaries. Lastly, HesabPay, a digital payments platform that was compatible across mobile network operators, facilitated transactions through its network of private merchants.  This division of labor is consistent with \citet{luo2019private}, who argue for-profit firms are best-positioned to innovate (as in the case of building a digital payments ecosystem), self-governing collectives have a comparative advantage in private ordering (as in determining which potential beneficiaries to prioritize), and nonprofit organizations are well-suited for fiduciary roles. All partners shared a common interest in testing whether digital payments were a viable and attractive channel for humanitarian aid. 

\subsection{HesabPay's technology and merchant network}

HesabPay is a digital payments platform that is interoperable across Afghanistan's mobile networks and transfers value using either a smartphone app or via a feature phone using a combination of transaction initiation via QR code and transaction verification via feature phone USSD.\footnote{HesabPay was founded in 2016 by entrepreneur Sanzar Kakar. For more details, see the Algorand Foundation case study \href{https://algorand.co/case-studies/how-hesabpay-became-the-first-and-only-interoperable-digital-payments-platform-in-afghanistan}{\textit{here}}.} HesabPay uses the Algorand blockchain as a settlement layer, facilitating finalization and recording of digital transactions; HesabPay users are automatically registered for a custodial Algorand wallet (e.g. HesabPay holds their private keys on their behalf) and all transactions are automatically recorded on the Algorand blockchain without disclosing user identities. HesabPay is licensed by Afghanistan's Central Bank as a financial service provider, similar to the regulatory framework for HesabPay's main competition from Afghanistan's mobile money operators.  

At the time of the study, HesabPay had an active and growing local merchant acceptance network in Afghanistan's major cities, composed of over one thousand pre-existing Afghan businesses providing basic foodstuffs and other household items for sale in the same neighborhoods as beneficiaries. In this program, aid payments took the form of a digital value voucher denominated in local currency that could be exchanged for any available goods from any HesabPay enrolled merchant. All beneficiaries conducted a test purchase with a nearby merchant immediately following their onboarding session. While many chose to return to that specific merchant for future purchases, we confirm broader engagement with the acceptance network using beneficiaries' transaction data from the digital payments provider. \ref{MerchByNahia} reports the total number of merchants serving beneficiaries in the transaction data, which is consistently larger than the number from onboarding sessions.\footnote{Note that this likely underestimates the true number of merchants in any given area, since i) we only contacted numbers with which our beneficiaries transacted more than 10 times, ii) among those numbers with which our participants transacted more than 10 times, we couldn't reach all of them, and iii) there are more merchants in the areas that accept the digital vouchers but our participants did not visit.} \ref{fig_mercLoc} maps the locations of these merchants for each of the three cities in which we conducted the intervention, visually demonstrating the decentralized distribution network.

\subsection{Beneficiary identification and onboarding} 

Our goal was to identify about 2400 vulnerable women in three Afghan cities (Kabul, Herat and Mazar-i-Sharif) to be part of the intervention. To do so, we worked with the Community Driven Development Organization (CDDO), an Afghan organization that assists CDCs in a wide-array of local activities.\footnote{The CDCs were established through local elections as part of the National Solidarity Program starting in 2004 \citep{beath2016electoral, beath2017direct, beath2013empowering, beath2017can, beath2018elected}, where their primary job was to oversee block grants of development funding, and they were given a much broader range of local administrative authorities under the Citizens' Charter, starting in 2016.} Local CDCs identified potential beneficiaries through a well-being analysis in which community members, elders and mullahs together categorize all community households into different socioeconomic groups (e.g., well-off, middle income, poor, very poor). Our participants come from the lowest group. Thus, participants are identified through a process relying on the community deliberation regarding who is most vulnerable. We note that ownership of a feature phone was not a binding constraint for any beneficiaries, reflecting high nationwide levels of mobile phone ownership.\footnote{Mobile phone ownership in Afghanistan has grown rapidly over the past two decades, from approximately 25,000 subscribers in 2002 to over 22 million subscribers in 2021 \citep{wbdata}. In a nationally-representative survey, 91\% of respondents reported at least one member of their household owned a mobile phone (66\% of respondents report personally using a mobile phone), while 46\% of that subgroup reported having an internet connection \citep{asiafoundation2019}.}

After identification, participants were invited to local onboarding sessions where four activities took place.\footnote{These took place during September 2022 in each of the three cities, with each session having between 24 and 80 participants.} First, the program was described to them. Second, if they wanted to participate, their informed consent was collected and a baseline survey was administered. Third, each woman opened an account with HesabPay, an Afghan commercial digital payments platform. Finally, each woman completed a test purchase with a nearby private merchant using the platform.\footnote{During pilots we conducted prior to the actual intervention, it became clear that participants needed help at first using the digital payments platform, as most of them were illiterate and had no prior experience with mobile money. This is why we decided to use the onboarding sessions to both help participants create their accounts and resolve any remaining questions and also to have participants complete a test purchase with a nearby merchant using the mobile payment platform.}  The CDDO, working together with local CDCs, ensured the safety of staff and beneficiaries by informing local authorities and maintaining independence from external interference in the beneficiary selection process.

\subsection{Intervention \& Randomization}
Uplift Afghanistan transferred digital value vouchers to beneficiaries' mobile phones that were redeemable at HesabPay's acceptance network of local merchants for any available goods. Specifically, the intervention transferred 4,000 AFN (approximately \$45 USD) every two weeks for two months to households in 16 urban neighborhoods in Kabul, Herat, and Mazar-i-Sharif. From our experimental sample of 2,409 households, we randomly assigned 1,208 households to an ``early'' group, which received benefits from November 6, 2022 until December 31, 2022 (henceforth, treatment group). The remaining 1,201 households formed the ``late'' group, which received benefits from January 1, 2023 (two weeks after the ``early'' group stopped receiving payments) until February 28, 2023 (henceforth, control group). The randomization was stratified on two variables: The nahia (neighborhood) in which they registered (our study includes beneficiaries from neighborhoods), and a measure of vulnerability.\footnote{Participants in the same nahia had access to the same merchant acceptance network, motivating this decision. We discuss the stratification procedure in detail in our pre-analysis plan.} As can be observed in \ref{baseBalCheck}, the treatment groups are balanced on 17 of the 18 outcome and heterogeneity variables we collected at baseline.\footnote{As we note on P. 15 of the PAP, with 18 balance variables and a 5\% significance level, ``we would expect to see p-values of less than 0.05 in 2 of the variables.'' We thus committed to implement the first randomization draw with less than two variables with a p-value below 0.05. This condition was met on the first random seed we tried, resulting in one unbalanced variable out of 18.}

While all participants understood they would eventually receive transfers, randomization took place after all onboarding sessions were completed and we only informed participants in both groups that they were going to start receiving their payments a few days before their first payment. Note that this is not a setting where consumption smoothing was feasible given widespread and acute hunger.\footnote{We pre-specified (on P. 13 of our Pre-Analysis Plan) that our sample would need to be credit constrained in order to interpret differences between our treatment and control groups.  Empirically, we asked if women could borrow - only 0.29\% of our sample indicated that they could - and we found no evidence of borrowing against future payments in the control group.} We identify causal impact by using the late group as a control for the treated early group during the two months in which the early group was receiving its payments and the late group was not receiving any payments. The study design and CONSORT flow diagram are depicted in Figure \ref{intFig}. Online Appendix \ref{ss:ethics} details the ethical considerations addressed in the design and implementation of the study.\footnote{We obtained Institutional Review Board approval on 4 May 2022 from the London School of Economics (\#89546). During the study, we submitted an amendment with plans to carry out a survey of experts (20 November 2022), which the IRB approved (Study \#145636 on 29 November 2022). A collaboration of practitioners, local grassroots organizations, the digital payments provider, and academics co-designed the study. Local- and internationally-based Afghans either led or worked with each of these collaborative organizations and fully participated in all decision-making, helping to ensure representation of the views of the participants, sensitivity to possible risks, and fair distribution of the program's benefits and costs. The study went through a due diligence phase of several months in which the team met weekly to assess the feasibility of implementing the program ethically. After launch, the entire team continued to meet every week to assess progress and implement any changes deemed necessary. The team was committed to early termination of the program, the evaluation, or both, if adverse events were to occur. Notably, although the research team conducted multiple rounds of surveying, the main NGO partner also conducted its own internal evaluations, which also involved interviews and surveys with participants, which they reported as independent checks on the research team's evaluation. See Online Appendix \ref{ss:ethics} for more details, including considerations specific to insecure, humanitarian crisis environments \citep{Wood2006,Campbell2017,PuriEtAl2017,Wolfe2020}.}

\subsection{Data collection}
 
Data were collected four ways. First, a baseline survey was completed during the onboarding session (see above). Second, we had access to participants' transaction data from the mobile payments provider, which we could link to participants' survey data. We obtained permission to do so during the consent process. Third, we conducted four rounds of follow-up surveys over the phone.\footnote{All surveys were conducted by female enumerators, consistent with local norms. Participants were informed that the survey was completely voluntary and would not affect their aid payments, and they could skip any question they did not want to answer.  Participants received a 350 AFA ($\sim$4 USD) payment for completing each survey (see Online Appendix \ref{ss:ethics}). Overall, we have response rates of about 99\% across survey rounds, with no difference in response rates between the treatment and control groups.} Fourth, we attempted to survey all 26 merchants who facilitated initial test transactions with beneficiaries, and succeeded in contacting 19, to confirm how beneficiaries had used their payments and to check whether they were asked to provide favors or extra-legal taxes to local authorities. 
 
We pre-specified all analyses in a PAP, including how the outcome variables would be constructed, and what our primary outcomes variables were going to be. We divided our outcome variables in three families: Basic needs (also called ``food security'' in the manuscript), well-being and informal taxation (also called ``diversion'' in the manuscript). In doing so, we tried to provide a level of detail consistent with the standard articulated in \citet{banerjee2020praise}. This standard requires that two research assistants can take the data and the PAP, and with only these two items separately produce identical analysis. For each of these three families of outcomes, we create a summary index following \citet{kling2007experimental}. 

\subsection{Methodology}

To estimate treatment effects, we use linear regressions (see Online Appendix~\ref{analysis}  for estimating equations). We study the intervention's effect by regressing the pre-specified outcome variables on an indicator for treatment assignment (baseline analysis). We control for the outcome variable's baseline value (if available), strata fixed effects, and survey round fixed effects. Standard errors are clustered at the level of treatment assignment (the participant). Note that due to very high response rates and no compliance issues, we do not conduct treatment-on-the-treated regressions. To avoid false discoveries due to multiple hypothesis testing, we control for the Family Wise Error Rate (FWER) for each family of primary outcomes and control for the False Discovery Rate for secondary outcomes as pre-specified in our PAP. 
 
We conduct several complementary analyses as well, which were also pre-specified. First, we study how the effects change over time by including an interaction term between the treatment assignment indicator and an indicator for the second survey round. We also study dynamic effects graphically by computing the means of given outcome variables at the survey week level for treatment and control groups separately, as shown in Figure~\ref{fig_longTerm}. Second, we study whether the effects are heterogeneous along several (pre-specified) variables, by interacting each of these variables with the treatment assignment indicator.
 
We check whether experimenter demand effects are present by randomly informing participants about the goals of the study (see Online Appendix~\ref{id_threats} for details). We analyze whether providing information about the goals of the study impacted survey responses in two ways: regressing each outcome on an indicator for the informational assignment indicator, and including an interaction term between an informational assignment indicator and the treatment assignment indicator (to see whether the information affects participants in the treatment and control groups separately).

\section{Results}

We organize our main results around the following primary Research Questions (RQs):
\begin{itemize}
\item RQ1: Can extremely poor, tech-illiterate women use digital transfers in Afghanistan?
\item RQ2: Do digital transfers improve food security and mental well-being?
\item RQ3: Are digital transfers diverted?
\item RQ4: How much does it cost to deliver digital aid?
\end{itemize} 

\subsection{Can extremely poor women use digital transfers?}

Can those in need use digital payments? This is a fundamental question for less technology-literate samples such as ours, where 63.3\% of the women  have no schooling and 33.9\% have at most primary education.\footnote{Using the 2015 DHS, we compare our sample to a representative sample of similarly-aged women in urban areas of Kabul, Balkh and Mazar. The DHS sample has higher educational attainment: 56.8\% have no schooling and 13.9\% have at most primary education.} The results indicate high levels of usage: Nearly all of the women in our treatment group (99.75\%) used their digital payments to buy goods.\footnote{Three women in the treatment group never used the funds transferred to their accounts. Tracking efforts indicate that each migrated to new cities during the period between enrollment and the start of payments.} 

Ninety-eight percent of the total value transferred in the four payments was spent in the first eight weeks. It is important to note that the degree of success in utilization was largely the result of an effective division of labor between program partners, an emphasis on human-centered design and extensive piloting (Online Appendix \ref{pilotSec} describes the piloting process). In particular, guiding beneficiaries through a test transaction during onboarding greatly increased comfort with digital payments. 

Beyond the high levels of usage, three additional results support the argument that this tech-illiterate population can use digital transfers. First, 20.9\% of the funds were spent at different merchants than those who facilitated an initial test transaction during onboarding, indicating that beneficiaries understood they could use the payment at any participating merchant. Second, following our PAP, we also checked whether impacts on food security outcomes vary by pre-intervention need, city, marital status, age, household size, education, and whether the recipient was the primary household financial decision-maker. We found no heterogeneity in impacts, consistent with the technology being roughly equally useful for the different groups in our study (\ref{fig_heterogeneity}). 

Finally, we asked participants in a hypothetical exercise whether they would prefer to receive their full 4000 AFN payments digitally, or $4000-X$ AFN in cash, where $X$ was either 100 AFN, 300 AFN, 500 AFN. These amounts might reflect the costs required to provide aid as physical cash (see Online Appendix \ref{analysis} for details). With a conservative fee of 100 AFN (2.5\%), 80\% of participants preferred digital aid over cash; with a higher fee of 300 AFN (7.5\%), the share choosing digital aid over cash increased to $\sim$95\% (\ref{fig_cashOut}). Collectively, the results indicate that digital payments are a viable option even when people have limited experience using digital technology. 

\subsection{Do digital transfers improve food security and mental well-being?}

\noindent The payments reduced all four pre-specified measures of food security by meaningful amounts (see Table \ref{ITTsumTableAbridged}, Panel A for estimates and control means). Beneficiaries, on average, skipped meals in 0.76  fewer days per week (SE = 0.051, $p$ $\leq$ 0.0001),  children were 11.7 percentage points less likely to have skipped meals over the past week (SE = 0.012, $p$ $\leq$ 0.0001), the share of households where everyone was able to eat regularly during the prior week increased by 9.3 percentage points (SE = 0.015, $p$ $\leq$ 0.0001), and beneficiaries reduced meals of only bread and tea by 1.608 (SE = 0.121, $p$ $\leq$ 0.0001). An index of these four measures -- constructed as the average of the standardized measures following \citet{kling2007experimental} -- improved by 0.5 SDs (SE = 0.032, $p$ $\leq$ 0.0001).

Under secondary outcomes, we find evidence of a more diverse diet (see Table \ref{ITTsumTableAbridged}, Panel B for estimates and control means). With a recall period of one week prior to the survey, beneficiaries ate rice on 0.6 days (SE = 0.035, $p$ $\leq$ 0.0001), beans on 0.49 days (SE = 0.029, $p$ $\leq$ 0.0001), chicken on 0.01 days (SE = 0.006, $p$ = 0.035), and dairy on 0.05 days (SE = 0.013, $p$ $\leq$ 0.0001) more than the control group.  Participants also report an increased ability to purchase medicine when needed. Consistent with pre-registered expectations in our PAP, we do not find any increases in outside income, employment, or agency over financial decisions.\footnote{Regarding income and employment, we write in our PAP: ``due to the existing restrictions on women's liberties in Afghanistan, we believe it is unlikely that [these] will change.''}

Primary food security outcomes kept improving for the two months during which beneficiaries were paid (Table \ref{dynEffectsMain}), and more modest improvements remain for at least two months after payments conclude (Figure~\ref{fig_longTerm}). Regarding dietary diversity, estimates reported in \ref{SvLSecTEs} indicate that consumption of beans and rice is increasing over time. 

To assess if these estimated impacts are in line with what we should expect given the size of the payments, we collected data from a subsample of households on what they were purchasing (e.g., wheat flour, cooking oil, and sugar) and price data from merchants. We calculate that the cost of a basket of popular goods to support a family for two weeks is consistent with the aid payment size plus the survey participation incentive \citep{Bete2022}. While not eliminated, skipped meals declined after each payment (Figure~\ref{fig_longTerm}, Panel A). Online Appendix \ref{ss:magnitude} provides further discussion of the magnitudes of needs results. 
 
Turning to mental well-being, since the Taliban takeover in 2021, Afghanistan consistently ranks as the country with the lowest levels of happiness \citep{gallup2022b,gallup2022c}. Our participants report extremely low levels of happiness and financial health. Treatment improved all four prespecified measures of mental and financial health, with an index of these measures improving by 1.5 SDs (Table \ref{ITTsumTableAbridged}, Panel B), although from a very low base and these impacts disappear as soon as payment stops (Figure~\ref{fig_longTerm}). Specifically, beneficiaries are 33.5 percentage points more likely to report that they feel the economic situation of their household has improved compared to 30 days ago (SE = 0.011, $p$ $\leq$ 0.0001), from a base of just 4.8\% in the control group. They are also 26.3 percentage points more likely to report being satisfied with their current financial situation (SE = 0.012, $p$ $\leq$ 0.0001). The intervention also increased measures of mental well-being: beneficiaries are 28 percentage points more likely to report being very or quite happy (SE = 0.014, $p$ $\leq$ 0.0001), and report a score 1.96 higher on the Cantril Self-Anchoring Striving Scale, an established measure of life satisfaction that runs from 1 to 10 (SE = 0.068, $p$ $\leq$ 0.0001), relative to the control group.\footnote{While these results may seem unrealistically large, this is attributable to low base levels, as is seen in other surveys in contemporary Afghanistan \citep{gallup2022b,gallup2022c}.} These effects are increasing over time, with the effects after receiving 3-4 aid payments between 40\% to 100\% larger than after 1-2 aid payments (Table \ref{dynEffectsMain}, Panel B).\footnote{Our food security results complement a broader literature demonstrating mostly positive impacts, whereas our mental well-being results contribute newer insights to a more nascent literature. As discussed in Online Appendix \ref{ExperimentalResultsLit}, our review of 23 cash-based programs in humanitarian contexts found that 21 included food security outcomes, with 16 of those 21 finding any positive effect. Less attention has focused on the mental impacts of cash-based programs, with only seven of the 23 studies considering a relevant outcome and six finding any positive effect.}

Our results are robust to common concerns in Randomized Control Trials (RCTs). First, we find no evidence of treatment beneficiaries providing part of their payments to control beneficiaries or of control beneficiaries borrowing against payments they would receive in the last two months of the program (see Online Appendix \ref{id_threats}). Second, given that participants had the option of skipping questions, and to keep the sample comparable, we re-estimate the results restricting the sample to those who answered all questions relevant for the analysis. Results using this restricted sample are almost identical to those estimated on the complete sample (see \ref{ITTsumTableRest}).

Finally, survey bias in the absence of objective measures of well-being (such as anthropometric or biometric measures) is a concern, especially since participants might respond strategically if they believe this could influence their benefits. To check for potential strategic reporting by participants, we primed half of our respondents with an explicit description of the study's purpose. Specifically, in the second round of follow-up surveys, we randomly assigned individuals into two groups: a ``primed'' group hears the following statement just before the questions related to needs: ``I would now like to ask you a few questions about how you and your family are doing. The goal of the CDDO and HesabPay program is to help you and your family meet basic needs, such as buying food, and we would like to see how you are doing in this regard. We will share what we learn from interviewing participants like yourself, with international organizations who are trying to help Afghans deal with these difficult times.'' Thus, this group is explicitly told what we are expecting to find. The ``not primed'' group hears this placebo statement instead: ``I would now like to ask you a few questions about how you and your family are doing.''  We find no evidence that the prime influenced responses (see Online Appendix \ref{id_threats} and \ref{ExpDemsumTableMain}).

\subsection{Are digital transfers diverted?}

An important concern about distributing aid in fragile settings is diversion, especially to regimes with poor human rights records and to those that sponsor international terror. Reports of aid diversion across multiple fragile countries have emerged in the past year \citep{odonnell2023,unNews2023,politico2023} and have been documented extensively for decades \citep{barnett2013} with scholarly evidence suggesting that diversion prolongs conflict \citep{findley2018,nunn2014us}. This is especially a concern in Afghanistan. In July of 2023, the US House of Representatives passed a bill that would bar the Secretary of State and the USAID Administrator from giving any funds to Afghanistan if they directly or indirectly supported the Taliban \citep{tolo2023}. Reports of Taliban aid requirements and infiltration of UN assistance \citep{odonnell2023, cbs2023} are deepening these concerns, adding further pressure to cut assistance. Therefore, the humanitarian mandate to address hunger depends partly on avoiding diversion.

There are at least four ways that hostile regimes might capture aid.  First, they might influence who is eligible, including by creating fictitious ``ghost'' beneficiaries. Second, they can ask recipients to hand over aid transfers. Third, they can capture aid while in transit. Fourth, they can ask merchants or payments platforms who are serving aid beneficiaries to pay bribes or additional tax. We organize the presentation of results around these four potential diversion strategies and discuss whether digital delivery might constrain them. Such diversion strategies are widely documented both in Afghanistan \citep{special2023quarterly,sopko2023examining} and in other countries (see Online Appendix \ref{CashHumanitarianLit} for a discussion of diversion of humanitarian aid in the literature).\\  

\paragraph{Influencing eligibility:} In our study, beneficiaries were identified by local elected community councils in consultation with communities using a ``well-being analysis'' described in Online Appendix \ref{main_appendix}. The prevalence of mobile phones enables phone surveys as a means of checking that beneficiaries are indeed vulnerable, even at large-scale. Our surveys confirm that beneficiaries were quite vulnerable (see \ref{baseBalComparison}). The digital payments platform also requires Know Your Customer identity verification, which can automatically compare beneficiary names against sanctions lists to further guard against capture by hostile actors. Evidence from a diverse set of global contexts \citep{GuggenheimPetrie2022,Samii2023} and the pre-2021 Afghanistan NSP program specifically \citep{beath2018elected,BurdeEtAl2023} also indicates that most diversion occurs at national and regional levels and primarily through bureaucratic and partisan channels. At the local level, in contrast, CDCs develop transparency and monitoring mechanisms to prevent diversion, making it more challenging for local authorities to interfere in influencing eligibility.\\

\paragraph{Taxing beneficiaries:} To check for taxation, we directly asked beneficiaries whether they have been asked for informal assistance. Specifically, we asked them whether local community leaders or government officials have asked them for any kind of assistance, such as food or money. Given that participants may be reluctant to disclose doing this themselves, we first asked them whether they know someone in their community who has been asked to do so \citep{Reinikka/Svensson/2006}, and then whether they themselves had been asked. Results are presented in the first four columns of Table \ref{infTaxResults}, Panel A, which show that the treatment group does not report informal payments to authorities in larger proportion than the control group regardless of how the question was framed (all coefficients are insignificant at traditional levels, and precisely estimated). When looking at the KLK Index combining the four individual questions (column 5), the coefficient is marginally significant at the 10\% level. However, this is driven by a single individual who answered yes to three of the four questions, which due to the way the index is constructed, receives an extremely high index value of 34 standard deviations. Removing this observation leads the results using the KLK Index to lose statistical significance at conventional levels (column 4 in \ref{infTaxResults_index}). 

We find very low levels of informal taxation: Overall, only 27 beneficiaries in the treatment group answered yes to any of these four questions since payments started, compared to 21 beneficiaries in the control group. When using an indicator for whether the respondent answered yes to any of the four informal taxation questions the difference is statistically insignificant (Table \ref{infTaxResults}, column 6).\footnote{Note that this categorical variable for whether the respondent answered yes to any of the four diversion questions was not pre-specified in our PAP. In Online Appendix \ref{ss:devPAP} we describe all deviations from the PAP.}

Estimates in Table \ref{infTaxResults}, Panel B, indicate that the results are also not increasing over time, when beneficiaries could have become more visible to local authorities. Based on qualitative debriefs with a subsample of our respondents, one potential reason for limited diversion is that participants are too poor for local authorities to ask them for payments. Moreover, it is politically and logistically costly to tax vulnerable beneficiaries after they have received their payments.

Other surveys in Afghanistan indicate that significant shares of Afghans are comfortable reporting corrupt behaviors, at least before the Taliban takeover \citep{asiafoundation2019}.  Even so, given the authoritarian context and the fact that these are questions that may be sensitive for participants \citep{BlairEtAl2020}, there could be a lack of positive responses due to fear of reporting inappropriate behavior by local authorities. We therefore conducted a list experiment, an established method to measure the presence of sensitive behaviors, with beneficiaries to gauge the extent of informal taxation (see Online Appendix \ref{analysis} for details). Table \ref{infTaxResults}, Panel C, shows the results of the list experiment. Regardless of whether we analyze the whole sample or either of the two treatment groups, individuals who receive the longer list, which includes the informal taxation statement, do not report experiencing more items on the list than those receiving the shorter list. The effects are precisely estimated and small in magnitude, supporting the argument that diversion was minimal.    

We also have access to the beneficiaries' transaction data from the digital payments provider, which coupled with the fact that they could not cash out their payments and that they spent almost all the money they received, provides a clear picture of how beneficiaries spent their funds. Beneficiaries spent 74.2\% of their funds at the merchants they visited during the beneficiaries' onboarding sessions, 20.9\% at other registered merchants, 3.5\% at individually-registered accounts, and 0.1\% on airtime purchases. Combined with the fact that almost no respondents indicate providing assistance to local authorities, this also supports the argument that there was little diversion. Moreover, only 6.7\% of beneficiaries reported that someone else decided how to use the aid payments -- all of whom were other household members.  

\begin{figure}[h!]
\begin{center}
    
\begin{subfigure}[t]{0.75\textwidth}
\includegraphics[width=\textwidth]{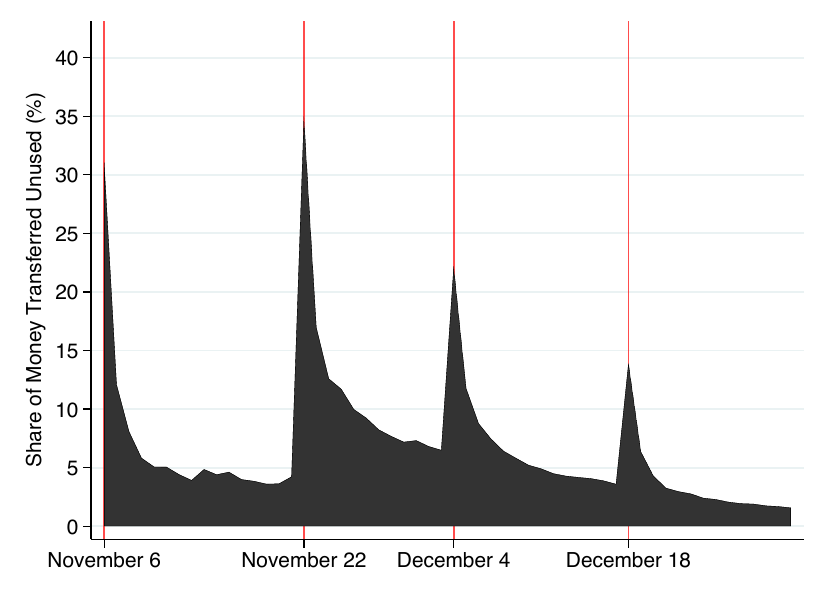}
\subcaption{Share of Money Unused}
\end{subfigure}
\begin{subfigure}[t]{0.75\textwidth}
\includegraphics[width=\textwidth]{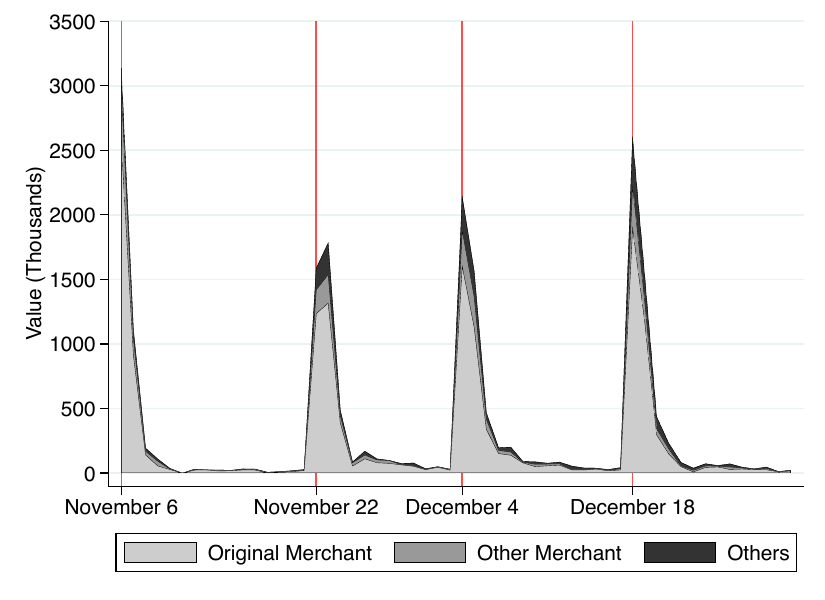}
\subcaption{Transactions Flow}
\end{subfigure}

\caption{Account Usage Over Time}
\label{accUsageOverTime}

\end{center}
\noindent\footnotesize{\textbf{Notes:} Panel A shows the cumulative share of money that participants in the treated group have not spent. Panel B shows where participants spent their funds at. ``Original merchant'' is the merchant a participant visited during the onboarding session to conduct their test purchase. ``Other merchant'' is another account that belongs to either a different test merchant or another merchant that did not participate in the onboarding sessions.}

\end{figure}

Importantly, once beneficiaries are onboarded, distribution does not require travel outside the home as it might with food or cash distributions. The distribution of digital aid is therefore less visible than cash or in-kind distribution. Finally, the distribution of digital payments is instantaneous, with recipients able to access their funds immediately. Figure \ref{accUsageOverTime}, Panel A displays the share of the cumulative funds participants have received that remains unspent over time, and shows that participants spent over 70\% of the funds they received the day they received their first payment. This drops quickly over the next few days to below 5\% of the received funds. A similar pattern holds every time participants receive their funds. Panel B displays how participants spent their money on a daily basis.\\

\paragraph{Capturing aid in transit:} Relative to in-kind or physical cash, which rely on intermediaries that staff convoys or physical distribution points, digital payments may reduce the opportunities for in-transit theft by transferring aid directly to beneficiaries. Also, with automatically generated transaction data, donors and humanitarian agencies gain increased transparency into the delivery and utilization of cash-based assistance programs that can be used for auditing and real-time programming adjustments. The digital payments platform in this study utilizes the Algorand blockchain, so all payments and purchases were automatically recorded on an immutable ledger that facilitates external auditing. Finally, phone surveys also provide a scalable means of confirming the delivery of benefits, as also demonstrated in \citet{MuralidharanEtAl2021}.\\

\paragraph{Taxing merchants or intermediaries:} The government has the legal authority to tax both merchants and digital payment platforms. To check for merchant taxation, we surveyed 26 merchants who assisted beneficiaries with an initial test transaction to ask about extra-legal taxes (merchants do need to pay ordinary sales tax on any transactions). None of the merchants reported paying extra-legal taxes, being asked to provide favors, or being aware of other merchants paying extra-legal taxes or providing favors.  

In our study the digital payments provider reported no attempts at extra-legal taxation. This could plausibly change for a larger initiative. In this scenario, a single tax would be levied on a single provider.  Economic theory suggests a single, negotiated tax is preferable to numerous, less-organized attempts at interception, potentially reducing overall diversion \citep{shleifer1993corruption}.

\subsection{How much does it cost to deliver digital aid?}

The cost-efficiency of digital aid, or the program costs required to serve a given number of beneficiaries, outperforms comparable estimates. The total cost per beneficiary (CPB) of delivering the entire \$180 (across four disbursements) to our beneficiaries is \$2.44 USD excluding recruitment costs or \$12 USD including recruitment costs. Correspondingly, the cost to deliver a single dollar including the transferred dollar, termed the total cost transfer ratio (TCTR), is either \$1.014 (\$182.44/\$180.00) or \$1.067 (\$192.00/\$180.00). Excluding the transferred dollar, the cost-transfer ratios (CTRs) are either 1.4 cents when recruitment costs are excluded or 6.7 cents when recruitment costs are included. For comparison, the World Food Programme's (WFP) global CTR for providing cash-based humanitarian aid is 17 cents per dollar \citep{wfp2023b}.  To contextualize this difference, we estimate that if WFP had delivered all \$357M of its 2022 cash-based assistance in Afghanistan digitally, the savings would be sufficient to support an additional 77,000 households or 538,075 individuals for the four-month lean season. 

We also computed the costs required to produce a given level of program impact, a measure of cost-effectiveness. Online Appendix~\ref{ss:costanalysis} details these cost-efficiency, scale, and cost-effectiveness estimates and reports ranges of estimates based on more or less conservative assumptions. The cost-efficiency estimates compare favorably to other cash-based transfer programs in humanitarian contexts. Although cost analysis is rarely performed for humanitarian cash programs \citep{gentilini2016revisiting,jeong2022cash}, we identified three organization-specific global estimates and 12 individual studies with estimates. When cost categories are comparable, all of these organizations and studies have higher costs of delivery than our digital aid intervention. 

Digital aid offers further advantages to donors and humanitarian agencies in terms of increased decentralization and transparency compared to status quo modalities of assistance. Delivery of in-kind food or physical cash incurs transport and mobilization costs associated with each distribution, and is subject to potential diversion or disruption at key chokepoints, such as airports and border crossings. Furthermore, beneficiaries bear the time and travel costs associated with reaching physical cash distribution points and may be subject to harassment or extortion during this process. 

By contrast, digital payments platforms enable instantaneous, private transfers of value, which can then be exchanged for goods at local merchants -- or potentially ``cashed-out'' for physical currency -- without need to visit a central distribution location each time. When payments are delivered digitally like in this study, there is less need to accumulate cash at intermediate locations to facilitate beneficiary cashout, further reducing delivery costs and opportunities for diversion. This is particularly important when considering women in food-insecure settings as existing research documents substantial benefits to channeling resources to women directly \citep{field2021her, riley2022resisting}. Beneficiaries in our study used their digital transfers across the merchant acceptance network, increasing convenience and privacy while also ensuring competitive pressure on any single merchant who might otherwise seek to engage in price gouging.

\section{Experts' beliefs and methodological safeguards}

Humanitarian operations are complex and expensive, and the stakes can involve life or death. If a new approach fails, the consequences can be severe, both financially and in terms of human suffering. This may lead to a reasonable reluctance to adopt new technologies, especially if they are untested or perceived as risky. To better understand these issues, we surveyed experts, asking them to predict the treatment impacts we would later estimate. To help address the same issues, we took several methodological steps to safeguard against false positives. 

We measure the expert views by surveying 55 international analysts, practitioners and policymakers working on global development issues (including many engaged specifically on hunger in Afghanistan) and 36 academics, many with expertise related to digital payment systems.\footnote{The policy organizations represented include the USAID, the UN World Food Program, UNICEF, the World Bank, GiveDirectly, Brookings and BRAC, and the academic institutions included Brown, Duke, Georgetown, the London School of Economics, Princeton, Stanford, the University of Chicago, UC Berkeley, and UC San Diego.} The survey provided a precise description of the intervention and asked respondents to predict the corresponding treatment impacts for several key outcomes we report here; see ~\ref{expSurveyQs} for the full text of the survey prompt. 

Table~\ref{expPreds} compares the actual values of the intervention's components  (column 1) to the respondents' predicted values (column 2), and shows the $p$-value of a test of equality (column 3). Across all four components, differences between expert predictions and the values from our data are statistically significant. These respondents predicted, on average, that only 43\% of the women in our study would be able to use digital payments (SD = 25.67pp). In practice, as described above, 99.75\% of women purchased goods at least once. They also predicted, on average, that local authorities would attempt to tax roughly 40\% of beneficiaries (SD = 27.28). In practice, less than 2\% of the treatment group reported any diversion attempts, with no statistically significant difference in the control group. Moreover, respondents had little agreement in their views, reflected in the large variance across predictions. A combination of skepticism and uncertainty about potential efficacy might partly explain an unwillingness to adopt innovations, especially when stakes are high. Experts did a better job of predicting cost-efficiency, guessing on average that delivery costs would be 10.65 cents on the dollar, while the actual delivery costs were 6.7 cents on the dollar, indicating a more common belief that digital aid delivery is relatively cheap.

Because of the potential stakes, we implemented three methodological safeguards against false positives (whereby the null hypothesis that an innovation has no impact is incorrectly rejected in favor of the alternative that it is effective). First, we conducted an RCT, which provides unbiased causal evidence by establishing an independent control group. This is particularly important in humanitarian settings, where the circumstances on the ground can change quickly and often worsen. Simpler analyses, such as comparing outcomes before and after an intervention, can therefore be highly misleading \citep{Wolfe2020}. Second, we strictly followed our PAP, which, while limiting our capacity to learn and adapt, helps ensure that $p$-values are correctly calculated, providing additional protection against false positives \citep{olken/2015, banerjee2020praise}. We also adjust our estimates to take into account the multiple hypotheses we are testing following the PAP. Third, as we rely on self-reported measures and not on anthropometrics or biometric markers due to surveying limitations, survey bias and strategic responses by participants are a source of concern. In our case, we followed the literature on testing and avoiding survey response bias \citep{de2018measuring} and took the most conservative approach we could think of to test for survey bias (see Online Appendix~\ref{id_threats}).


\section{Discussion} 
\label{conclusion}

We evaluate a digital aid program and show that extremely poor, tech-illiterate women can receive humanitarian aid through digital payment systems to meet basic food security and mental well-being outcomes. We also find no evidence of diversion. Accounting for all aspects of facilitation, the estimated cost of delivery at 6.7 cents per dollar is less than 40\% of the WFP's global figure of 17 cents per dollar for cash-based transfers. Our estimates are also better than the predictions of policy and research experts. Why did digital aid outperform these expectations?

The hybrid partnership that delivered the program was carefully designed to address common issues working with poor, tech-illiterate populations in a politically fragile context. Vulnerable households were identified by locally-elected CDCs, which ensured the safety of staff and beneficiaries by informing local authorities and maintained independence from interference in the beneficiary selection process. A U.S.-based nonprofit provided fiduciary oversight. The digital payments platform was compatible across mobile operators and did not require a smartphone, ensuring beneficiaries could use existing phones to access transfers. In addition, user-centered design principles helped minimize potential confusion: most importantly, beneficiaries were trained to use the payments platform by completing a test transaction with a merchant.

We expect that whether the findings obtained from this context and population generalize to other settings will depend most critically on three key enabling conditions. First, vulnerable households must have access to phones. Second, there must be enough merchants who accept digital payments to allow convenient and competitive shopping. And third, markets must have enough goods to meet demand. Transitioning from in-kind to cash and digital support can increase competition for goods and so may increase market prices and thereby displace hunger to other vulnerable groups. Studies of famine highlight that in extreme cases of sudden and widespread deprivation -- as in Gaza at the time of writing \citep{chotiner2024gaza} -- increasing food supply is critical, but that response to most famines historically required complementary strategies to redistribute resources \citep{dreze1990hunger,sen2014development}.

Our study is motivated by the challenges facing populations who are isolated by the state in complex crisis situations like Afghanistan, where basic human survival - not sustained poverty reduction - is the immediate policy goal. Evidence on whether digital humanitarian aid can help achieve this objective is limited. Most cash-based programs are carried out in more stable contexts, contributing to a rich evidence base on outcomes like poverty reduction, education, health, financial inclusion, and social protection \citep{BastagliEtAl2019,gd2024}. 

Based largely on this research, humanitarian cash-based programming is growing and digital delivery is increasingly popular \citep{UrquhartEtAl2023}, but research during humanitarian crises with populations facing extreme cultural and political constraints is still nascent. In Online Appendix~\ref{CashHumanitarianLit}, we review the literature and find 23 RCTs of cash for humanitarian applications, 10 such RCTs in contexts categorized as ``not free'' by Freedom House, three of those 10 having some digital component, and none, that we can find, targeting populations that the government actively oppresses. This knowledge gap is consequential as these contexts and groups are increasingly affected by acute hunger \citep{fcdo2023}. As our expert survey demonstrates, our results are not readily inferred from prior studies: illiterate Afghan women could have been unable to utilize digital payments, local merchants could have refused to accept digital transactions, and the Taliban authorities could have interfered. Our study thus provides an ``existence result'' on the feasibility of digital aid in humanitarian response, with insight into the organizational and technological factors that contributed to success.

The exigencies of this context and the fact that we work with a hard-to-reach group, however, necessitated three design choices that affect our generalizability. First, our experimental data cover only two months before the early group is treated. This time frame, however, is consistent with our learning objective: can digital payments cost-effectively reduce acute hunger. As such, we pre-specified this as the relevant time frame, and relatedly that we did not expect changes in outcomes like income or employment. Such impacts were unlikely -- and did not realize -- given the large humanitarian needs, prohibitions on women's employment, and relatively modest transfer size. Second, working with our sample required approval of the \emph{de facto} authorities. Our partners could not justify seeking Taliban permission to onboard women to the digital payments app without eventually providing humanitarian assistance.  We thus opted for a short, staggered program where every participant received some assistance during the lean season. While the control group knew they would receive transfers, they were unaware when these would start and we find no evidence of borrowing or other anticipatory changes in behavior. Our late group, thus, provides a valid counterfactual. Third, because we could not visit our participants, and to avoid survey fatigue over the phone, we focused on a small set of questions and could not take measurements that would yield deeper insights (such as anthropometrics). Thus, other relevant topics (e.g., within household dynamics, validated mental health measures, and impacts on other household members) were not covered. 

Nonetheless, this exercise provides proof-of-concept that digital aid represents a potential cost-effective complement to existing modalities. We see five important directions for future research to help humanitarians select strategies to address hunger. First, relative to physical goods and cash, digital payments enable humanitarians to more easily vary when aid is delivered, its frequency, and the amount. Research can help identify how to optimally take advantage of this added flexibility.  Second, more evidence is needed on the costs and benefits of digital relative to cash delivery. Some prior work addresses this question \citep{AkerEtAl2016}, but we think more of this work in fragile contexts controlled by hostile actors and where the vulnerable are hard-to-reach will be relevant given the rise in hunger among such groups. This is best pursued in experiments which randomize both digital delivery and cash delivery against a control group. Third, humanitarians need further evidence on how key enabling conditions -- such as mobile phone penetration and the availability of merchants who accept digital payments -- affect these relative costs and benefits, and how best to organize future hybrid partnerships based on differences in enabling conditions. Fourth, additional focus is needed on approaches to targeting, onboarding and monitoring beneficiaries in hostile environments that minimize risks to participants \citep{aiken2022machine, JeongTrako2022}. In particular, the one point of physical contact in our study between humanitarians and beneficiaries was during onboarding. Further innovations might eliminate the need for contact between humanitarians and those in need altogether and apply automated screening of data from digital transactions and phone surveys to monitor user experience and identify potential fraudulent patterns. And fifth, implementing digital payments at scale may affect diversion strategies, market prices, and the broader use of digital payments technologies for other financial transactions. It may also spur the transformation of humanitarian organizations into leaner organizational models to best complement the capacities of private and community organizations. Future work with large-scale experiments should explore such changes. 

Encouragingly, aid agencies are increasing their use of digital payments. The WFP is using insights from our study to scale digital payments to over 100,000 households in Afghanistan in partnership with HesabPay, which has been able to maintain its license to operate despite the dynamic strategic and political conditions in the country. As this program scales, it will need to grapple with the complex interdependencies between humanitarian agencies, digital payment platforms, and governments in fragile states. While local authorities could still choose to block digital aid provision, doing so might impede future humanitarian commitments and draw increased public attention to the issues of aid diversion and vulnerable populations. The centralized management of digital payments relative to cash or food distribution also reduces the need to engage with local actors who might complicate aid delivery and increases humanitarians’ bargaining power. The existence of key enablers for the success of a program like this in other fragile settings (e.g., the presence of CDCs, mobile phone availability, and digital payment platforms), as shown in~\ref{CountryCharacteristics}, where Afghanistan ranks as a particularly challenging case, suggests that this approach could increase food security in other fragile states with high levels of hunger.


\clearpage

\section*{Tables and figures}

\begin{figure}[h!]
\centering
\includegraphics[width=0.8\textwidth]{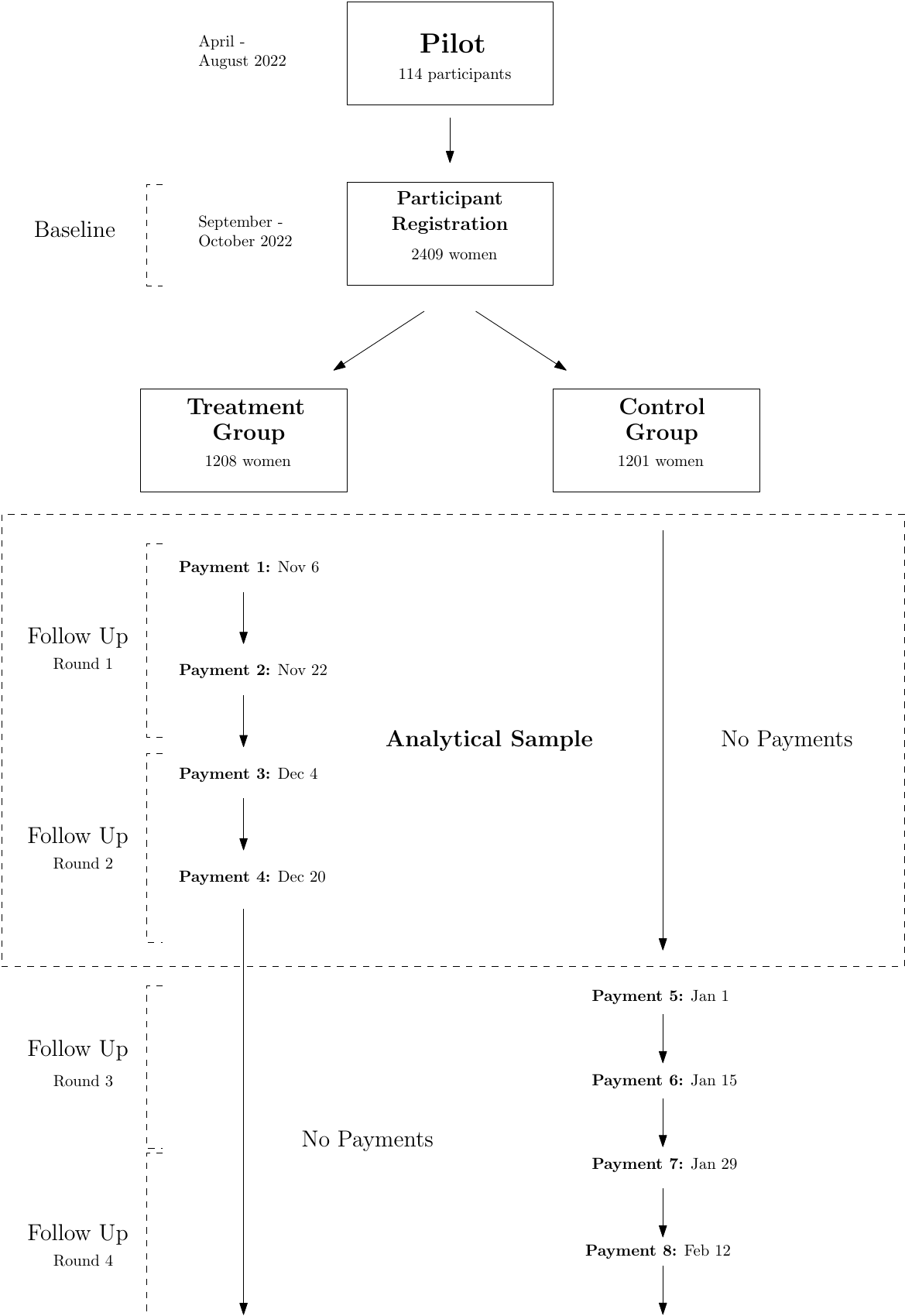}
\caption{Intervention Timeline and Design}
\label{intFig}
\end{figure}

\clearpage
\begin{figure}[h!]
\begin{center}
\begin{subfigure}[t]{0.49\textwidth}
\includegraphics[width=\textwidth]{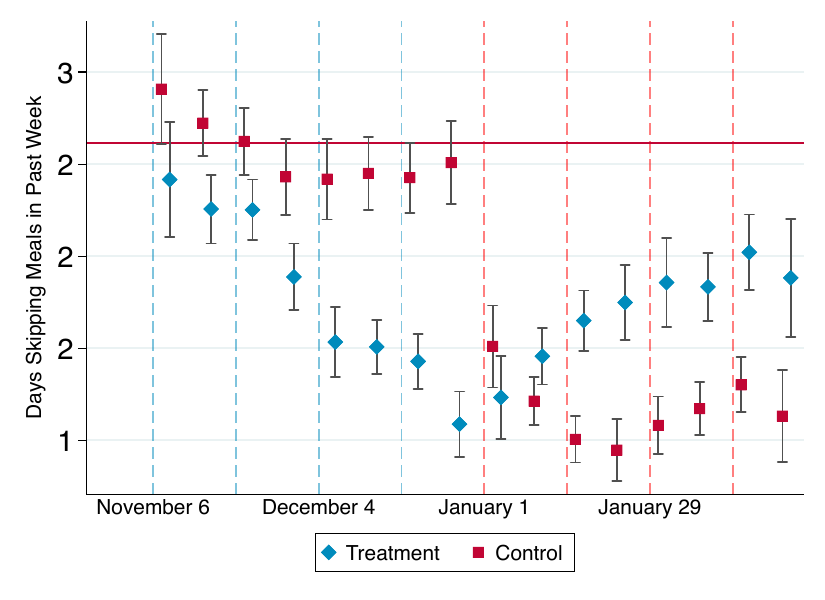}
\subcaption{Skipped Meals}
\end{subfigure}
\begin{subfigure}[t]{0.49\textwidth}
\includegraphics[width=\textwidth]{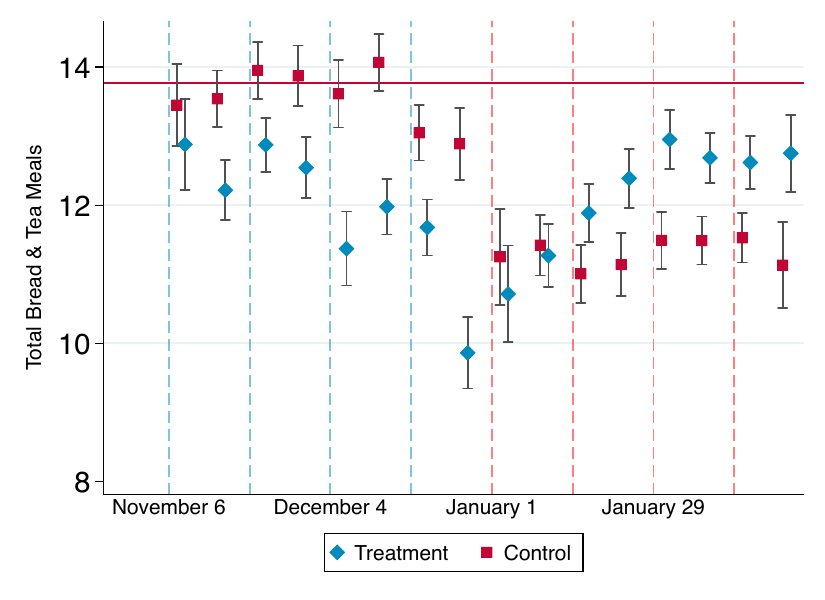}
\subcaption{Total Bread and Tea Meals}
\end{subfigure}
\begin{subfigure}[t]{0.49\textwidth}
\includegraphics[width=\textwidth]{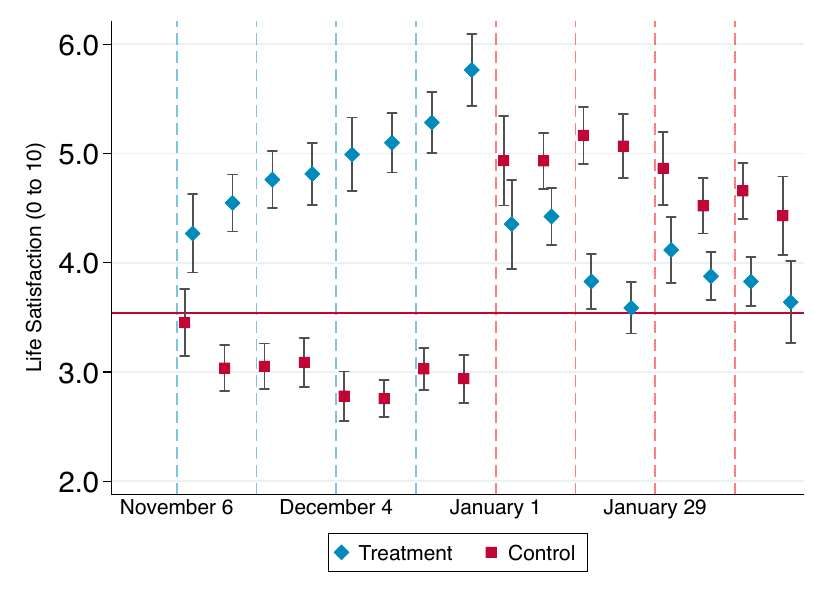}
\subcaption{Life Satisfaction}
\end{subfigure}
\begin{subfigure}[t]{0.49\textwidth}
\includegraphics[width=\textwidth]{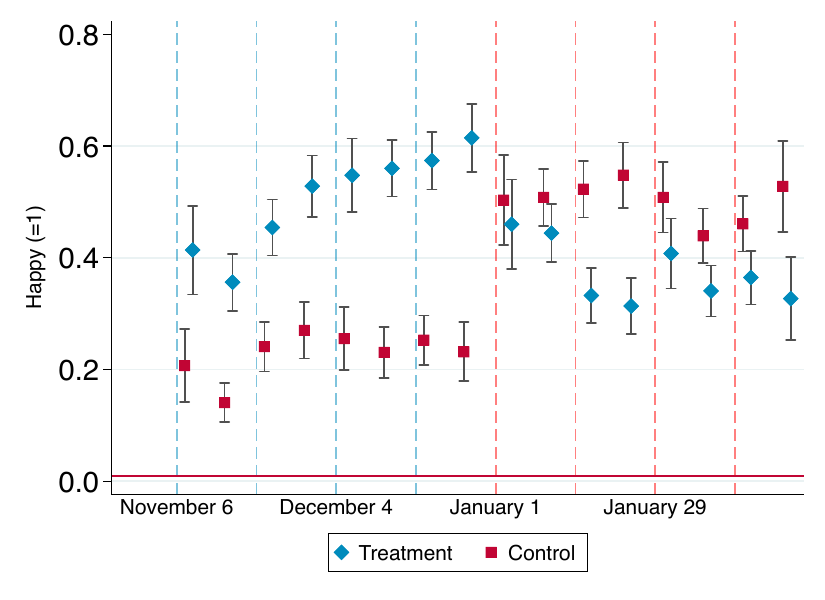}
\subcaption{Happy}
\end{subfigure}

\caption{Effects over Entire Duration of Program}
\label{fig_longTerm}

\end{center}
\noindent\scriptsize{\textbf{Notes:} The red horizontal line corresponds to the mean of the variable at baseline across all individuals in the sample. Red squares (blue diamonds) show the mean value of the variable among control group (treatment group) respondents divided in two-week bins since the start of the payments. Blue (red) vertical lines represent the dates in which treatment group (control group) participants received their payments. Note that this is the only analysis in which we are using/showing data for the period after the control group starts receiving payments (i.e. while the rest of the paper uses only data from survey rounds 1 and 2, this shows in addition data from survey rounds 3 and 4, after the control group starts receiving payments). Bars show 95\% confidence intervals.}

\end{figure}

\clearpage

\clearpage

\begin{table}[h!] \centering
\newcolumntype{R}{>{\raggedleft\arraybackslash}X}
\newcolumntype{L}{>{\raggedright\arraybackslash}X}
\newcolumntype{C}{>{\centering\arraybackslash}X}

\caption{Summary Table -- Treatment Effects}
\label{ITTsumTableAbridged}
\begin{adjustbox}{max width=\linewidth, max height=\textwidth}\begin{tabular}{lccccccc}

\toprule
& Control & Control & Treatment & Standard & Naive & Adjusted &  \tabularnewline & Mean & SD & Effect & Error & \textit{p}-value & \textit{p}-value & N \tabularnewline 
{}&{(1)}&{(2)}&{(3)}&{(4)}&{(5)}&{(6)}&{(7)} \tabularnewline
\midrule \addlinespace[\belowrulesep]
\midrule \textbf{Panel A. Primary Outcomes}&&&&&&& \tabularnewline
Days skipping meals (past week)&2.569&1.717&--0.76&0.051&0&0.0002&4761 \tabularnewline
Children skipping meals (=1)&0.873&0.333&--0.117&0.012&0&0.0002&4750 \tabularnewline
Regularly eat twice a day&0.501&0.5&0.093&0.015&0&0.0002&4736 \tabularnewline
Total bread and tea meals (past week)&13.639&3.549&--1.608&0.121&0&0.0002&4763 \tabularnewline
\textit{Food Security - KLK Index}&0&1&0.501&0.032&0&&4763 \tabularnewline
&&&&&&& \tabularnewline
Better economic situation&0.048&0.213&0.335&0.011&0&0.0002&4762 \tabularnewline
Satisfied with fin. situation&0.133&0.34&0.263&0.012&0&0.0002&4755 \tabularnewline
Happy&0.154&0.361&0.28&0.014&0&0.0002&4732 \tabularnewline
Life satisfaction&3.179&1.707&1.963&0.068&0&0.0002&4763 \tabularnewline
\textit{Economic/Wellbeing - KLK Index}&0&1&1.498&0.042&0&&4763 \tabularnewline
&&&&&&& \tabularnewline
\textbf{Panel B. Secondary Outcomes}&&&&&&& \tabularnewline
Days eating rice (past week)&0.698&1.058&0.597&0.035&0&0.001&4763 \tabularnewline
Days eating beans (past week)&0.518&0.836&0.493&0.029&0&0.001&4763 \tabularnewline
Days eating vegetables (past week)&1.394&1.346&--0.003&0.041&0.934&0.453&4763 \tabularnewline
Days eating chicken (past week)&0.021&0.147&0.012&0.006&0.035&0.037&4763 \tabularnewline
Days eating dairy (past week)&0.074&0.383&0.047&0.013&0.001&0.002&4763 \tabularnewline
Able to buy medicine&0.051&0.221&0.034&0.01&0.001&0.002&3582 \tabularnewline
Involved in fin. decisions&0.666&0.472&0.017&0.015&0.272&0.185&4757 \tabularnewline
Total household income (past month)&876.683&1581.531&128.482&90.93&0.158&0.118&4763 \tabularnewline
Household's head employed (past month)&0.192&0.394&--0.008&0.014&0.578&0.311&4741 \tabularnewline
\bottomrule \addlinespace[\belowrulesep]

\end{tabular} \end{adjustbox}
\\ \parbox{\linewidth}{\scriptsize \noindent \justifying \(Notes \): Stratification fixed effects, survey round fixed effects, and baseline values of dependent variables, if available, are included. Standard errors are clustered at the individual level. Primary outcomes show FWER-adjusted p-values within each family outcome (following Romano \& Wolf, 2005, using 5000 repetitions), while secondary outcomes show FDR-adjusted p-values (following Anderson, 2008). The KLK Index is created following Katz, Kling, \& Liebman (2007), and is the equally-weighted sum of the standardised component variables. Better economic situation is an index that equals 1 if the respondent answered that her economic situation compared to 30 days ago is slightly or much better, and 0 otherwise. Satisfied with financial situation is a dummy that equals 1 if the respondent answered that she agrees a lot or somewhat with the statement that she is highly satisfied with her current financial condition, and 0 otherwise. Happy is a dummy that equals 1 if respondent said that she was very happy or quite happy, and 0 otherwise. Life satisfaction is the score from 1 (dissatisfied) to 10 (satisfied) in terms of how satisfied the respondent is with her life as a whole these days. Total household income excludes the aid payments.}

\end{table}

\begin{table}[h!] \centering
\newcolumntype{R}{>{\raggedleft\arraybackslash}X}
\newcolumntype{L}{>{\raggedright\arraybackslash}X}
\newcolumntype{C}{>{\centering\arraybackslash}X}

\caption{Short vs. Long Treatment Effects -- Food Security \& Wellbeing Outcomes}
\label{dynEffectsMain}
\begin{adjustbox}{max width=\linewidth, max height=\textwidth}\begin{tabular}{lccccc}

\toprule
{}&{(1)}&{(2)}&{(3)}&{(4)}&{(5)} \tabularnewline
\midrule \addlinespace[\belowrulesep]
\midrule \textbf{Panel A. Food Security}&Days Skipping&Children Skipping&Regularly Eat&Total Bread&KLK Index \tabularnewline
&Meals&Meals (=1)&Twice a Day (=1)&\& Tea Meals& \tabularnewline
\midrule \( \beta_1 \): Treated \( \times \) Round 2&--0.580***&--0.096***&--0.005&--0.948***&0.309*** \tabularnewline
&(0.094)&(0.021)&(0.028)&(0.219)&(0.056) \tabularnewline
\( \beta_2 \): Round 2&--0.195***&0.025*&0.072***&--0.322**&0.096** \tabularnewline
&(0.072)&(0.013)&(0.020)&(0.151)&(0.038) \tabularnewline
\( \beta_3 \): Treated&--0.470***&--0.069***&0.096***&--1.134***&0.346*** \tabularnewline
&(0.072)&(0.015)&(0.020)&(0.162)&(0.042) \tabularnewline
Control Mean&2.569&0.873&0.501&13.639&0.000 \tabularnewline
Observations&4,761&4,750&4,736&4,763&4,763 \tabularnewline
(\( \beta_1 \) + \( \beta_2 \) + \( \beta_3 \)) / \( \beta_3 \) &2.650&2.040&1.700&2.120&2.170 \tabularnewline
&&&&& \tabularnewline
\midrule \textbf{Panel B. Wellbeing}&Better Economic&Satisfied with&Happy&Life Satisfaction&KLK Index \tabularnewline
&Situation&Financial Sit.&&& \tabularnewline
\midrule \( \beta_1 \): Treated \( \times \) Round 2&0.219***&0.200***&0.102***&0.842***&0.837*** \tabularnewline
&(0.021)&(0.024)&(0.026)&(0.127)&(0.078) \tabularnewline
\( \beta_2 \): Round 2&--0.012&--0.036***&0.030*&--0.223***&--0.073* \tabularnewline
&(0.009)&(0.014)&(0.017)&(0.075)&(0.040) \tabularnewline
\( \beta_3 \): Treated&0.226***&0.163***&0.229***&1.542***&1.079*** \tabularnewline
&(0.015)&(0.017)&(0.019)&(0.093)&(0.055) \tabularnewline
Control Mean&0.048&0.133&0.154&3.179&0.000 \tabularnewline
Observations&4,762&4,755&4,732&4,763&4,763 \tabularnewline
(\( \beta_1 \) + \( \beta_2 \) + \( \beta_3 \)) / \( \beta_3 \) &1.910&2.010&1.580&1.400&1.710 \tabularnewline
\bottomrule \addlinespace[\belowrulesep]

\end{tabular} \end{adjustbox}
\\ \parbox{\linewidth}{\scriptsize \justifying \noindent \(Notes \): This table reports estimated impacts of treatment separately for the first and second survey round. Households were surveyed once per month for two months. Each of these months constitutes a survey round. All specifications control for stratum fixed effects and the baseline value of the dependent variable, if available. Standard errors are clustered at individual level. The outcome variables follow the primary outcomes shown in Table \ref{ITTsumTableAbridged}.}
\end{table}

\clearpage

\begin{table}[h!] \centering
\newcolumntype{R}{>{\raggedleft\arraybackslash}X}
\newcolumntype{L}{>{\raggedright\arraybackslash}X}
\newcolumntype{C}{>{\centering\arraybackslash}X}

\caption{Are Digital Payments Diverted?}
\label{infTaxResults}
\begin{adjustbox}{max width=\linewidth, max height=\textwidth}\begin{tabular}{lcccccc}

\toprule
& Gov. Off. & Comm. Leader & Gov. Off. & Comm. Leader & KLK & Yes to \tabularnewline & Others & Others & You & You & Index & Any Question \tabularnewline 
{}&{(1)}&{(2)}&{(3)}&{(4)}&{(5)}&{(6)} \tabularnewline
\midrule \addlinespace[\belowrulesep]
\midrule \textbf{Panel A. Baseline}&&&&&& \tabularnewline
Treated&0.001&0.002&0.002&0.002&0.074*&0.003 \tabularnewline
&(0.001)&(0.002)&(0.001)&(0.003)&(0.042)&(0.003) \tabularnewline
Observations&4,558&4,596&4,631&4,626&4,648&4,509 \tabularnewline
Control Mean&0.002&0.002&0.001&0.007&0.012&0.009 \tabularnewline
&&&&&& \tabularnewline
\midrule \textbf{Panel B. Long-Run}&&&&&& \tabularnewline
\( \beta_1 \): Treated \( \times \) Round 2&--0.000&0.006*&0.002&--0.001&0.091&0.004 \tabularnewline
&(0.003)&(0.003)&(0.003)&(0.005)&(0.084)&(0.006) \tabularnewline
\( \beta_2 \): Round 2&--0.002&--0.003&--0.003*&--0.008**&--0.137***&--0.010** \tabularnewline
&(0.002)&(0.002)&(0.002)&(0.003)&(0.045)&(0.004) \tabularnewline
\( \beta_3 \): Treated&0.001&--0.007&--0.001&0.003&--0.062&--0.004 \tabularnewline
&(0.005)&(0.005)&(0.005)&(0.009)&(0.140)&(0.011) \tabularnewline
(\( \beta_1 \) + \( \beta_2 \) + \( \beta_3 \)) / \( \beta_3 \) &--0.770&0.500&1.880&--1.680&1.740&2.550 \tabularnewline
Observations&4,558&4,596&4,631&4,626&4,648&4,509 \tabularnewline
&&&&&& \tabularnewline
\midrule \textbf{Panel C. List Experiment}&All Sample&Late Sample&Early Sample&&& \tabularnewline
Treated List&0.011&0.019&0.008&&& \tabularnewline
&(0.023)&(0.032)&(0.032)&&& \tabularnewline
Observations&2,358&1,171&1,187&&& \tabularnewline
Mean Items Mentioned&0.746&0.691&0.800&&& \tabularnewline
\bottomrule \addlinespace[\belowrulesep]

\end{tabular} \end{adjustbox}
\\ \parbox{\linewidth}{\scriptsize \justifying \noindent \(Notes \): In Panels A and B, stratification fixed effects, survey round fixed effects, and baseline values of dependent variables, if available, are included. These are answers to questions of the type "Have you/someone in your community been asked to provide informal assistance (for example money or food) to local community leaders/government officials in the past month?". The outcome in column 6 was not pre-specified. Panel C shows the results of a list experiment where the treatment group received the following additional statement: "I have been approached by government officials or community leaders to provide them with any kind of assistance, like food or money, in the past month". This includes a control for surveyor fixed effects. Standard errors are clustered at the individual level.}
\end{table}

\clearpage

\begin{table}[h!] \centering
\newcolumntype{C}{>{\centering\arraybackslash}X}

\caption{Actual Values vs. Experts' Predictions}
\label{expPreds}
\begin{adjustbox}{max width=\linewidth, max height=\textwidth}\begin{tabular}{lccc}

\toprule
Variable & Actual Values & Predicted Values & p-value Predicted = Actual \tabularnewline
{}&{(1)}&{(2)}&{(3)} \tabularnewline
\midrule \addlinespace[\belowrulesep]
\midrule 
Percent Able to Use Digital Payments&99.75&43.82&\(<\) 0.001 \tabularnewline
&&(2.75)& \tabularnewline
Percent Reporting Diversion Attempts&1.99&39.85&\(<\) 0.001 \tabularnewline
&&(2.98)& \tabularnewline
Delivery Costs (cents per \$1 delivered)&6.7&10.65&\(<\) 0.001 \tabularnewline
&&(0.74)& \tabularnewline
How Many Bread and Tea Meals in Past Week?&11.96&10.32&\(<\) 0.001 \tabularnewline
&&(0.45)& \tabularnewline
\bottomrule \addlinespace[\belowrulesep]

\end{tabular} \end{adjustbox}
\\ \parbox{\linewidth}{\scriptsize \justifying \noindent \(Notes\): The first column shows the actual values of different elements of the intervention. The second column shows the mean predicted value by our sample of experts (with standard deviations in parentheses). The third column shows the p-value of a test of the mean predicted value being equal to the actual value (i.e. a test of equality of columns 1 and 2). Results are similar when looking at predictions by academics and practitioners separately and are available on request.}
\end{table}

\clearpage

\bibliography{DirectAid_References}

\clearpage 

\renewcommand{\thefigure}{A\arabic{figure}}
\setcounter{figure}{0}
\renewcommand{\thetable}{A\arabic{table}}
\setcounter{table}{0}

\clearpage
\appendix
\phantomsection
\chead{\textsc{Online Appendix}}
\addcontentsline{toc}{section}{Online Appendix}
\section*{Online Appendix}
\label{sec:appendix}
\renewcommand{\thesubsection}{\arabic{subsection}}
\setcounter{subsection}{0}

\section{Related work} \label{RelatedWork}

\subsection{Cash-based transfers in humanitarian contexts} \label{CashHumanitarianLit}

Cash-based programs have become one of the most common foreign aid modalities, with nearly 17\% of the world's population (1.36 billion people) receiving some form of cash assistance during the pandemic period alone \citep{Gentilini2022}. The evidence base for cash programming is extensive, with documented positive impacts in almost all categories of development outcomes, including poverty reduction, education, improving financial outcomes, promoting human development, empowering the vulnerable, and social cohesion \citep{BastagliEtAl2019,gd2024}. While the vast majority of cash-based programs are carried out in developing, but stable contexts, recent years have seen tremendous growth in humanitarian cash-based programming. The amount of cash- and voucher-based humanitarian assistance has more than doubled since 2017 (3.3 billion USD in 2017 to 7.9 billion USD in 2022), now comprises 20\% of all humanitarian assistance globally, and appears to be growing \citep{UrquhartEtAl2023}. To date, there has been very little evaluation of the impact of cash-based humanitarian assistance. In recent systematic reviews of thousands of humanitarian cash-based assistance programs, only a very small set included rigorous evaluations of impact \citep{PuriEtAl2017,TappisDoocy2018,JeongTrako2022}, the evidence is especially thin in contexts facing the most severe and protracted humanitarian crises \citep{KurtzEtAl2021}.  

At the same time that humanitarians are increasing their use of cash-based programming, they are simultaneously shifting from physical to digital distribution \citep{WFP2023c,WFP2020,WfpJordan2022}. Although physical cash-based programs are unlikely to disappear entirely, digital is clearly becoming the preferred delivery mechanism, with growing policy support \citep{ODI2015,IRC2018,SmithEtAl2018,AgurEtAl2020}. The evidence on digital cash-based programs also comes primarily from retrospective performance evaluations or synthetic reviews, and appears primarily in policy reports rather than peer-reviewed outlets. 

To understand the prevalence of humanitarian, cash-based programs that have been rigorously evaluated, we examined the most comprehensive repository of systematic evaluations (N=332 as of 2 January 2024) of cash-based programs: GiveDirectly's \textit{Cash Evidence Explorer}. We text-mined the abstracts and target beneficiary fields of the 328 papers for the following relevant keywords (and variations on these words): women, vulnerability, conflict, violence, war, humanitarian, crisis, emergency, repression, mobile, digital, phone, ATM, bank, e-card, and electronic. Restricting the sample to studies with any of those keywords (Boolean \texttt{OR} operator) resulted in 97 papers, of which 90 presented evaluation results, of which there were approximately 36 unique programs (i.e., there could be multiple evaluations of the same program, such as with Progresa / Oportunidades in Mexico, though we note that the database does not have a unique program identifier and so we also coded the abstracts for program titles to code uniqueness). We then hand-coded the resulting subset of studies and found a number of false positives, for example, the keyword ``violence'' returned some studies that addressed intimate partner violence as opposed to government, rebel, or structural violence. In our review of the database, we found only four digital cash-based programs in humanitarian contexts.

To further map the evidence base specific to our study, we culled details of relevant cash-based transfer programs from systematic academic reviews \citep{BastagliEtAl2019, TappisDoocy2018, PuriEtAl2017, gentilini2016revisiting}, policy evidence reviews  \citep{JeongTrako2022, WFP2021a, Mikulak2018, DFID2017, MaunderEtAl2016, WorldBank2016, BaileyHarvey2015}, evaluation databases and clearinghouses \citep{gd2024, ALNAP2023, Gentilini2022}, and broader policy discussions \citep{WFP2023c, UrquhartEtAl2023, IRC2016, SmithEtAl2018, ODI2015}, which together covered hundreds of cash-based transfer programs. For purposes of coding relevant cash-based programs, our inclusion criteria consisted of (a) cash-based transfers (including vouchers) needed to comprise a core part of the intervention, (b) the program needed to be carried out during a humanitarian crisis, and (c) the program needed to be evaluated using a method that established a credible counterfactual (i.e., randomized controlled trial, natural experiment, regression discontinuity, and difference-in-differences). 

We focused on conflict- and natural-disaster-based humanitarian crises because rigorous evaluations of cash-based programs during health outbreaks have rarely been conducted \citep{JeongTrako2022}. With the onset of COVID-19, many cash-based programs were implemented around the world, but overwhelmingly in contexts where governments and humanitarians alike had a shared interest in mitigating negative impact.  The overwhelming majority of nearly 4,000 cash-based programs occurred in high- and middle-income countries and, in most cases, as part of existing government response systems \citep{Gentilini2022}. Furthermore, although the evidence base is growing, rigorous evaluation appears to be following patterns in programming, overwhelmingly confined to wealthier (middle-income) countries. 

In all, we identified 23 relevant cash-based transfer programs summarized in \ref{23ProgramsSummary}. Peer-reviewed academic studies account for 16 of the 23, five are academic working papers, and two are evaluation reports. Column 1 reports the country in which the cash-based intervention took place and the countries are ordered by the current level of fragility of the country in which they occur, moving from the least fragile (top) to most fragile (bottom) \citep{OECD2022}. Ecuador and Philippines appear at the top because neither country is listed in the OECD's list of fragile countries. The next two, Lebanon and Sri Lanka, are not on the core OECD list either, but are flagged as showing key early-warning signs. Sometimes the target beneficiaries (Column 2) were refugees from another country, but located in a host country, and sometimes the intended beneficiaries were local to a country. Regardless, target beneficiaries were high need or otherwise vulnerable. Column 3 reports the type of cash-based assistance, specifically identifying cash (C) and vouchers (V), and the most relevant alternative, food assistance (F). Some of these programs include other/combined treatments, which we note in the extended discussions. Column 4 reports the delivery mechanism, whether some form of digital, physical, or a combination.

Sixteen of the 23 programs occurred in conflict-affected countries (seven targeting displaced populations and nine targeting poor/vulnerable households) and seven in countries facing high natural disaster risk. Collectively, the programs cluster in 13 countries, but only three of which currently rank among the world's most food insecure.  Although the populations covered in these 23 programs are highly vulnerable, all have been located in contexts where host governments do not actively interfere in the provision of humanitarian assistance. Programs that address conflict-based humanitarian crises, for example, frequently occur in displacement camps in which host governments and humanitarians typically attempt to cooperate \citep{deHoopEtAl2019}. Our collective understanding of cash-based programming in humanitarian crises has little to say about protracted crisis contexts that sit at the confluence of extreme need, high insecurity, and oppression \citep{KurtzEtAl2021}, where social protection systems are also extremely underdeveloped \citep{Peachey2020}. 
   
Across the 23 programs, 12 distributed cash physically, three distributed cash through ATM cards, three distributed through mobile money accounts, four distributed through some combination of digital and physical depending on the program mandate, and one program did not specify the delivery mechanism. In the 23 programs we reviewed, digital programs were implemented mostly in the least insecure/oppressive countries, whereas physical transfers occurred primarily in contexts characterized by greater insecurity and oppression. 

%

Of the 23 RCTs of humanitarian cash-based transfers, 10 are in Freedom House 'not free' countries, three of those 10 have some digital component, and none, that we can find, target large populations that the government actively oppresses? We coded the Freedom House rating for the first year of the respective program if the program occurred in 2013 or later given that the Freedom House ratings only go to 2013. For programs occurring before 2013 (Ecuador 2011; Philippines 2010; Sri Lanka 2005; Uganda 2009; Niger 2009 and 2011; DRC 2011; Yemen 2011), we used the Freedom House 2013 value \citep{FreedomHouse2023}.
%

\subsection{Work related to outcomes: Usage, food security, well-being, diversion} \label{ExperimentalResultsLit}

To contextualize our experimental findings on usage, food security and mental well-being, \ref{23ProgramsSummary} reports whether the 23 reviewed studies empirically examined usage at any level, denoting examination with a check mark (\checkmark), the same (or similar) outcome categories that we did, including whether they found positive, null, or negative effects. Few studies carried out systematic tests for diversion, so we only include a check mark (\checkmark) if the study made any observation about levels of diversion in their context. To supplement the summary table, \ref{LiteratureFindings1} provides details about specific study context and findings.   

\paragraph{Usage} Because tech-illiterate populations are largely poor, rural, elderly, and female \citep{Peachey2020}, it is important to assess whether digital transfers will reach those most vulnerable and be used effectively. If they cannot, then digital cash-based approaches may hurt intended beneficiaries relative to what may have occurred with physical cash or in-kind food distribution. In India's PMGKY program intended to provide COVID relief, for example, the government transferred benefits to the accounts of over 200 million beneficiaries, but being female, illiterate, and living in a household without a smartphone resulted in extremely low use of digital payments, which ranged from 1\% to 3.9\% \citep{Gentilini2022}. Of the 10 studies  in our review with a digital component, five relied at least in part on mobile platforms, meaning that the beneficiary needed to use the phone for receipt/withdrawal of funds. The remaining five relied primarily on ATM cards, which are less demanding technologically, and also confer fewer advantages such as decentralization and transparency. Of these five, only one reported on usage, noting that most of the time the funds were withdrawn immediately and, where not, they were used up eventually \citep{LyallEtAl2020}. The remaining four did not report usage explicitly, but descriptions were suggestive of substantial usage, though we acknowledge that this assessment may be overly generous. Our own results in one of the most challenging environments, and with an arguably more vulnerable demographic than most other studies, are encouraging, though we emphasize the importance of user-centered design in our registration and training procedures. Near 100\%  usage by a largely illiterate and vulnerable population in a context of extreme oppression and insecurity suggests genuine potential for technology to ``flatten access'' to otherwise marginalized populations \citep{GrossmanEtAl2014}. (See \ref{23ProgramsSummary}, Column 5).

\paragraph{Food Security} Most of the studies (21 of 23) considered a food security outcome in some form, although there is considerable variation in the indicators each examines. Of the 21, 16 find that cash-based transfers improved food security outcomes, five found no relationship, and none of the studies found any negative effects. Of the 21, 20 examined at least one indicator of food quantity (e.g., food expenditures or calorie consumption). Of the 21, 18 examined at least one indicator of food quality (e.g., dietary diversity). And 16 of 21 considered at least one indicator of negative coping strategies (e.g., meal skipping). These patterns add support to a growing body of evidence documenting that humanitarian cash-based programs have positive impacts on a variety of food security outcomes. Existing studies most frequently considered basic needs outcomes. In a recent systematic review, eight of 20 studies considered food security (primarily food consumption) and four of 20 studies examined dietary diversity, most of which demonstrated positive impacts \citep{JeongTrako2022}. Moreover, five studies examined reductions in negative coping strategies (e.g., meal skipping) and three found positive impacts. (See \ref{23ProgramsSummary}, Column 6). 

\paragraph{Mental well-being} In contrast to food security outcomes, \ref{23ProgramsSummary} shows that limited attention has been given to the possible mental benefits (or costs) of cash-based programs. Only seven of the 23 studies considered some mental well-being outcomes and six found positive effects. An evaluation of a voucher program in DRC  found a positive impact (+0.32sd) on an overall mental well-being index at the end of six weeks, which attenuated to some extent, but remained remarkably durable at the end of one year (+0.18sd and statistically significant) \citep{QuattrochiEtAl2022}. They also measured life satisfaction, where they found a positive impact at six weeks (18\% higher in treatment; control: 3.29 on 10-pt scale), but the result disappeared by the one-year mark. Perhaps the strongest impact of cash on mental well-being, with documented short-, medium-, and long-term effects of cash relative to control, occurred in Niger \citep{BossuroyEtAl2022}. Importantly, like the other treatment arms, cash was accompanied by coaching, savings, training, and market facilitation, making the overall bundle difficult to compare to other cash-only programs. The dearth of studies examining mental well-being mirrors that of prominent systematic reviews  \citep{JeongTrako2022}. They only identified two (of 20) studies reporting subjective well-being results: the DRC voucher study that we just discussed \citep{QuattrochiEtAl2022} and a study of Syrian refugees in Jordan (that did not meet our inclusion criteria) that found isolated improvement on a single measure of self-esteem, but notably no effect on a measure of satisfaction, the closest to our life satisfaction indicator. Similarly, in another systematic review, there were no rigorously evaluated studies that addressed mental well-being though some mixed-methods evaluations have found isolated positive impacts \citep{AttahEtAl2016, DoocyTappis2016, HagenZankerEtAl2017}.  (See \ref{23ProgramsSummary}, Column 7).

\paragraph{Diversion} The extent of diversion in these studies is essentially unknown as four of the 23 studies addressed diversion, but only three attempted to measure its extent. Studies measured the amount of the cash and vouchers actually received and found little leakage \citep{Aker2017}, how much of the grant recipients had to give to other household and community members and participants who reported less than 1\% of the grant \citep{BlattmanEtAl2016}, and the occurrence of robbery  where participants reported no occurrence \citep{LehmannMasterson2020}. Another study mentioned that the implementer feared diversion, and therefore hired security to guard cash during transport, which comprised the largest share of their costs, but did not report measuring diversion \citep{AkerEtAl2016}. Many policy reports have downplayed concerns about diversion, suggesting that diversion risks are likely overstated and that cash is no less prone to diversion than other types of aid \citep{Sossouvi2013,Gordon2015,ODI2015,EckerEtAl2019}. 
As far as we are aware, no systematic tests of diversion in a cash-based context have been undertaken. In our study, we experimentally tested for diversion, both using direct and indirect questions, and found no evidence of its presence. We even interviewed key stakeholders, such as merchants, to understand alternative points where diversion could have occurred, and found little indication. This finding is especially informative given how extensively the scholarly and policy literatures have identified diversion as a core concern for in-kind assistance. Indeed, anecdotal evidence of diversion has been extensively documented across the globe \citep{perlez1992, bryercairns1997, dewaal1997, anderson1999, goodhand2002, lischer2003, maren2009, polman2010, barnett2013, zurcher2019, odonnell2023} and cross-national inquiries also suggest that aid prolongs conflict in part due to diversion \citep{nunn2014us, narang2015, woodsullivan2015, woodmolfino2016, findley2018, findleyetal2023}. (See \ref{23ProgramsSummary}, Column 8). 


\subsection{Costs of Delivery} \label{AdditionalAdvantagesLit}

The last three columns of \ref{23ProgramsSummary} report whether the programs (a) reported on and delivered aid cost-efficiently or cost-effectively, (b) decentralized their aid distribution, and (c) could transparently track aid receipt and usage. 

\paragraph{Cost Analysis} Of the 23 reviewed programs, 12 report cost-efficiency or cost-effectiveness metrics. (\ref{23ProgramsSummary} reports the derivable total cost transfer ratios (TCTRs) for nine of the 12; the remaining three, which we tag with a check mark (\checkmark), either reported on bundled programs for which we could not disaggregate the costs or did not include the precise estimates in their reports.) The cost efficiency of our digital aid program is the lowest of any of these programs when comparably counting cost categories. Although all of the reviewed studies focus on variable operational costs, they vary in the specific cost categories they count as relevant. In particular, very few studies include targeting/recruitment costs, and are thus comparable to our TCTR of \$1.014 (1.4 cents required to deliver 1 dollar of aid), though we think the inclusion of targeting/recruitment costs is more appropriate (and conservative) given that these costs are likely to be present in most crisis contexts. 

The cost efficiency of our direct aid intervention is 40\% of the World Food Programme's global average, 48\% of IRC's (International Rescue Committee's) most cost-efficient program, and 45\% of ECHO's (European Commission Directorate General for Humanitarian Aid and Civil Protection) most cost-efficient program. We do not have access to the disaggregated data used to compute WFP's or IRC's cost efficiency estimates, but the report of ECHO-funded projects reports a variety of disaggregated statistics \citep{MaunderEtAl2016}. The average total cost transfer ratios range from \$1.15 (cash for refugee response) to \$2.81 (cash for complex emergencies). Their TCTR estimates of cash for slow-onset and sudden-onset crises are \$1.64 and \$1.39 respectively. Their estimates for voucher projects are \$1.54 (slow onset), \$1.81 (refugee response), \$2.11 (complex emergency), and \$2.72 (sudden onset). They disaggregate their cash projects by delivery mechanism, reporting average TCTRs of \$1.32 (ATM card; 2 projects), \$1.64 (mobile phone; 7 projects), \$1.66 (mix of bank transfer and cash in envelope; 3 projects), \$1.97 (cash in envelope; 22 projects), and \$2.03 (bank transfer; 13 projects).  They also disaggregate vouchers and report TCTRs of \$1.31 (electronic), \$1.76 (paper), and \$2.25 (voucher fair). Their lowest TCTR is more than double that of our digital aid program in Afghanistan, and their most relevant categories are many times higher than ours (electronic voucher: \$1.31; mobile phone: \$1.64). 

The cost-efficiency of digital aid in Afghanistan is considerably lower than all other humanitarian programs for which we could find estimates. The cost-efficiency of our study was substantially lower, in part, because our approach was consistent with the ``High-Level Panel on Humanitarian Cash Transfers'' guidance to work with the private sector \citep{ODI2015}, which, among other benefits, reduced potentially significant start-up costs. Others have argued that mobile transfers may only be more efficient than physical cash when a mobile network infrastructure is available, convenient, and clear \citep{AkerEtAl2016,JeongTrako2022}, which the private digital payments platform provided in our context.

These cost-efficiency estimates also compare favorably to non-humanitarian social transfer programs --- Kenya CT-OVC, Nigeria CDG, Mexico PROGRESA, and Kenya HSNP --- which reported TCTRs at each year of 2--5 year programs \citep{white2013guidance}, estimates that, from the beginning to the end of these programs, TCTRs ranged from $2.63$ to $1.34$ (Kenya CT-OVC), $2.04$ to $1.40$ (Nigeria CDG), $2.34$ to $1.05$ (Mexico PROGRESA), and $2.41$ to $1.21$ (Kenya HSNP). Only Mexico PROGRESA had a lower TCTR than our study, but this lower TCTR was only obtained in year 4 of operation, once the program had gone to scale. Furthermore, these four programs were carried out in more favorable implementation environments, where costs should be expected to be lower than in a context such as Afghanistan. 


\paragraph{Decentralization} Because most cash-based assistance has been centralized with a single donor or small set of donors and can depend critically on a small set of distribution points, which can be costly to establish and maintain, decentralizing assistance has core implications for costs of delivery. Only five of the 23 programs that we reviewed could be considered decentralized, capable of facilitating quicker access and use of funds. Of these five, three distributed cash on ATM cards that could be withdrawn at any ATM, any number of times, and in any amount, which were less constrained than ATM cards only redeemable at artificial NGO-operated mobile ATMs but still dependent on ATM penetration, which is unlikely to be high in most crisis contexts. It is unclear whether funds from other organizational or individual donors could add funds to these ATM cards. Two of the five programs were based on mobile money platforms that did not strictly require cash out, but in practice, nearly all recipients cashed out quickly and fully, thus making the aid dependent on mobile agent penetration. In contrast, our digital value voucher approach could be agnostic as to the donor, and could be redeemed at a variety of merchants relatively close to beneficiaries.   

\paragraph{Transparency} All 23 studies we reviewed report only survey-based measures of activity, which can be expensive to collect and sustain. Still, some of the approaches could, in theory, lend themselves to greater transparency of receipt/usage of funds. We thus coded programs as transparent if their delivery technology could potentially capture non-survey-based receipt/usage data. Only three of the programs could provide transparency beyond the point of funds distribution or withdrawal. For physical cash, ATM cards, and mobile transfers with required cash-out, subsequent usage cannot be traced. Of the three approaches in which mobile transfers did not require cash out (though nearly all recipients still cashed out), activity could be tracked to understand the uses of assistance and also detect diversionary efforts. Policy makers have long cited the transparency gains of moving to digital cash \citep{ODI2015}, but in this context, there remains almost no evidence of entirely mobile transfer systems, which can ensure end-to-end tracing from deposit to purchase.   

\section{Intervention and analysis details}
\label{main_appendix}

The research protocol was approved by the London School of Economics' Institutional Review Board (study number 89546) and preregistered in the AEA RCT Registry (study number 0010189). In Online Appendix \ref{ss:ethics} we discuss ethical considerations. The Pre-Analysis Plan (PAP) is registered at: \\ \href{https://www.socialscienceregistry.org/versions/160809/docs/version/document}{https://www.socialscienceregistry.org/versions/160809/docs/version/document}. 

\subsection{Research design} 
\label{pilotSec}




\paragraph{Local context}

Afghanistan is in its third year of a complex humanitarian crisis combining economic contraction, political repression, and ongoing concerns over stability and security. Its humanitarian appeal of \$4.6 billion in 2023 represented the world's largest single-country appeal to date \citep{un2023}. It is also the WFP's largest recipient by value of cash-based transfers and commodity vouchers \citep{wfpdashboard} and though multiple financial service providers exist in the country, digital aid programs are still nascent \citep{uncdf2023}. Afghanistan's GDP contracted by 30-35\% since the Taliban took over in 2021, and ever since the country has been under the threat of ``winters of famine'' \citep{wb2022,wfpafg2022}. 

Given the Taliban's draconian restrictions on education, employment, and freedom of movement, Afghan women are particularly vulnerable. Taliban edicts prevent women from travelling more than 75 kilometers without a male guardian and ban them from visiting public baths, restaurants and parks \citep{un2023c}. In December 2022, the Taliban banned female Afghan employees from working in non-governmental organizations, prompting major foreign aid groups to suspend operations \citep{roberts2023taliban}. The ban was expanded in April 2023 to include female Afghan employees of the United Nations \citep{un2023a}. The Afghan central bank's reserves have been frozen since 2021, requiring the United Nations to fly in approximately 40 million U.S. dollars per week to support humanitarian operations \citep{unama23}. Western donors remain concerned that diversion of these substantial flows is enabling the world's most gender-repressive regime \citep{odonnell2022,sigar2023} and that the Taliban are providing safe haven to international terrorist groups \citep{centcom2023}.

Mobile phone ownership has grown rapidly over the past two decades, from approximately 25,000 subscribers in 2002 to over 22 million subscribers in 2021 \citep{wbdata}. In a nationally-representative survey, 91\% of respondents reported at least one member of their household owned a mobile phone (66\% of respondents report personally using a mobile phone), while 46\% of that subgroup reported having an internet connection \citep{asiafoundation2019}. Other studies have documented a substantial gender gap in women's access to mobile phones \citep{bbg2015} and access to the internet \citep{gallup2023b}. For an overview of empirical research on digital payments in Afghanistan over the last decade, see \citep{blumenstock2015promises, blumenstock2018defaults, blumenstock2021violence, blumenstock2023strengthening}.

\paragraph{Piloting activities}

Prior to starting the experiment, we conducted three small pilots (N$<$50) to i) refine our survey instruments, ii) work out logistical processes including how to enroll beneficiaries and iii) identify patterns that needed to be taken into account before the full scale up of the intervention.

The first pilot involved around 30 women in Kabul. Our initial idea was to conduct the experiment without any face-to-face interaction. Thus, participants were contacted over the phone, invited to participate, and explained how to open accounts with the digital payments provider. They received smaller payments (800 AFA) than in the actual experiment. A second, similar pilot was conducted a few weeks after the first one. These two initial pilots were intended to evaluate the survey instruments and sort out the logistics for the eventual scale up. From these pilots, it was apparent that participants were struggling to open accounts with the digital payments provider and use their funds, as almost all participants had never used mobile money services or apps similar to the one used in this program, had never been part of the formal banking system, and mostly had feature phones. While the digital payments platform can be used with a feature phone, the process of creating an account is more complicated than when using a smartphone. 

Due to these pilots, we decided to organize in-person registration sessions with around 50 women each, where potential participants would be introduced to the program, helped by the Community Driven Development Organization (CDDO) and representatives from the digital payments provider. We conducted a third pilot with 52 women in Kabul to test the logistics of the full scale up and revise the last versions of the survey instruments before conducting the registration sessions with all participants. This included conducting the in-person registration session and several rounds of phone follow-up surveys. This also allowed us to check whether congregating women in a given place would cause problems. We observed much higher rates of usage of the funds sent to women and no meaningful problems during the onboarding process.

\paragraph{Beneficiary identification} 

Our goal was to identify  $\sim$2400 vulnerable women in three Afghan cities (Kabul, Herat and Balkh) to be part of the intervention. To do so, we worked with the CDDO, an Afghan organization that assists Community Development Councils (CDCs) in a wide-array of local activities. The CDCs were established through local elections as part of the National Solidarity Program starting in 2004 \citep{beath2016electoral, beath2017direct, beath2013empowering, beath2017can, beath2018elected}, where their primary job was to oversee block grants of development funding, and they were given a much broader range of local administrative authorities under the Citizens' Charter, starting in 2016. Local CDCs identify potential beneficiaries through a community-based exercise (``Well-Being Analysis'') in which community members, elders and mullahs together categorize all community households into different socioeconomic groups (e.g., well-off, middle income, poor, very poor). Our participants come from the lowest group. Thus, participants are identified through a process relying on the community's consensus of who is most vulnerable. After participants were identified, they were onboarded as described in the main text.

\paragraph{Timeline and randomization}

During the onboarding session, 2,422 women agreed to participate. Due to technical and logistical issues, the final experimental sample consisted of 2,409 women. The 13 remaining women had issues with their phone numbers, including mismatches between phone numbers in the survey and transaction data or multiple women registered with the same phone number. While these 13 women were dropped from the experimental sample, we contacted each of them, resolved the issues, and transferred them the humanitarian payments as we did with the women in the experimental sample. Women in the experimental sample were assigned to one of two groups. The treatment group received four bi-weekly 4,000 Afghani (roughly 45 USD) aid payments via the digital payments platform first, between November 6, 2022 and December 20, 2022. The control group received the same payments between January 1, 2023 and February 12, 2023, after the treatment group had received all four of its payments. Figure \ref{intFig} shows the timeline of the project. We believe that given the abject situation of participants, the ethical thing to do was for all participants to receive the aid payments eventually, and that is why we settled for this staggered intervention. During the onboarding sessions, all participants were told that they would receive the payments eventually. 

Participants were randomly assigned to the treatment and control groups, stratifying on two variables. First, they were stratified on the nahia (neighborhood) in which they registered. There were 16 nahias in total across Kabul, Mazar and Herat. Second, they were stratified on a measure of vulnerability. During the baseline, we asked participants in the past seven days, how many of their meals (breakfasts, lunches and dinners) had been only bread and tea, a measure of vulnerability and poor dietary diversity. We then created a categorical variable that indicated whether the participant was above or below the median number of bread and tea meals, which was used as the second stratification variable. We assigned ``misfits'' independently across strata. Given that we had few strata and a single treatment arm, there were few misfits, so the risk of harming treatment fractions by independently assigning misfits across strata was low \citep{carril2017dealing}. We pre-specified this outcome as one of our primary food security outcomes in the PAP. As can be observed in our PAP, the treatment groups are balanced in 17 of the 18 outcome and heterogeneity variables we collected at baseline. The only unbalanced variable, whether the individual has been asked for any kind of assistance by local community leaders in the past month, is because only four individuals answered yes to this question, and they all ended up in the control group by chance.




\paragraph{Sample characteristics}

\ref{baseBalComparison} reports summary statistics for the experimental sample. Most participants are widows (66\%), have no education (63\%) and are the main financial decision-maker in the household (66\%), which on average has 6.31 members, indicating that the women in the study are highly vulnerable. Using the 2015 Demographic and Health Survey (DHS), we compare our sample to a representative sample of similarly-aged women in urban areas of Kabul, Balkh and Mazar. The DHS sample has higher educational attainment: 56.8\% have no schooling and 13.9\% have at most primary education. 

The women are also poor and food-insecure: Out of 21 possible meals in the last 7 days, 13.76 consisted of just bread and tea. Mean household income in the past 30 days was just 357.97 Afghanis (roughly 4 USD), and any kind of employment was basically non-existent (only 3 reported that the head of the household had worked in the past 30 days). Among those who had a medical emergency in the last month, only 1\% could afford the medicine needed. Unsurprisingly, these facts translated into extremely low levels of happiness (just 1\% report being very or quite happy) and life satisfaction (mean score 3.53 on the Cantril Self-Anchoring Striving Scale which runs from 1 to 10). This is in line with recent evidence: A nationwide Gallup survey conducted just after the Taliban took control in August 2021 indicates that 94\% of Afghans rate life satisfaction below 4 \citep{gallup2022}. 

In addition, the digital payments provider conducted a concurrent program in which women who already had an account with them or clicked on an advertisement in social media received four 800 AFA payments over a two month period. This helps benchmark the vulnerability and poverty of our sample. As expected given that this other sample is more tech-savvy, participants were significantly more likely to have some education (88\%) and were, on average, younger (28.44 years) than those in our study. They were also less likely to be the main financial decision-maker of the household (31\%), had much higher incomes (5,325 AFA) and employment levels, and only 5.31 of their meals in the past week had been bread and tea only.

In addition to being poor, vulnerable, and mostly uneducated, our sample also had had almost no experience with mobile money or other financial mechanisms. Only 2 women reported having transferred airtime in the past month, none reported having transferred money via a digital payments platform, and only 2 already had accounts with the digital payments provider before the start of the program. Just four of them report that anyone in their household had ever had a bank account. 

\paragraph{Data collection}

Data were collected through three ways. First, a baseline survey was completed during the onboarding session (see above). Second, we had access to participants' transaction data from the digital payments provider, which we could link to participants' survey data. We obtained permission to do so during the consent process. Third, we conducted four rounds of follow-up surveys over the phone, from after the treatment group received its first payment to a couple of weeks after the control group received its last payment (see Figure \ref{intFig}). 

A team of 12 female enumerators were tasked with contacting each participant once every month. Participants were randomly assigned a date to be contacted. Enumerators contacted participants on said date, and if the survey was not completed (because the participant did not pick up or was busy), they would attempt again on a different date and time. Participants received a 350 AFA payment as compensation for their time for completing their survey. Overall, for the two survey rounds that correspond to the experimental sample, completion rates were around 99\%. In the first survey round, 29 interviews could not be completed (17 treatment, 12 control). In the second round, 26 interviews could not be completed (10 treatment, 16 control). Attrition is not differential by treatment status, as seen in \ref{attrition}, but there is a slightly higher non-response rate among control participants in the second round of surveys (the difference is 6 surveys). 

We pre-specified all analyses, including how the outcome variables would be constructed, and what our primary outcomes variables were going to be. We divided our outcomes variables in three families: Basic needs, wellbeing and informal taxation. In doing so, we provided a level of detail consistent with that articulated in \citep{banerjee2020praise}, that two research assistants could take the data, and the PAP, and separately produce identical analyses.  

\subsection{Analysis}
\label{analysis} 

\paragraph{Estimation strategy of baseline results}

We estimate intent-to-treat treatment effects for pre-specified outcomes based on the following specification: 
\begin{equation}
\label{eq_baseline}
Y_{itn} = \gamma_0 + \gamma_1 \mathbb{1}[\textit{Treatment Group}]_{in} + \gamma_2 X_{i0n} + \gamma_3 Y_{i0n} + \gamma_4 \mathbb{1}[t = 2] + \varepsilon_{itn}
\end{equation}

where $Y_{itn}$ is the outcome of woman $i$ in survey round $t$ in nahia $n$. Only the first two months of intervention are used for all primary analyses as those are the months in which we have clear experimental variation.  $X_{i0n}$ are the stratification variables (nahia fixed effects, and baseline needs). $\mathbb{1}[t=2]$ is a dummy for the second survey round period. For variables for which we have values at baseline, we control for the baseline values $Y_{i0n}$. Standard errors are clustered at the individual level, the unit of randomization. 

We conduct our analysis in accordance with our PAP. For each of the three families of outcomes, we specified four different primary outcome variables. For needs, the four primary outcomes are the number of days in which the person skipped meals in the past week, an indicator for whether children in the household skipped meals in the past week, an indicator for whether household members have regularly eaten at twice a day for the past week, and the total number of meals that have consisted solely of bread and tea in the past week. For the informal taxation outcomes, the four primary outcomes are indicators for whether the participant has been approached by government officials to provide them with any kind of assistance (such as food or money) in the past month,  whether the participant has been approached by local community leaders to provide them with any kind of assistance (such as food or money) in the past month, and the same two questions but for anyone else in the community. For wellbeing, the four primary outcomes are an indicator for whether the participant feels that the overall economic situation of her household is slightly or much better than 30 days ago, an indicator for whether the participant agrees a lot or somewhat with the statement ``I am highly satisfied with my present financial condition'', an indicator for whether the participant says that taking all things together she is very or quite happy, and a score on how satisfied she is with her life as a whole these days (from 1, dissatisfied, to 10, satisfied). For each of these questions, participants were allowed to not answer the question if they did not want to.

The results for needs and wellbeing are presented in Table \ref{ITTsumTableAbridged}. These also include other variables that we denoted as secondary, which include measures on participants' diets (whether they consumed rice, beans, vegetables, chicken, or dairy in the past week), whether they were able to buy medicine if they had an emergency, household income, and employment status of the head of the household. The medicine question is asked only to those who said that they had had any medical needs to purchase medicine in the past 30 days. Given that attrition could be differential by treatment group (e.g., if the treatment group is doing better health-wise due to the aid payments), we also calculate Lee bounds for this variable following \citep{lee2009training}. The 95\% confidence interval for the Lee bounds goes from 0.014 to 0.068. 

To increase power and precision, for each family of outcomes, we also summarized the primary measures into a single summary index following \citep{kling2007experimental}, which we denote the Kling, Liebman, and Katz ``KLK'' index. The index is created as follows: First, each measure is standardized by the pre-intervention values of the variable in the control group. Second, for those observations with missing values, these are imputed as the mean in the participant's treatment group. For missing values in the baseline data, the imputation is done with the values at baseline, and for missing values in the follow-up data, the imputation is done with the values at the follow-up rounds. Third, all variables are aligned in the same direction, such that higher values indicate ``better'' outcomes. The final index is the equally-weighted average of z-scores of the index's individual component variables. The final measure is then standardized (relative to the control group) to assist interpretation. Given that our main experimental hypothesis is that direct aid payments will reduce immediate humanitarian needs and improve wellbeing, we control for the Family Wise Error Rate (FWER) for each family of primary outcomes, as our primary concerns relate to falsely rejecting the null that the program had no impact on humanitarian needs or wellbeing. We therefore take the more conservative approach of controlling the FWER rather than the False Discovery Rate (FDR). For secondary outcomes, we control instead for the FDR. These results using these multiple hypothesis corrections are shown in column 6 of Table \ref{ITTsumTableAbridged}. 

The results for informal taxation are presented in Table \ref{infTaxResults}. Panel A shows the corresponding results for the informal taxation questions as in Table \ref{ITTsumTableAbridged}. Column 5 includes a dummy for whether the participant responded yes to any of the four informal taxation questions, which better captures the prevalence of informal taxation in these communities. However, note that this outcome was not pre-specified. Controlling for the FWER, none of the adjusted p-values for these four outcomes is below 0.1 (results not shown for brevity).

Given that participants were allowed to skip certain questions if they wanted, this means that the sample in Table \ref{ITTsumTableAbridged} changes in each regression. We also provide results restricting the sample to only those individuals who answered all questions necessary to construct our primary and secondary outcomes, ensuring a constant sample across regressions. These results are presented in \ref{ITTsumTableRest}, which follows the same structure as Table \ref{ITTsumTableAbridged}. Results using this alternative approach are consistent with the baseline results in Table \ref{ITTsumTableAbridged}.

\paragraph{Estimation strategy of results over time}

We also evaluate how the results change in the second round of follow up surveys, after the treatment group has received 3-4 aid payments, compared to the first round of follow up surveys, after the treatment group has received only 1-2 aid payments. We do so by estimating the following specification:
\begin{equation}
\label{shortLongEq}
\begin{aligned}
Y_{itn} =& \beta_0 + \beta_1 \mathbb{1}[\textit{Treatment Group}]_{in} \times \mathbb{1}[t = 2] + \beta_2 \mathbb{1}[t = 2] \\ &+ \beta_3 \mathbb{1}[\textit{Treatment Group}]_{in} + \beta_4 X_{i0n} + \beta_5 Y_{i0n} +  u_{itn}
\end{aligned}
\end{equation}

and testing ($H_1$) whether the effect of the treatment in the second round is statistically different from that of the control group, with $H_0: \beta_1 + \beta_3 = 0$. Panel B of Table \ref{infTaxResults} shows the results for the informal taxation outcomes, while Table \ref{dynEffectsMain} shows the results for the needs (Panel A) and wellbeing (Panel B) primary outcomes.  

\paragraph{List experiment}

A key question, given the context, is whether the Taliban government managed to capture any of the aid payments. Participants might feel uncomfortable answering questions about informal taxation by community leaders and government officials, and so they might answer our informal taxation questions falsely. To provide additional evidence that the lack of informal taxation we observe is not due to fears of answering yes to our questions, we conducted a list experiment with participants. List experiments have been used extensively in the political science literature to gauge the prevalence of sensitive behaviors or situations, without individuals having to disclose that they have indeed done so. The basic idea is to create two lists of statements which are identical but for the fact that one of the list (``treated list'') has an additional statement that is the one the researchers want to learn about. Participants are asked how many of the different statements apply to them, and then by comparing the number of reported statements in the short and long lists we can estimate differences in the shares of respondents reporting they experienced a situation/behavior of interest. The basic idea is that respondents are provided cover as they are only signaling how many behaviors they undertook, not which specific behaviors \citep{boudreau2023monitoring}.  

We implemented a list experiment to gauge the prevalence of informal taxation in the population. There are two items to note. First, we decided to conduct the list experiment as an additional check after we had analyzed some of the follow up data and thus this analysis was not pre-specified. However, we follow the standard approach in the literature for analyzing list experiments. Second, we conducted the list experiment in the very last round of surveys, when the control group was receiving their \nth{3} and \nth{4} aid payments. Thus, the timeframe is different from that of all the other analysis. 

Concretely, we asked participants how many of  4 (control) or 5 (treatment) situations have happened to them in the past month. Both lists include the same initial four statements: ``I have received some form of financial support from local authorities'', ``I have borrowed money from a friend or family member'', ``I have participated in an informal savings group'' and ``I have borrowed money from an informal loan provider''. The treatment list included in addition the statement ``I have been approached by government officials or community leaders to provide them with any kind of assistance, like food or money''. Participants were told that they only needed to give the total number of these situations that had happened to them in the past month, not which of the situations applied to them.

We randomly assigned individuals to the treatment and control lists, stratifying the randomization by surveyor and treatment status. We estimate the following specification to evaluate the results of the list experiment:
\begin{equation}
\textit{Number of Statements}_{isn} = \rho_0 + \rho_1 \textit{Treated List}_{isn} + \rho_2 X_{isn} + v_{isn}
\end{equation}

where $s$ corresponds to the surveyor in charge of the survey. We control for strata fixed effects, $X_{isn}$. We do this combining the whole sample, for the control group only (which has been receiving aid payments for over a month) and for the treatment group only (which has not received any aid payments in over a month). Panel C of Table \ref{infTaxResults} shows the results of this exercise. Expanding the regression equation by adding an interaction term between the main treatment assignment (early vs. late aid payments) and the list experiment's treatment assignment does not change results, with the coefficient on the interaction term equal to 0.0338 (SE = 0.058).

\paragraph{Checking treatment impact heterogeneity} 

\begin{figure}[h!]
\begin{center}

\begin{subfigure}[t]{0.48\textwidth}
\includegraphics[width=\textwidth]{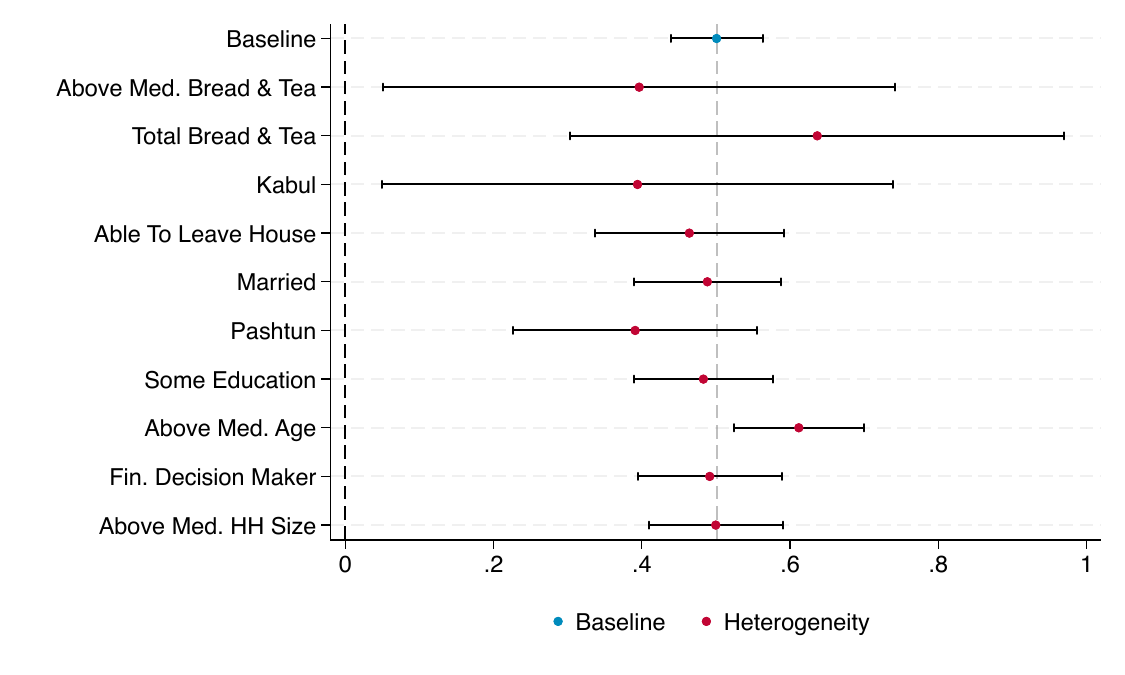}
\subcaption{Food Insecurity}
\end{subfigure}
\begin{subfigure}[t]{0.48\textwidth}
\includegraphics[width=\textwidth]{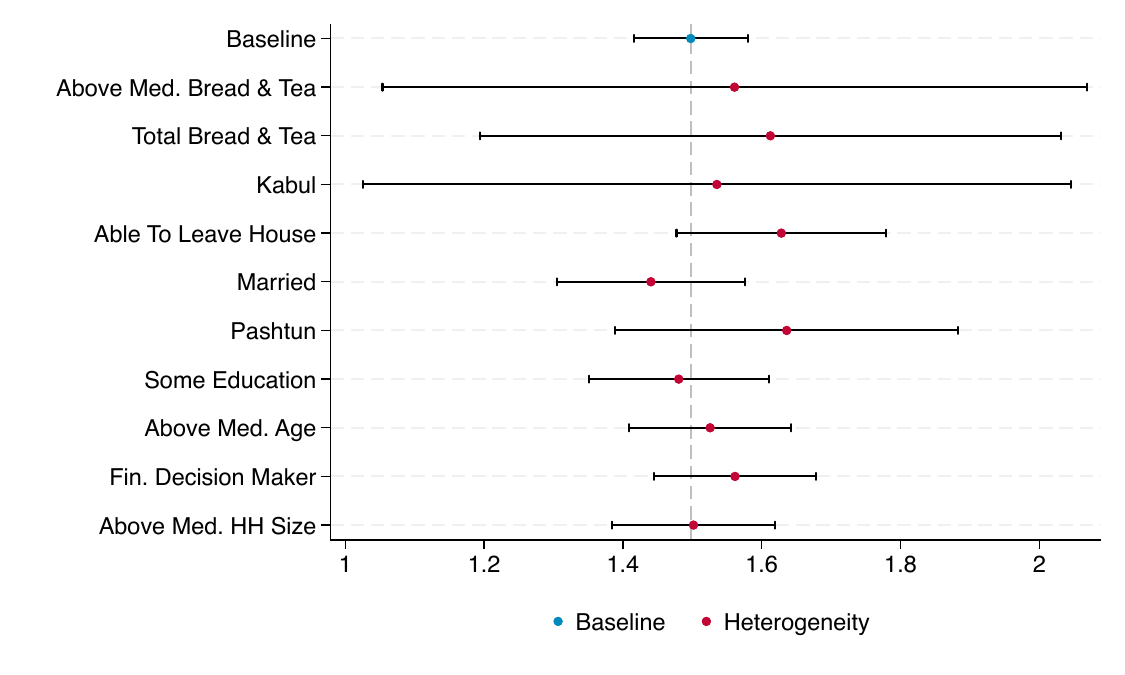}
\subcaption{Wellbeing}
\end{subfigure}
\begin{subfigure}[t]{0.48\textwidth}
\includegraphics[width=\textwidth]{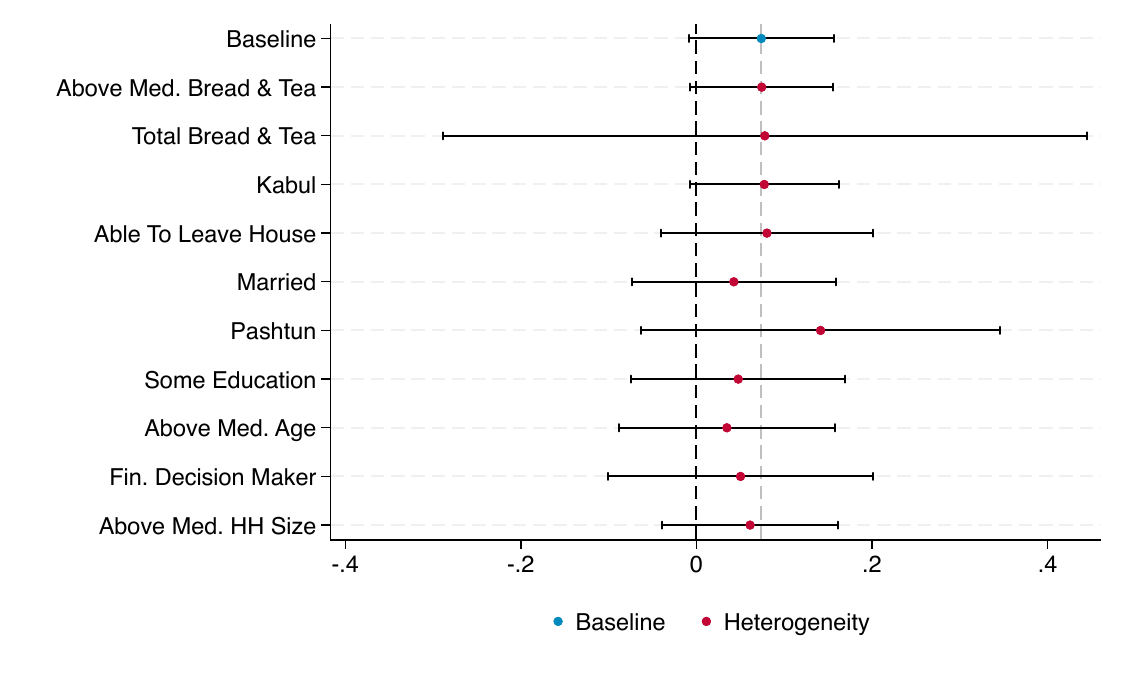}
\subcaption{Informal Taxation}
\end{subfigure}

\caption{Heterogeneity of Treatment Effects}
\label{fig_heterogeneity}

\end{center}
\noindent\scriptsize{\textbf{Notes:} The blue dots and the gray-shaded vertical lines correspond to the baseline estimate $\hat{\gamma}_1$ from Equation \eqref{eq_baseline}. The red dots correspond to $\hat{\mu}_1 + \hat{\mu}_2 + \hat{\mu}_3$} from Equation \eqref{eq_heterogeneity}. Bars show 95\% confidence intervals.

\end{figure}

We look for differential treatment effects on outcomes by subgroup by estimating the following specification:
\begin{equation}
\label{eq_heterogeneity}
\begin{aligned}
Y_{itn} =& \mu_0 + \mu_1 \mathbb{1}[\textit{Treatment Group}]_{in} \times \mathbb{1}[\textit{Heterogeneity}]_{in} + \mu_2 \mathbb{1}[\textit{Heterogeneity}]_{in} \\ &+ \mu_3 \mathbb{1}[\textit{Treatment Group}]_{in}
+ \mu_4 X_{i0n} + \mu_5 Y_{i0n} + \mu_6 \mathbb{1}[t = 2] + \nu_{itn}
\end{aligned}
\end{equation}

where $\mathbb{1}[\textit{Heterogeneity}]_{in}$ is a dummy for whether woman $i$ in Nahia $n$ belongs to a certain subgroup. We check for heterogeneity along the following dimensions: basic needs (total meals composed of only bread and tea), city of residence (Kabul vs. Mazar and Herat), whether the woman is able to leave the house at baseline, whether the woman is married, whether the woman is Pashtun, whether the woman has some education, whether the woman is above the median age, whether the woman is the household's financial decision-maker, and whether the household is above the median household size. For brevity, we only present results for the KLK indices of the three main outcome families in \ref{fig_heterogeneity}. 

\paragraph{Hypothetical cash versus digital aid}

During the fourth survey round, we asked participants a hypothetical question to measure their willingness to pay to receive their aid payments in cash rather than digital. The question asked ``We are hoping to use what we have learned from these surveys and from your experience with these payments to try to expand the program. While we do not have funding to do so at the moment, we are working to find it. In the future, we are also considering whether to give recipients the option to exchange the voucher for cash, rather than for goods at merchants. If we provide a cash out option, however, the fortnightly payments would be smaller because we have to pay a fee to make physical cash available.'' and then proceeded by asking participants ``If the fee was $X$ AFN, would you prefer $4000-X$ AFN in physical cash or $4000$ AFN in HesabPay credit?'', where $X\in\{100, 300, 500\}$. 

\begin{figure}[h!]
\begin{center}

\includegraphics[width=\textwidth]{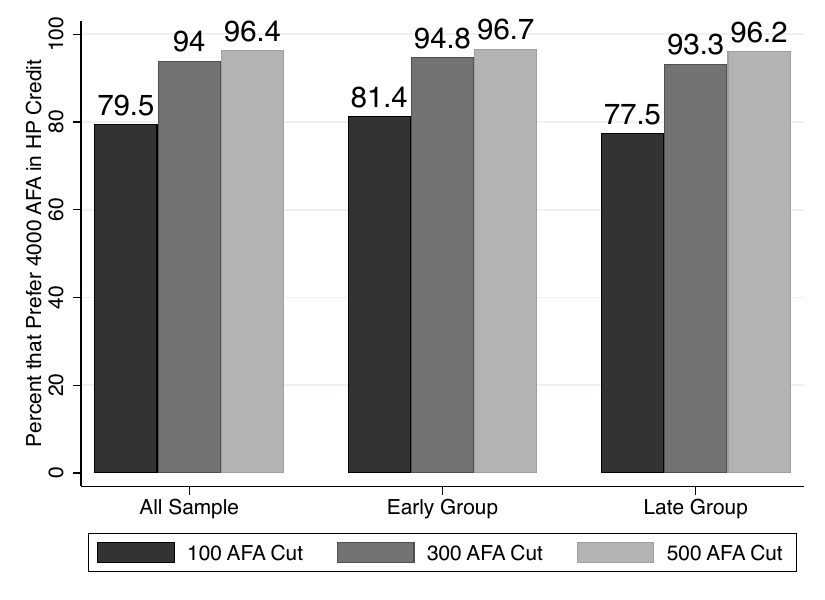}

\caption{Hypothetical Question on Preference Between Digital vs. Cash Aid}
\label{fig_cashOut}

\end{center}

\noindent\scriptsize{\textbf{Notes:} During the fourth survey round, we asked participants a hypothetical question to measure their willingness to pay to receive their aid payments in cash rather than digital. The question asked ``We are hoping to use what we have learned from these surveys and from your experience with these payments to try to expand the program. While we do not have funding to do so at the moment, we are working to find it. In the future, we are also considering whether to give recipients the option to exchange the voucher for cash, rather than for goods at merchants. If we provide a cash out option, however, the fortnightly payments would be smaller because we have to pay a fee to make physical cash available.'' We then proceeded by asking participants ``If the fee was $X$ AFN, would you prefer $4000-X$ AFN in physical cash or $4000$ AFN in HesabPay credit?'', where $X\in\{100, 300, 500\}$. The Figure presents the share of participants who preferred $4000$ AFN in HesabPay credit over 3900 AFN (black bar), 3700 AFN (dark gray bar), or 3500 AFN (light gray bar) in cash, for the whole sample (first three bars), or divided by treatment group.}

\end{figure}

Results of this exercise are presented in \ref{fig_cashOut}. Even with a fee as small as 100 AFN (2.5\% of the total payment), almost 80\% prefer receiving the aid digitally rather than in cash. A fee of 300 AFN that more accurately reflects the true costs of delivering the aid in cash results in 94\% of respondents preferring digital aid. The treatment group has a slightly higher preference for digital aid than the control group, potentially because they have had more experience receiving the aid digitally, although responses are very similar across the two groups.

Outside the study, the digital payments platform applied a 0.5\% commission for customers to withdraw cash from centralized agents, and a 3-5\% commission for community-based cash-out events (similar to a WFP cash distribution, but with individual digital vouchers) depending on the remoteness of the site. By contrast, purchases at registered merchants did not involve a fee.

\subsection{Identification Threats} 
\label{id_threats}

\paragraph{Experimenter demand effects}

Our primary outcomes are self-reported survey data. Because of social restrictions at the time of the study in Afghanistan, we could not send (even female) enumerators to interview beneficiaries in-person. Moreover, subjects cannot be blinded to their treatment status. As such, there is potential for experimenter demand effects (i.e. the participants answering what they believed we wanted to hear, not their true answers). In particular, it is plausible that respondents might either indicate that they are doing worse than they are in fact in order to influence the experimenter to send more aid. It is also plausible that respondents might want to overstate their wellbeing in order to provide more favorable evidence that might encourage policymakers to scale the program. We remain agnostic as to whether participants would under or overreport across measures. A priori, this could go either way, and could differ across treatment and control groups. Thus, we present results across the whole population and divided by treatment group.

To assess whether this is a problem in this setting, in the last round of follow-up surveys ($t = 2$) we ``primed'' participants by telling them the purpose of the study quite explicitly and checking whether that information affects their responses. This exercise is similar in spirit to the work by \citep{de2018measuring}. More specifically, we randomly assigned individuals into two groups: a ``primed'' group hears the following statement just before the questions related to needs: ``I would now like to ask you a few questions about how you and your family are doing. The goal of the CDDO and HesabPay program is to help you and your family meet basic needs, such as buying food, and we would like to see how you are doing in this regard. We will share what we learn from interviewing participants like yourself, with international organizations who are trying to help Afghans deal with these difficult times.'' Thus, this group is explicitly told what we are expecting to find. The ``not primed'' group hears this placebo statement instead: ``I would now like to ask you a few questions about how you and your family are doing.'' We stratified the random assignment by treatment status and the enumerator that will conduct the survey.

We run two types of specifications. First, to evaluate whether primed individuals give different answers than not primed individuals, we estimate the following specification:
\begin{equation}
\textit{Y}_{isn} = \psi_0 + \psi_1 Primed_{isn} + \psi_2 X_{isn} + \psi_3 Y_{is0n} + \omega_{isn}
\end{equation}

where we control for strata fixed effects (surveyor and treatment group status), $X_{isn}$ and the baseline value of the dependent variable (when available), $Y_{is0n}$. Standard errors are clustered at the individual level, the unit of randomization. 

In addition, we test whether the prime affected participants' responses differently depending on whether they belonged to the treatment or control group, by estimating the specification:
\begin{equation}
\begin{aligned}
\textit{Y}_{isn} =& \eta_0 + \eta_1 Primed_{isn} + \eta_2 Primed_{isn} \times \mathbb{1}[\textit{Treatment Group}]_{isn}+ \eta_3 \mathbb{1}[\textit{Treatment Group}]_{isn}  \\
&+ \eta_4 X_{isn} + \eta_5 Y_{is0n} + \omega_{isn}
\end{aligned}
\end{equation}

The results of the experimenter demand effect analysis are shown in \ref{ExpDemsumTableMain}. Column 2 shows the baseline estimates, $\hat{\gamma}_1$, column 3 shows the estimates for the overall experimenter demand effects, $\hat{\psi}_1$, column 4 shows the experimenter demand effects only for the control group, $\hat{\eta}_1$, and column 5 shows the (overall) experimenter demand effects for the treatment group, $\hat{\eta}_1 + \hat{\eta}_2$. 

\paragraph{Borrowing and spillovers}

There are two additional concerns that might affect our results. First, given that we told participants that they would eventually receive the treatment (however we did not tell them when this would happen), it is possible that they could have borrowed money at the time, alleviating needs in the short run. However, it is unlikely that this affects our results. When asked at baseline how difficult it would be for them to raise 1,500 Afghanis within a month in case of an emergency, only 7 women answered that this would be somewhat or very easy. Moreover, as explained above, our sample seems to have very little experience with financial instruments, and only one participant in the treatment group reported using her aid payments to repay debt. 

The second concern is that there might be spillovers from treated to control households. Given that our sample comes from only 16 Nahias and they met other participants during the onboarding sessions, it could be that the treatment group women, when they were receiving the aid payments, helped the control group women. However, our data indicate this is not likely to be a problem. Only around 30\% of our sample reports knowing another woman receiving aid payments from our program. Among the subsample that reports knowing another participating woman, just 2\% reports receiving any aid from another participating household (17 women in the control group).

\subsection{Deviations from PAP}
\label{ss:devPAP}

As mentioned before, our PAP is registered at \\ \href{https://www.socialscienceregistry.org/versions/160809/docs/version/document}{https://www.socialscienceregistry.org/versions/160809/docs/version/document}. All the analysis was conducted after the submission of this PAP.

We adhered to the PAP as closely as possible, although there were a few instances in which we deviated. We do not report Treatment on the Treated (ToT) estimates because there were no issues with non-compliance and survey response rates were extremely high. Given that attrition was so low and not different across treatment groups, we also do not show Lee Bounds results. We do not report results on participants' satisfaction with the digital payments platform, HesabPay, but there were only 2 instances of participants stating that they were ``somewhat dissatisfied'' with HesabPay (out of 2393 responses in the treatment group), with 2358 reporting being ``very satisfied''. In addition to testing for experimenter demand effects for the whole primed sample (pre-specified), we also show results disaggregated by treatment group (not pre-specified). We also don't report results on the ``Response timing'' section of the PAP, as there were no significant results (results are available upon request). Initially, we had described a broader set of questions for the experts' survey, but we decided to cut the number of questions from six to four in order to reduce the burden on respondents. The four questions we did ask match the topics in the PAP. We created a table summarizing all the main (pooled) results, in which the sample was restricted to the participants who answered all the relevant questions across the two months of surveying, to allay concerns of comparability (see \ref{ITTsumTableRest}). As a robustness check, we also include an alternative way of creating summary indices following \citep{anderson2008multiple}, another popular approach for constructing indices. This was also not pre-specified.

We deviate slightly from the PAP in the diversion results. Initially, we had pre-specified presenting results for each of the four measures of diversion, as well as an index created by combining the four measures following \citep{kling2007experimental}, the KLK index. Upon further consideration, and after computing the results, we decided to additionally show results using a dummy for whether the respondent answered yes to any of the four diversion tables (column 6 in Table \ref{infTaxResults}), which was not pre-specified. This is because, due to the lack of positive responses to the informal taxation questions and the way in which the index is created (standardising first each component variable, and then the final index) leads to extremely large values for those who answered yes to multiple questions. For reference, out of the 59 instances of participants answering yes to any of the four questions in any of the two survey rounds, 49 said yes to only one question, 9 said yes to two questions, and only one said yes to three questions (e.g. index value for this woman is 34, a clear outlier), with no one saying yes to all four. For completeness and robustness, \ref{infTaxResults_index} shows the results  in which the two measures of diversion of others are combined into a single KLK index (column 1), the two measures of diversion of the participants themselves are combined into a single KLK index (column 2), and an index of the four measures together following \citep{anderson2008multiple} (column 3). Moreover, we also present results dropping the one outlier that answered yes to three of the taxation questions in one round and received a KLK Index value above 34 standard deviations (column 4). None of these alternative indices was pre-specified. As can be observed, while the results using the pre-specified KLK index in Table \ref{infTaxResults} are only significant at the 10\% level, the other four alternative indices are all statistically insignificant at conventional levels. This suggests that the one outlier is what leads to the significant results in the main results. We also do not report the $p$-values adjusted for multiple hypothesis testing for the individual diversion measures, but they are all above 0.25.


\subsection{Interpreting magnitudes of needs results}
\label{ss:magnitude}

The Afghanistan Cash and Voucher Working Group (CVWG) estimates a Basic Food Basket for a family of seven costs approximately 96 USD per month \citep{Bete2022}, roughly equivalent to two biweekly direct aid transfers of 4000 AFA and a monthly survey incentive of 350 AFA. This basket was composed of 89 kg wheat flour, 21 kg domestic rice, 7 liters vegetable oil, 9 kg puSlses, and 1 kg salt at prevailing exchange rates in August 2022. For reference, the Minimum Expenditure Basket including food, healthcare, shelter, and other components totaled almost twice as much at 181.36 USD. In mid-December 2022, we contacted 25 merchants serving beneficiaries to better understand spending patterns and realized prices. We  confirmed that the most popular purchases were wheat flour, cooking oil and sugar, and using merchant-specific estimates of prices and volumes, verified that the cost of a typical basket of goods matched the total value of the aid payments and survey incentive. 

It is natural to ask whether the estimated impacts on basic needs are in line with what we should expect given the size of payments -- and more specifically, why hunger appears to persist throughout the length of the intervention. As noted above, we observe large reductions in each of the four primary food security outcomes, and these measures kept improving throughout the two months of payments (Table \ref{dynEffectsMain}, Panel A). While not eliminated, the number of days in which a person skipped meals in the past week declined after each payment (Figure~\ref{fig_longTerm}, Panel A). To better understand the explanatory factors for this persistence, we completed a brief survey on January 2 \& 3, 2023 with 52 randomly selected treatment group participants who reported any skipped meals in the second month of surveys (ie. December 2022). Consistent with the gradual phase-in of impacts, 31 respondents (60\%) reported no adults had skipped meals in the last week. The remaining respondents provided explanations for persistent needs consistent with a wide range of potential economic factors including larger-than-average needs, inter-temporal substitution, inter-household risk-sharing, and non-consumption substitution. 9 respondents (17\%) indicated the payment size was insufficient for everyone in their household to eat at least bread and tea at each meal, with several highlighting the difficulty of feeding larger families. 12 respondents (23\%) indicated saving some food for future consumption, but typically only mentioned having one or two weeks' worth on hand. 4 respondents (8\%) reported sharing food with individuals outside their household, and 3 respondents (6\%) reported bartering food for other expenses like medicine or rent.

For a comparison of magnitudes, albeit in a very different context, in 2013--14 UNHCR gave 100 USD per month for nearly six months (575 USD total) to Syrian refugee households in Lebanon and found that cash reduced the number of days/week that adults decreased their daily meal intake by 0.6 relative to the control mean of 3.2 \citep{LehmannMasterson2020}. Two other programs found positive effects, but the differing outcome measurements complicate effect size comparisons  \citep{PopleEtAl2021,BedoyaEtAl2019}. In four other programs, cash did not result in any improvements in similar measures of food security. Unconditional cash in South Sudan and cash-for-work in Central African Republic did not reduce the number of days with skipped meals \citep{AlikLagrangeEtAl2019,ChowdhuryEtAl2017}, vouchers in D.R.C. did not reduce days relying on a variety of coping strategies that included meal skipping \citep{QuattrochiEtAl2022}, and varying the number of cash disbursements (lump sum vs. three disbursements along with financial education and nudging) in Philippines did not produce any differential impacts on similar coping strategies \citep{MercyCorps2022}.

\section{Cost analysis}  \label{ss:costanalysis}

\subsection{Cost-efficiency}

We estimated three cost-efficiency quantities: the total cost-transfer ratio (TCTR), the cost-transfer ratio (CTR), and the cost-per-beneficiary (CPB). The TCTR is the ratio of total program costs to the transfer value, the CTR is the ratio of distribution/administrative costs to the transfer value, and the CPB is the total program costs for a household \citep{riegg2015cost}. 
All estimates are based on the variable operational costs of the digital payments provider, NGO, \&  community development organization. We excluded fixed costs, which consist of the digital provider's platform origination and standard maintenance costs. Consistent with standard practice, we also excluded evaluation costs given that future operation of the program would not depend on evaluation costs. Due to the security environment and the vulnerability of our subjects, we were not able to collect additional data on indirect costs. 

\ref{Ta:costefficiency} reports raw costs and estimated cost-efficiency quantities. Panel A reports raw costs (a) disaggregated by category, (b) including/excluding onboarding, and (c) by treatment and control group status, which both received transfers but at different times. 
Panel B reports the CPB, including/excluding the transfer value, and including/excluding onboarding. The total CPB is $192.00$ USD consisting of four approximately $45$ USD transfers amounting to $180$ USD and the variable operations costs totaling $12$ USD. Excluding the costs of onboarding, the total CPB is $182.44$ USD. Here, the transfer amount is the same ($180$ USD) but the administrative costs are $2.44$ USD. Panel C reports the TCTR and CTR, first including  onboarding, and then excluding onboarding costs. When including all costs, the TCTR is $1.067$, meaning that $6.7$ cents is required for each dollar transferred to a beneficiary (CTR). When excluding the costs of onboarding, the TCTR is $1.014$, meaning that $1.4$ cents is required for each dollar transferred (CTR). 

To contextualize our cost-efficiency estimate of $6.7$ cents, we estimate how many more individuals could have been served by the WFP in 2022 had their cash-based portfolio been delivered at $6.7$ cents rather than $17$ cents for each dollar. Using reasonable assumptions, we estimate that had WFP delivered all \$357M digitally, the savings would have been sufficient to support an additional 77,000 households or 538,075 individuals for the four month lean season. More conservatively, if we assume that WFP's costs were lower by 1 cent (CTR=0.16) and ours were higher by the same amount (CTR=0.077), then 432,250 individuals would have been supported through the four month lean season. More ambitiously, if we assume that WFP's costs were higher by 1 cent (CTR=0.18) and ours were lower by the same amount (CTR=0.057), then 641,613 additional individuals would have been supported. The replication materials contain our inputs, assumptions, calculations, and can be adjusted to explore different scenarios. 

\subsection{Cost-effectiveness}

We compute cost-effectiveness ratios (CERs) by dividing costs by treatment effects aggregated over participants and time \citep{riegg2015cost}, which are interpreted as the number of dollars required to achieve a defined impact. All CER estimates appear in \ref{Ta:costeffectivenessmealscalories}.  

We first estimated the cost required to achieve a reduction of a day with skipped meals, one of our key preregistered outcomes. The estimates rely on the following survey question: ``Over the past seven days, how many days did you or any other adults in your household skip meals because there were not enough resources for food?''. When a respondent reports the number of days, we assume that all meals are skipped for all adults in the household for those days. This corresponds to what we observe in the data. That is, there are 21 meals a week, 14 of which are bread and tea meals. Participants rarely eat any of the other food groups, which leaves seven meals unaccounted for. At baseline, they report skipping meals on average 2.5 days, which means that assuming all the meals in a given day are skipped by all adults translates roughly into the seven missing meals that we observe. Most conservatively, the average cost to achieve a reduction of a day with skipped meals is \$31.58 USD: \$231,936.08 $/$ ($1208$ beneficiaries $*$ $0.760$ fewer days with skipped meals per week $*$ 8 weeks). Panel A shows that across total possible days $(67,648)$, program participants skipped $7,344.64$ fewer days of meals, a 10.857pp decrease. Because there were four bi-weekly transfers, estimates from the program's second half (Weeks 5--8) and final week (Week 8) may better approximate a steady-state program effect. When based on  estimates from Weeks 5--8 (Panel B), \$22.91 USD results in a reduction of a day with skipped meals, aggregating to 14.964pp fewer days of skipped meals for treated households.
When based on estimates from Week 8, arguably the most appropriate given that the beneficiaries had now received the full \$180 transfer, \$16.90 USD results in a reduction of a day with skipped meals, which aggregates to 20.286pp fewer days of skipped meals for treated households (Panel C). 

Reducing days with skipped meals is important for many reasons, especially for improving the health of household members.
We draw on the CVWG estimates of calorie consumption per person-household-day \citep{Bete2022} to estimate the increase in calorie intake associated with a reduction of a day with skipped meals, which allows the CER to be represented as the cost required to increase calorie consumption by a given amount. CVWG estimates that a day of meals amounts to $2,100$ calories/person/day. 
In our sample, the average household has 6.31 members, making 13,251 the total possible calories/household/day. Based on these inputs and assumptions, a household of $6.31$ should consume $92,757$ calories per week, but on average skips $34,042$ each week (Panel D, control mean). With this benchmark, we can estimate the cost required to increase the number of calories by a fixed amount, here  by $1,000$. Averaging across the full eight-week program period (Panel D), we estimate that $2.38$ USD in digital aid results in an increase of 1,000 calories: \$231,936.08 $/$ (1208 beneficiaries $*$ (97,323,825 fewer forgone calories $*$ 8 weeks)). Based on estimates from the second half of the program (Panel E), \$1.73 USD results in a 1,000 calorie increase.
Based on estimates only from the program's final week (Panel F), \$1.28 USD results in a 1,000 calorie increase. 

Because digital aid continued to have an effect after the program ended (see Figure~\ref{fig_longTerm}), CERs can reasonably include longer-run effects in which 
costs remain the same, but are now scaled by the cumulative reduction in days with skipped meals across all sixteen weeks. For the post-program period, the treatment group is compared to the control group's baseline value (2.615) because in weeks 9--16 the control group is now receiving transfers and no longer constitutes a control. The treatment group's 12- and 16-week values could be compared to those of the control group just before receiving treatment (2.546), but by that point the control group could have experienced other changes, including updated expectations about entering the program or any effect of receiving the monthly 350 AFA survey incentive, which almost all households received given the high response rates. Alternatively, they could be compared to the treatment group's own baseline value (2.626) or an average of the treatment and control groups' baseline values. In practice, all three are similar and do not qualitatively change the inferences. 
Based on these inputs and assumptions, the approximately $180$ USD in digital aid to 1208 beneficiaries resulted in a reduction of 15,844.13 days that households skipped meals (7,344.64 during the eight-week program period and 8,499.49 in the eight-week post-program period) out of a possible $135,296$ possible days across the sixteen weeks. This is a 11.71 percentage point reduction relative to not receiving assistance, making 14.64 USD the cost to reduce a day with skipped meals (\$231,936.08/15,844.13). 
The corresponding, cumulative calorie intake estimate is 209,950,567 fewer forgone calories for the treatment group (15,844.13 fewer days * 13,251 calories per household-day), making 1.10 USD the cost to increase calorie intake by 1,000 (\$231,936.08/(209,950,567/1000)).   

By improving the food security and well-being of beneficiary households, it is likely that program beneficiaries experienced related improvements, including higher household productivity, and that the community experienced related improvements, such as better market conditions. Cost-benefit analysis characterizes the relationship between all direct/indirect benefits and all direct/indirect costs, both for the present as well as the future, by monetizing all benefits and then calculating a benefit-cost ratio, net present value, or an internal rate of return. Given the security context and the vulnerability of our subjects, unfortunately we were unable to collect sufficiently detailed information to assess these possible benefits and any other indirect costs that would make possible a formal cost-benefit analysis. Studying a vulnerable demographic in an oppressive environment such as Afghanistan means that certain types of data will always be difficult to obtain \citep{ClarkeDarcy2014}.

\section{Ethical considerations} \label{ss:ethics} 



We obtained Institutional Review Board approval on 4 May 2022 from the London School of Economics (\#89546). There is no local IRB in Afghanistan. After submitting our initial application with accompanying consent/instrument forms (22 April 2022), the IRB asked for one round of revisions (27 April 2022), which we resubmitted (3 May 2022) and then received approval (4 May 2022).

As part of the approval process, the IRB requested: clarifications about the digital payment provider, the qualifications of enumerators, the inclusivity of the recruitment process, privacy/confidentiality of the subjects, attention to a careful assessment of digital technologies in research outputs, COVID safety measures, a suggestion for the survey instrument, and a post-approval written update of how the registration process proceeded in practice. We provided a comprehensive response to each of the IRB's questions and requests, including providing the enumerator non-disclosure agreement and updated instruments. Additionally, the IRB asked us to provide a written update after beneficiary registration procedures were clarified (but before the program began), which we submitted on schedule (30 June 2022) along with results of three small-scale pilots. During the study, we submitted an amendment with plans to carry out a survey of experts (20 November 2022), which the IRB approved (Study \#145636 on 29 November 2022). 

A consortium of practitioners, local grassroots organizations, the digital payments provider, and academics co-designed the study. Local- and internationally-based Afghans either led or worked with each of these collaborative organizations and fully participated in all decision-making, helping to ensure representation of the views of the participants, sensitivity to possible risks, and fair distribution of the program's benefits and costs. Because we were using a new technology in an insecure and uncertain operating environment, we also designed the study to adhere faithfully to key international principles governing digital engagement: fair treatment, protection and accessibility of funds, prioritization of women, safeguarding of data,  designing for individuals, transparency, interoperability, responsive, and accountability in the value-chain \citep{Burton2020,WFP2021a,UnPrinciplesResponsibleDigital2021}. 

The study went through a due diligence phase of several months in which the team met weekly to assess the feasibility of implementing the program ethically. After launch, the entire team continued to meet every week to assess progress and implement any changes deemed necessary. The team was committed to early termination of the program, evaluation, or both, if adverse events were to occur. Notably, although the research team conducted multiple rounds of surveying, the main NGO partner also conducted its own internal evaluations, which also involved interviews and surveys with participants, which they reported as independent checks on the research team's evaluation. In what follows, we discuss the ethical dimensions of our study, which we organized around Belmont Report principles \citep{Belmont1979}, and further address considerations specific to insecure, humanitarian crisis environments \citep{Wood2006,Campbell2017,PuriEtAl2017,Wolfe2020}.


\subsection{Respect for persons}

All participants were adults and, due to low literacy, were verbally informed about the study (and provided their verbal consent) in either Dari or Pashtu. The consent process occurred during the onboarding/baseline survey as well as with each successive survey wave. Specifically, when beneficiaries were registered for the program during the onboarding sessions described in Online Appendix~\ref{pilotSec}, they were asked to participate in a baseline survey in which detailed information about the study was provided. In particular, the informed consent process included discussion of possible risks and benefits, their right to skip questions or opt out entirely without penalty or loss of benefits, our commitment to data confidentiality, and advance notice that they would be invited to participate in follow-up surveys conducted by phone. Moreover, each time beneficiaries were invited to participate in follow-up surveys during and after the program, they were reminded about opt-out, skipping questions, and data confidentiality.

When a program provides large financial resources to potential participants, subjects could feel pressure to participate even if they have reservations. If that is the case, then informed consent to participate in the program could be insufficient. Thus, it was important that the implementation and research teams fully commit themselves to early termination of the program if any possible signs of harm were to arise. The entire team met weekly throughout the design and execution of the project and evaluated this possibility in each session. Notably, the local community council liaison and the two local leaders of the enumeration team attended these calls and weighed in from their perspective as those closest to implementation and evaluation. No significant issues were identified during the study.

For the surveys, although subjects were informed that they could opt out of the survey or skip specific questions without any penalty related to the program (or survey incentives), we nonetheless designed the instrument to avoid objectionable material. We vetted the instrument with all partners and made important modifications based on input from all partners, especially the local Afghan contributors. In particular, the survey did not ask sensitive questions, with the exception of the diversion questions. To reduce researcher-subject power differentials, which can undermine subject autonomy, the enumerator team was made up entirely of local Afghan women. The all-women enumerator team helped ensure greater sensitivity to positionality concerns, which shapes whether and how well enumerators can respect subject autonomy or recognize diminished subject autonomy.

Treatment and control group beneficiaries who elected to participate in surveys received compensation for their time, which was transferred using the same digital payments platform described in the study. The in-person baseline survey and each of the first three monthly surveys entailed one-time transfers of 350 AFA, and the final phone survey entailed a one-time transfer of 800 AFA. The amount was calibrated to be economically meaningful, and corresponded roughly to the average reported monthly income of households in our surveys. At the same time, as 350 AFA represented only 5\% of the size of the monthly direct aid transfers, the incentive payment was intended to be small enough that beneficiaries would not feel pressured to participate in surveys. 

\subsection{Beneficence}

\noindent \textit{Risks:} We prioritized the long-standing principle of ``do no harm'' \citep{Belmont1979,Wood2006}. In addition to designing the study to minimize the risk of potential harm (See Online Appendix~\ref{pilotSec}), we also proactively monitored the program and evaluation for evidence of any realized harm. We identified several key risks. 

First, we anticipated that the biggest risk stemmed from the recent rise of the Taliban (August 2021), including the possibility that the Taliban would try to divert aid to themselves, or threaten subjects or local implementation partners for their participation. We took several steps to mitigate the risk, which included: (a) implementation through CDCs comprised of social workers, social organizers, and other community advocates (many of whom are women) and evaluation by local Afghan enumerators, which could ensure sensitivity to the context and improved ability to detect interference, (b) distribution of program benefits of sufficient size to meet humanitarian needs responsibly, but not so large that they would attract unnecessary attention, and (c) readiness to engage in public messaging including with the de facto authorities to address possible concerns. In the surveys, less than 2\% of participants reported any attempts at diversion. We quickly followed up on samples of those reports and learned that they were negligible and, in some cases, likely misreported.  We also contacted participating merchants, and they confirmed that they had not been approached for funds by government or community entities. Because we had full access to the participants' transaction data, we also confirmed where and how the funds were spent, which confirmed little risk of diversion.  

Second, diversion, threats, and retaliation would be possible primarily if the privacy and confidentiality of the participants were compromised. We sought to maximize privacy by conducting surveys by phone so that door-to-door presence would not unnecessarily expose participants. We sought to maximize confidentiality by being extremely cautious about what information we collected in the surveys and how we protected that information. We developed a data management plan, which included provisions to fully anonymize all personally identifiable information for any use outside of the research team. Although merchants were not direct research subjects, we also took steps to preserve their anonymity, including how precisely they are represented in resulting data and maps. As noted above, we were careful not to collect other sensitive information.

Third, a frequently cited concern with cash-based programs is intimate partner violence (IPV), although we note that recent rigorous studies and research syntheses indicate that IPV risks are in practice extremely small and confined to certain types of individuals and types of IPV \citep{HidroboEtAl2016,BullerEtAl2018,BastagliEtAl2019,BaranovEtAl2021,BlofieldEtAl2022}. For example, there may be isolated negative impacts for women with similar or higher education than their partner \citep{HidroboFernald2013} or when transfers are relatively large \citep{Angelucci2008}. They may also be more pronounced for emotional violence and threats rather than physical violence \citep{Bobonis2011,BobonisEtAl2013}. We deliberately targeted female-headed households as this would help mitigate the possibility of IPV by program design. (Two-thirds of participants reported not having a partner and two-thirds also reported that they are the household's financial decisionmaker.) Given the challenging context and vulnerability of our subjects, we could not ask about intimate partner violence directly. Instead, we looked to possible observable indicators, including self-reported measures of happiness and within-household diversion, which did not suggest reasons to be concerned. As with possible government- or community-level diversion, 98\% of women reported that they had not been approached by anyone for money, and 93\% reported that they alone made the decision on how to spend the funds. We did not receive any (even informal) reports of intimate partner violence whether from the survey enumerators or from the NGO's own internal evaluation team, which were conducted separately.

Fourth, although research ethics have largely focused on direct human subjects, we took seriously the minimization of risks to the implementers and research teams \citep{McDermottHatemi2020,APSA2020}. The implementer was an established community development organization, which had a long-time presence in all of the study sites, fully understood how to work with CDCs and community members, and had full legal approval to operate. It collected data for the baseline survey as part of the registration and onboarding. For follow-up surveys, the enumeration team operated entirely remotely, reducing the security risks associated with having a physical presence in communities and at the homes of beneficiaries. Given that enumerators were not going door to door, remote data collection also helped insure participants' privacy.   

From the study's inception, we sought the input of all partners, including especially local implementers and enumerators, and verified that they did not see any other risks of harm. Throughout the study, we repeatedly and proactively evaluated any realization of harm. Across all discussions with implementers and enumerators, we received no reports of retaliatory activity by the regime or others in the community. We also see no patterns in any of the data analysis that indicate other forms of harm that may have befell beneficiaries or merchants, or more broadly the local implementation or evaluation teams. 

\noindent \textit{Benefits:} Although minimizing the risk of harm is essential, the principle of beneficence entails maximizing possible benefits while at the same time minimizing the risks to subjects and society \citep{Belmont1979}. In practice, quantifying direct and indirect benefits can be challenging and perhaps even misleading \citep{Baele2013,Humphreys2015}, necessitating extreme caution about relying on a benefit-risk ``balance sheet approach''. As such, we were careful not to over promise, acknowledging in our informed consent communication that: ``There may or may not be any direct benefits to you from participating in this study.'' That said, because the program consisted of a cash-based transfer, redeemable at nearby merchants selling food and household items, the program reasonably provides some valuable humanitarian relief directly to participants, which is also an important justification for conducting a randomized trial during a humanitarian crisis \citep{Wolfe2020}. Although any subsequent scale up of digital aid should be considered prospective and, therefore, not weighed heavily (or at all) in calculations of potential societal benefits, we note that there could be downstream benefits to delivering aid in a decentralized, transparent, and cost-effective way to address the challenges of operating in such a difficult political and security environment. \\

\subsection{Justice}

Extreme need in Afghanistan far exceeded available program resources, making it critical that the program reach the most vulnerable. The program was implemented through CDCs, which were integrated in the communities, and had established practices for identifying households with extreme needs who were not already benefiting from other aid programs. Before launching the study, we conducted three small-scale pilots and found that nearly 80 per cent had reported skipping a meal or cutting down on meals in the past 30 days. Data collected in the full study further validated these levels of need. 

Beneficiaries were randomly assigned to treatment and control using a phase-in/waitlist design so that program benefits were not denied to anyone deemed eligible through the targeting and recruitment process. Although all study participants eventually received the program, a relevant question is whether half of those deemed eligible should be required to wait during a crisis. Put differently, is a randomized design appropriate in the midst of a humanitarian crisis? We addressed this in Section~\ref{conclusion} and further note that in contrast to the relatively long timeline of many randomized programs, the wait time was two months and both treated and control households received their assistance during the lean or hungry season. Given the high need in the entire sample, randomization was the most equitable approach for deciding on whether a household received assistance in the first versus the second two-month period.  Regardless of whether a household was assigned to the treatment or control group, they all received the survey incentive each time that they participated. As such, those in the control group did not accrue costs to their time that the treatment group did not, which is important among other reasons because if they were to attrit before receiving the program transfers then they would have paid higher costs.  

Because researchers and society might derive benefit from the study, we also informed possible participants so that they were aware. In particular, after noting that they ``may or may not benefit directly'' we acknowledged that researchers might benefit saying ``The investigator, however, may learn more about the usage of digital payment platforms in Afghanistan and how to deliver humanitarian aid'' and then raised the possibility of broader benefits for society saying ``and society may benefit from this knowledge.'' 

Finally, although not acknowledged to program participants, we note here that the funds provided by the donor were used entirely for programmatic purposes with the intended beneficiaries receiving nearly all of those program funds, as the cost-efficiency section discussed. The evaluation costs were funded entirely by external resources (in particular a grant from J-PAL's CVI initiative, study/award GR-1969, with funding provided by UK FCDO).

\clearpage 

\section{Supplementary material figures and tables}

\begin{figure}[h!]
\begin{center}

\begin{subfigure}[t]{0.9\textwidth}
\includegraphics[width=\textwidth]{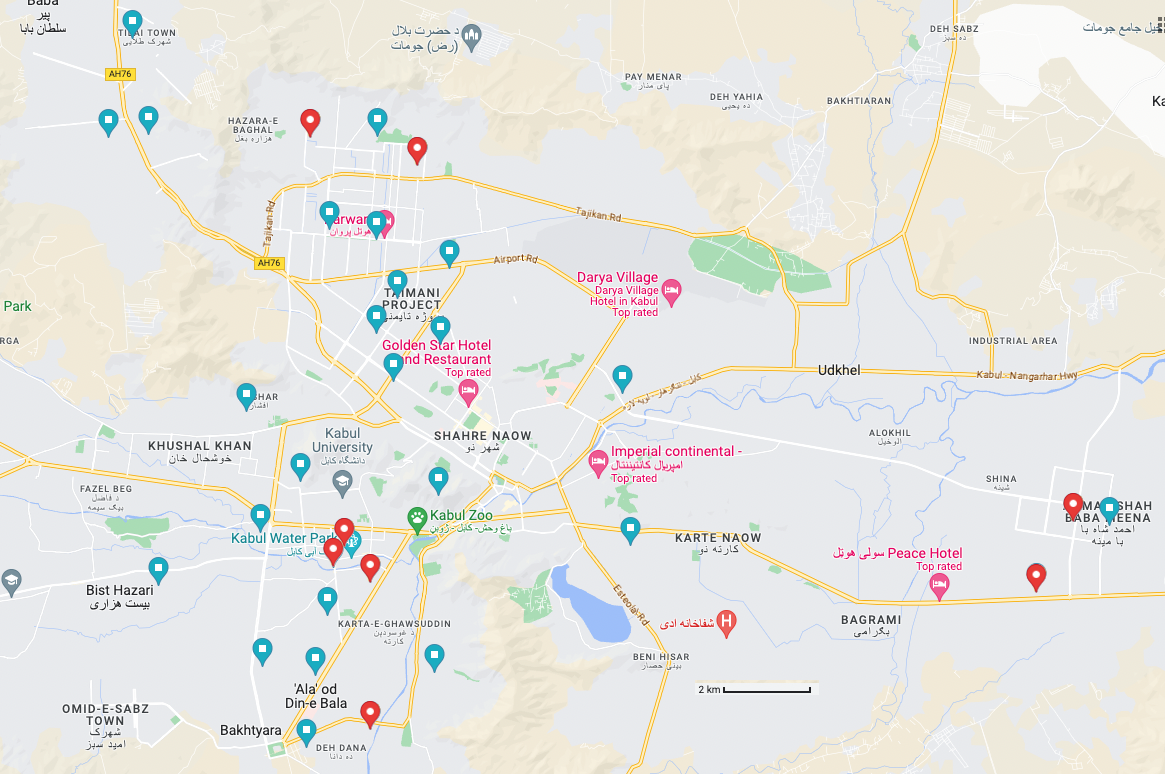}
\subcaption{Kabul}
\end{subfigure}
\begin{subfigure}[t]{0.49\textwidth}
\includegraphics[width=\textwidth]{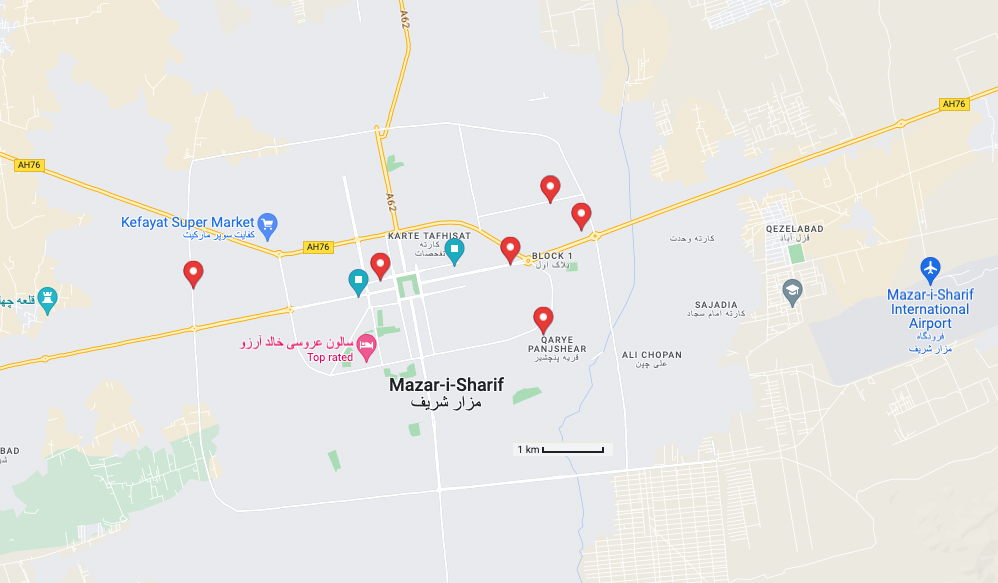}
\subcaption{Balkh}
\end{subfigure}
\begin{subfigure}[t]{0.49\textwidth}
\includegraphics[width=\textwidth]{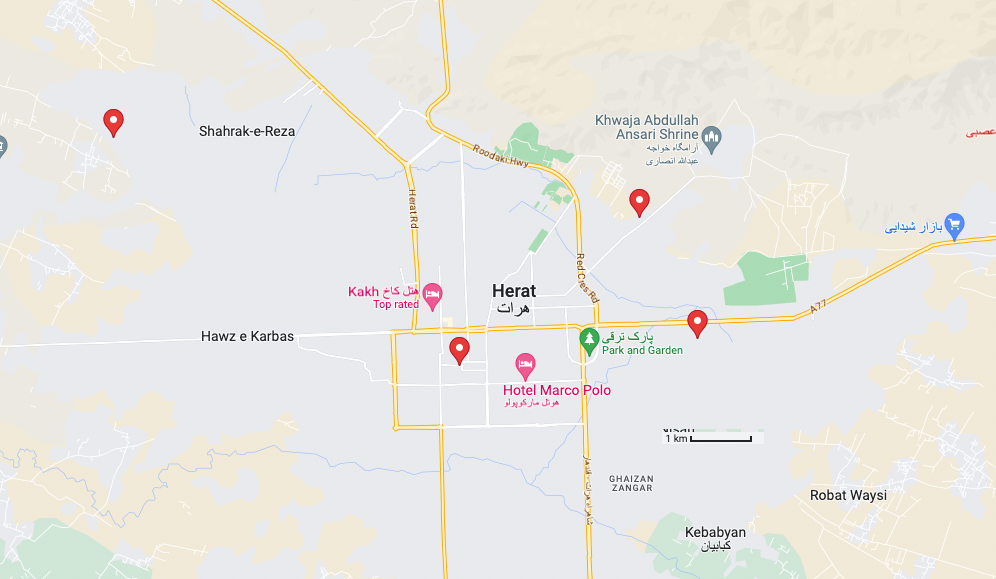}
\subcaption{Herat}
\end{subfigure}

\caption{Location of Merchants}
\label{fig_mercLoc}

\end{center}
\noindent\scriptsize{\textbf{Notes:} The map shows the location of merchants that participants visited at some point between the start of the program and December 31, 2022. Red pins are those that belong to merchants who participated in the onboarding sessions, while blue pins are those of other merchants. Note that not all merchants could be contacted/located. Coordinates have been jittered.}

\end{figure}

%
%
%

\clearpage

\begin{table}[h!] \centering
\newcolumntype{C}{>{\centering\arraybackslash}X}

\caption{Number of Merchants by Nahia}
\label{MerchByNahia}
\begin{adjustbox}{max width=\linewidth, max height=\textwidth}\begin{tabular}{lcccc}

\toprule
&  & \# & \# Onboarding & \# All \tabularnewline City & Nahia & Participants & Merchants & Merchants \tabularnewline
{}&{}&{(1)}&{(2)}&{(3)} \tabularnewline
\midrule \addlinespace[\belowrulesep]
\midrule Kabul&6&82&3&22 \tabularnewline
Kabul&7&83&2&19 \tabularnewline
Kabul&8&51&1&8 \tabularnewline
Kabul&11&137&1&23 \tabularnewline
Kabul&12&99&1&10 \tabularnewline
Kabul&13&152&3&25 \tabularnewline
Kabul&15&85&1&26 \tabularnewline
Kabul&17&98&1&15 \tabularnewline
Kabul&22&30&1&8 \tabularnewline
Herat&1&100&3&4 \tabularnewline
Herat&11&100&3&5 \tabularnewline
Herat&13&597&5&7 \tabularnewline
Balkh&4&200&1&8 \tabularnewline
Balkh&6&196&1&6 \tabularnewline
Balkh&7&200&3&4 \tabularnewline
Balkh&8&199&2&6 \tabularnewline
\bottomrule \addlinespace[\belowrulesep]

\end{tabular} \end{adjustbox}
\\ \parbox{\linewidth}{\scriptsize \noindent \justifying \(Notes \): The number of own test merchants reflects the merchants that participants visited during the onboarding session to conduct their test purchase. The number of all merchants includes all numbers that we can identify as merchants in the transaction data.}
\end{table}

\begin{table}[h!] \centering
\newcolumntype{R}{>{\raggedleft\arraybackslash}X}
\newcolumntype{L}{>{\raggedright\arraybackslash}X}
\newcolumntype{C}{>{\centering\arraybackslash}X}

\caption{Baseline Balance Check}
\label{baseBalCheck}
\begin{adjustbox}{max width=\linewidth, max height=\textwidth}\begin{tabular}{l|ccccc}

\toprule
  & \multicolumn{2}{c}{Whole Sample} & Treatment & Control & p-value \tabularnewline & Mean & SD & Mean & Mean & Difference \tabularnewline 
{Variable}&{(1)}&{(2)}&{(3)}&{(4)}&{(5)} \tabularnewline
\midrule \addlinespace[\belowrulesep]
\midrule \textbf{Panel A. Outcome Vars}&&&&& \tabularnewline
1. Total bread-tea meals&13.76&2.56&13.74&13.77&0.79 \tabularnewline
2. Skipped meals&2.62&1.52&2.63&2.62&0.86 \tabularnewline
3. Inf tax gov (others)&0.003&0.058&0.002&0.004&0.474 \tabularnewline
4. Inf tax leaders (others)&0&0.02&0&0.001&0.317 \tabularnewline
5. Inf tax gov (you)&0&0&0&0&1 \tabularnewline
6. Inf tax leaders (you)&0.002&0.041&0&0.003&0.045 \tabularnewline
7. Life satisfaction&3.53&1.17&3.51&3.54&0.59 \tabularnewline
8. Happy&0.01&0.1&0.01&0.01&0.99 \tabularnewline
&&&&& \tabularnewline
\textbf{Panel B. Heterogeneity Vars}&&&&& \tabularnewline
1. Married&0.34&0.47&0.34&0.34&0.88 \tabularnewline
2. Pashtun&0.1&0.3&0.11&0.09&0.15 \tabularnewline
3. Some education&0.37&0.48&0.37&0.36&0.37 \tabularnewline
4. Balkh&0.33&0.47&0.33&0.33&0.77 \tabularnewline
5. Herat&0.33&0.47&0.33&0.33&0.85 \tabularnewline
6. Kabul&0.34&0.47&0.34&0.34&0.95 \tabularnewline
7. Above median age&0.56&0.5&0.56&0.56&0.81 \tabularnewline
8. Fin. decision maker&0.66&0.47&0.66&0.66&0.99 \tabularnewline
9. Above median HH size&0.64&0.48&0.64&0.65&0.6 \tabularnewline
10. Able to leave house&0.65&0.48&0.64&0.66&0.56 \tabularnewline
Number of individuals&2409&&1208&1201& \tabularnewline
\bottomrule \addlinespace[\belowrulesep]

\end{tabular} \end{adjustbox}
\\ \parbox{\linewidth}{\scriptsize \justifying \noindent \(Notes \): The table shows, for different pre-specified variables at baseline, the overall mean and standard deviation (columns 1 and 2), the mean in the treatment group (column 3) and the mean in the control group (column 4). Column 5 shows the p-value of the difference between the means in the treatment and control group, adjusting for robust standard errors.}
\end{table}

\begin{table}[h!] \centering
\newcolumntype{R}{>{\raggedleft\arraybackslash}X}
\newcolumntype{L}{>{\raggedright\arraybackslash}X}
\newcolumntype{C}{>{\centering\arraybackslash}X}

\caption{Treatment Effects by Survey Round -- Secondary Outcomes}
\label{SvLSecTEs}
\begin{adjustbox}{max width=\linewidth, max height=\textwidth}\begin{tabular}{l|ccccc}

\toprule
{}&{(1)}&{(2)}&{(3)}&{(4)}&{(5)} \tabularnewline
\midrule \addlinespace[\belowrulesep]
\midrule \textbf{Panel A. Types of Food}&Rice&Beans&Vegetables&Chicken&Dairy \tabularnewline
\midrule \( \beta_1 \): Treated \( \times \) Round 2&0.366***&0.240***&0.014&0.001&0.032 \tabularnewline
&(0.061)&(0.050)&(0.071)&(0.011)&(0.023) \tabularnewline
\( \beta_2 \): Round 2&--0.051&--0.124***&0.164***&--0.008&--0.065*** \tabularnewline
&(0.040)&(0.030)&(0.051)&(0.006)&(0.015) \tabularnewline
\( \beta_3 \): Treated&0.413***&0.373***&--0.011&0.012&0.031 \tabularnewline
&(0.049)&(0.041)&(0.056)&(0.009)&(0.021) \tabularnewline
Control Mean&0.698&0.518&1.394&0.021&0.074 \tabularnewline
Observations&4,763&4,763&4,763&4,763&4,763 \tabularnewline
R Squared&0.088&0.084&0.013&0.012&0.014 \tabularnewline
&&&&& \tabularnewline
\midrule \textbf{Panel B. Other Economic Outcomes}&Income&Employed&Fin. Decision-Maker&Medicine Purchase& \tabularnewline
\midrule \( \beta_1 \): Treated \( \times \) Round 2&36.972&0.050**&0.014&0.008& \tabularnewline
&(151.108)&(0.023)&(0.022)&(0.019)& \tabularnewline
\( \beta_2 \): Round 2&--165.374**&0.032*&0.030**&--0.046***& \tabularnewline
&(64.319)&(0.017)&(0.015)&(0.012)& \tabularnewline
\( \beta_3 \): Treated&109.985&--0.033*&0.010&0.031**& \tabularnewline
&(149.582)&(0.018)&(0.019)&(0.015)& \tabularnewline
Control Mean&876.683&0.192&0.666&0.051& \tabularnewline
Observations&4,763&4,741&4,757&3,582& \tabularnewline
R Squared&0.012&0.024&0.064&0.015& \tabularnewline
\bottomrule \addlinespace[\belowrulesep]

\end{tabular} \end{adjustbox}
\\ \parbox{\linewidth}{\scriptsize \noindent \justifying \(Notes \): This table reports estimated impacts of treatment separately for the first and second survey round. Households were surveyed once per month for two months. Each of these months constitutes a survey round. All specifications control for stratum fixed effects and the baseline value of the dependent variable, if available. Standard errors are clustered at individual level.
\\   \textit{Levels of significance}: *$ p<0.1$ , **$ p<0.05$ , ***$ p<0.01$.}
\end{table}

\begin{table}[h!] \centering
\newcolumntype{R}{>{\raggedleft\arraybackslash}X}
\newcolumntype{L}{>{\raggedright\arraybackslash}X}
\newcolumntype{C}{>{\centering\arraybackslash}X}

\caption{Summary Table -- Experimenter Demand Effects}
\label{ExpDemsumTableMain}
\begin{adjustbox}{max width=\linewidth, max height=\textwidth}\begin{tabular}{lccccccccc}

\toprule
 &  &  & \multicolumn{4}{c}{\textbf{Experimenter Demand}} \tabularnewline  & Control & Baseline & Overall & FWER & Control & Treatment \tabularnewline  & Mean & Estimate & Estimate & \(p\)-value & Estimate & Estimate \tabularnewline 
{Variable}&{(1)}&{(2)}&{(3)}&{(4)}&{(5)}&{(6)} \tabularnewline
\midrule \addlinespace[\belowrulesep]
\midrule \textbf{Panel A. Food Security}&&&&&& \tabularnewline
Days skipping meals (past week)&2.615&--0.76***&--0.031&0.7225&--0.082&0.041 \tabularnewline
&&(0.051)&(0.062)&&(0.087)&(0.077) \tabularnewline
Children skipping meals (=1)&0.873&--0.117***&0.016&0.3913&0.028&0.009 \tabularnewline
&&(0.012)&(0.016)&&(0.021)&(0.024) \tabularnewline
Regularly eat twice a day&0.501&0.093***&--0.001&0.925&--0.004&--0.001 \tabularnewline
&&(0.015)&(0.014)&&(0.02)&(0.02) \tabularnewline
Total bread and tea meals (past week)&13.639&--1.608***&--0.186&0.3897&--0.059&--0.267 \tabularnewline
&&(0.121)&(0.161)&&(0.212)&(0.217) \tabularnewline
\textit{Food Security - KLK Index}&0&0.501***&0.005&&--0.011&0.007 \tabularnewline
&&(0.032)&(0.039)&&(0.05)&(0.051) \tabularnewline
&&&&&& \tabularnewline
\textbf{Panel B. Informal Taxation}&&&&&& \tabularnewline
Inf. tax gov. off. (others)&0.004&0.001&--0.002&0.4468&--0.002&--0.001 \tabularnewline
&&(0.001)&(0.002)&&(0.002)&(0.003) \tabularnewline
Inf. tax comm. leader (others)&0.001&0.002&--0.002&0.4513&0&--0.005 \tabularnewline
&&(0.002)&(0.003)&&(0.002)&(0.005) \tabularnewline
Inf. tax gov. off. (you)&0&0.002**&--0.001&0.4513&--0.001&--0.001 \tabularnewline
&&(0.001)&(0.001)&&(0.001)&(0.003) \tabularnewline
Inf. tax comm. leader (you)&0.003&0.002&--0.006**&0.2026&--0.006&--0.006 \tabularnewline
&&(0.003)&(0.003)&&(0.004)&(0.004) \tabularnewline
\textit{Informal tax. - KLK Index}&0&0.074*&--0.1*&&--0.075&--0.127 \tabularnewline
&&(0.042)&(0.057)&&(0.05)&(0.093) \tabularnewline
&&&&&& \tabularnewline
\textbf{Panel C. Wellbeing}&&&&&& \tabularnewline
Better economic situation&0.048&0.335***&0.006&0.9754&0.016&--0.012 \tabularnewline
&&(0.011)&(0.019)&&(0.017)&(0.026) \tabularnewline
Satisfied with fin. situation&0.133&0.263***&--0.003&0.9754&0.005&--0.017 \tabularnewline
&&(0.012)&(0.018)&&(0.02)&(0.026) \tabularnewline
Happy&0.009&0.28***&--0.005&0.9754&--0.014&--0.003 \tabularnewline
&&(0.014)&(0.017)&&(0.02)&(0.022) \tabularnewline
Life satisfaction (std)&0.011&1.682***&0.002&0.9754&0.043&--0.081 \tabularnewline
&&(0.058)&(0.089)&&(0.098)&(0.111) \tabularnewline
\textit{Economic/Wellbeing - KLK Index}&0&1.498***&0.004&&0.027&--0.059 \tabularnewline
&&(0.042)&(0.071)&&(0.064)&(0.086) \tabularnewline
\bottomrule \addlinespace[\belowrulesep]

\end{tabular} \end{adjustbox}
\\ \parbox{\linewidth}{\scriptsize \justifying \noindent \(Notes \): Control for stratification fixed effects, survey round fixed effects (baseline estimates), and baseline value of dependent variable, if available. Standard errors clustered at individual level in parenthesis. Control mean is the mean in the baseline if available, or across follow up rounds otherwise, for the control group. The KLK Index is created following Katz, Kling, \& Liebman (2007), and is the equally-weighted sum of the standardised component variables. The baseline effect is the (pooled) ITT effect of the main treatment (receiving the aid payments). Primary outcomes show FWER-adjusted p-values within each family outcome (following Romano \& Wolf, 2005, using 5000 repetitions). The overall effect is the coefficient on the prime treatment. The control effect is the coefficient on the prime treatment in a regression where the prime treatment and the main treatment are interacted, while the effect on the treated is the sum of the prime treatment and the interaction term between the two treatments from the same regression. Better economic situation is an index that equals 1 if the respondent answered that her economic situation compared to 30 days ago is slightly or much better, and 0 otherwise. Satisfied with financial situation is a dummy that equals 1 if the respondent answered that she agrees a lot or somewhat with the statement that she is highly satisfied with her current financial condition, and 0 otherwise. Happy is a dummy that equals 1 if respondent said that she was very happy or quite happy, and 0 otherwise. Life satisfaction is the score from 1 (dissatisfied) to 10 (satisfied) in terms of how satisfied the respondent is with her life as a whole these days (standardised). Total household income excludes the aid payments. \\   \textit{Levels of significance}: *$ p<0.1$ , **$ p<0.05$ , ***$ p<0.01$.}
\end{table}

\begin{table}[h!] \centering
\newcolumntype{R}{>{\raggedleft\arraybackslash}X}
\newcolumntype{L}{>{\raggedright\arraybackslash}X}
\newcolumntype{C}{>{\centering\arraybackslash}X}

\caption{Summary Table -- Treatment Effects, Restricted Sample}
\label{ITTsumTableRest}
\begin{adjustbox}{max width=\linewidth, max height=\textwidth}\begin{tabular}{lccccccc}

\toprule
& Control & Control & Treatment & Standard & Naive & Adjusted & \tabularnewline & Mean & SD & Effect & Error & \textit{p}-value & \textit{p}-value & N \tabularnewline 
{}&{(1)}&{(2)}&{(3)}&{(4)}&{(5)}&{(6)}&{(7)} \tabularnewline
\midrule \addlinespace[\belowrulesep]
\midrule \textbf{Panel A. Primary Outcomes}&&&&&&& \tabularnewline
Days skipping meals (past week)&2.579&1.813&--0.785&0.053&0&0.0002&4412 \tabularnewline
Children skipping meals (=1)&0.87&0.336&--0.113&0.012&0&0.0002&4412 \tabularnewline
Regularly eat twice a day&0.493&0.5&0.1&0.015&0&0.0002&4412 \tabularnewline
Total bread and tea meals (past week)&13.644&3.921&--1.628&0.124&0&0.0002&4412 \tabularnewline
\textit{Food Security - KLK Index}&--0.016&1.01&0.511&0.033&0&&4412 \tabularnewline
\textit{Food Security - Anderson Index}&0.002&1.058&0.494&0.034&0&&4412 \tabularnewline
&&&&&&& \tabularnewline
Inf. tax gov. off. (others)&0.002&0.043&0.001&0.001&0.563&0.6818&4412 \tabularnewline
Inf. tax comm. leader (others)&0.002&0.043&0.003&0.002&0.116&0.108&4412 \tabularnewline
Inf. tax gov. off. (you)&0.001&0.03&0.002&0.001&0.163&0.1524&4412 \tabularnewline
Inf. tax comm. leader (you)&0.005&0.071&0.001&0.002&0.609&0.6818&4412 \tabularnewline
\textit{Informal tax. - KLK Index}&--0.007&0.923&0.075&0.041&0.067&&4412 \tabularnewline
\textit{Informal tax. - Anderson Index}&--0.016&0.763&0.048&0.031&0.115&&4412 \tabularnewline
&&&&&&& \tabularnewline
Better economic situation&0.047&0.212&0.34&0.011&0&0.0002&4412 \tabularnewline
Satisfied with fin. situation&0.133&0.339&0.264&0.013&0&0.0002&4412 \tabularnewline
Happy&0.235&0.424&0.275&0.014&0&0.0002&4412 \tabularnewline
Life satisfaction&2.969&1.888&2&0.07&0&0.0002&4412 \tabularnewline
\textit{Economic/Wellbeing - KLK Index}&0.001&1&1.51&0.043&0&&4412 \tabularnewline
\textit{Economic/Wellbeing - Anderson Index}&0.096&1.167&1.326&0.045&0&&4412 \tabularnewline
&&&&&&& \tabularnewline
\textbf{Panel B. Secondary Outcomes}&&&&&&& \tabularnewline
Days eating rice (past week)&0.672&1.05&0.616&0.036&0&0.001&4412 \tabularnewline
Days eating beans (past week)&0.517&0.841&0.491&0.03&0&0.001&4412 \tabularnewline
Days eating vegetables (past week)&1.377&1.331&0&0.042&0.991&0.415&4412 \tabularnewline
Days eating chicken (past week)&0.018&0.138&0.014&0.006&0.021&0.022&4412 \tabularnewline
Days eating dairy (past week)&0.073&0.387&0.05&0.014&0&0.001&4412 \tabularnewline
Able to buy medicine&0.078&0.269&0.031&0.01&0.002&0.004&3344 \tabularnewline
Involved in fin. decisions&0.672&0.47&0.019&0.016&0.228&0.15&4412 \tabularnewline
Total household income (past month)&1093.634&1792.27&156.196&94.148&0.097&0.07&4412 \tabularnewline
Household's head employed (past month)&0.284&0.451&--0.008&0.014&0.573&0.274&4412 \tabularnewline
\bottomrule \addlinespace[\belowrulesep]

\end{tabular} \end{adjustbox}
    \\ \parbox{\linewidth}{\scriptsize \justifying \noindent \(Notes \): Controls for stratification fixed effects, survey round fixed effects, and baseline values of dependent variables, if available, are included. Standard errors are clustered at the individual level. Primary outcomes show FWER-adjusted p-values within each family outcome (following Romano \& Wolf, 2005, using 5000 repetitions), while secondary outcomes show FDR-adjusted p-values (following Anderson, 2008). The KLK Index is created following Katz, Kling, \& Liebman (2007), and is the equally-weighted sum of the standardised component variables. The Anderson Index is created following Anderson (2008), and weights the component variables by the inverse of their variance-covariance matrix. Better economic situation is an index that equals 1 if the respondent answered that her economic situation compared to 30 days ago is slightly or much better, and 0 otherwise. Satisfied with financial situation is a dummy that equals 1 if the respondent answered that she agrees a lot or somewhat with the statement that she is highly satisfied with her current financial condition, and 0 otherwise. Happy is a dummy that equals 1 if respondent said that she was very happy or quite happy, and 0 otherwise. Life satisfaction is the score from 1 (dissatisfied) to 10 (satisfied) in terms of how satisfied the respondent is with her life as a whole these days. Total household income excludes the aid payments. In Round 1, there are 207 women who did not respond to every single question (out of 2381 respondents), while in Round 2 there are 145 (out of 2383 respondents). }
\end{table}

\begin{table}[h!] \centering
\newcolumntype{C}{>{\centering\arraybackslash}X}

\caption{Summary Statistics}
\label{baseBalComparison}
\begin{adjustbox}{max width=\linewidth, max height=\textwidth}\begin{tabular}{l|ccc}

\toprule
& Mean & SD & N \tabularnewline
{Variable}&{(1)}&{(2)}&{(3)} \tabularnewline
\midrule \addlinespace[\belowrulesep]
\midrule \textit{Panel A. Outcome Vars}&&& \tabularnewline
1. Bread-tea breakfast&6.67&0.89&2409 \tabularnewline
2. Bread-tea lunch&3.58&1.34&2409 \tabularnewline
3. Bread-tea dinner&3.51&1.49&2409 \tabularnewline
4. Total bread-tea meals&13.76&2.56&2409 \tabularnewline
5. Afford Medicine&0.01&0.09&2300 \tabularnewline
6. Income&357.92&291.75&2409 \tabularnewline
7. Life satisfaction&3.53&1.17&2409 \tabularnewline
8. Skipped meals&2.62&1.52&2408 \tabularnewline
9. Happy&0.01&0.1&2408 \tabularnewline
10. Employed&0&0.03&2405 \tabularnewline
&&& \tabularnewline
\textit{Panel B. Demographic Vars}&&& \tabularnewline
1. Married&0.34&0.47&2409 \tabularnewline
2. Some education&0.37&0.48&2404 \tabularnewline
3. Fin. decision maker&0.66&0.47&2409 \tabularnewline
4. Age&43&12.81&2409 \tabularnewline
5. Number family members&6.31&2.05&2409 \tabularnewline
6. Has had bank account&0&0.04&2409 \tabularnewline
7. Able to leave house&0.65&0.48&2409 \tabularnewline
8. Has transf. airtime&0&0.02&2409 \tabularnewline
9. Has transf. money&0&0&2409 \tabularnewline
10. Credit constrained&1&0.05&2409 \tabularnewline
\bottomrule 

\end{tabular} \end{adjustbox}
\\ \parbox{0.6\linewidth}{\footnotesize \noindent \justifying \(Notes \): Table shows the value of different variables at baseline, collected during the onboarding sessions, for the whole sample. Column 1 shows the mean of the variable, column 2 the standard deviation, and column 3 the number of respondents for this question at baseline.}
\end{table}

\begin{table}[h!] \centering
\newcolumntype{R}{>{\raggedleft\arraybackslash}X}
\newcolumntype{L}{>{\raggedright\arraybackslash}X}
\newcolumntype{C}{>{\centering\arraybackslash}X}

\caption{Are Digital Payments Diverted? Indices}
\label{infTaxResults_index}
\begin{adjustbox}{max width=\linewidth, max height=\textwidth}\begin{tabular}{lcccc}

\toprule
& KLK Index & KLK Index & Anderson Index & Dropping \tabularnewline & Others & You & All & Outlier \tabularnewline 
{}&{(1)}&{(2)}&{(3)}&{(4)} \tabularnewline
\midrule \addlinespace[\belowrulesep]
\midrule \textbf{Panel A. Baseline}&&&& \tabularnewline
Treated&0.034&0.043&0.049&0.059 \tabularnewline
&(0.026)&(0.031)&(0.036)&(0.039) \tabularnewline
Observations&4,611&4,647&4,648&4,647 \tabularnewline
Control Mean&--0.003&0.015&0.011&0.012 \tabularnewline
&&&& \tabularnewline
\midrule \textbf{Panel B. Long-Run}&&&& \tabularnewline
\( \beta_1 \): Treated \( \times \) Round 2&0.072&0.024&0.051&0.062 \tabularnewline
&(0.053)&(0.061)&(0.071)&(0.079) \tabularnewline
\( \beta_2 \): Round 2&--0.048&--0.096***&--0.135***&--0.137*** \tabularnewline
&(0.029)&(0.037)&(0.046)&(0.045) \tabularnewline
\( \beta_3 \): Treated&--0.073&0.007&--0.027&--0.034 \tabularnewline
&(0.085)&(0.109)&(0.125)&(0.137) \tabularnewline
Observations&4,611&4,647&4,648&4,647 \tabularnewline
(\( \beta_1 \) + \( \beta_2 \) + \( \beta_3 \)) / \( \beta_3 \) &0.670&--9.410&4.090&3.230 \tabularnewline
\bottomrule \addlinespace[\belowrulesep]

\end{tabular} \end{adjustbox}
\\ \parbox{\linewidth}{\scriptsize \justifying \noindent \(Notes \): In Panels A and B, controls for stratification fixed effects, survey round fixed effects, and baseline values of dependent variables, if available, are included. Columns 1-2 create indices following Kling, Liebman \& Katz (2007). Column 1 uses the two measures on whether others in their community have been asked to provide informal assistance, while column 2 uses the two measures on whether participants themselves have been asked to provide informal assistance by political actors. Column 3 includes the four measures, but creates the index following Anderson (2008). Column 4 uses the KLK Index composed of the four informal taxation questions (as in the baseline results), but dropping the one observation with the highest KLK Index value. Control for surveyor fixed effects are included. Standard errors are clustered at the individual level.}
\end{table}

\begin{table}[]
\caption{Experts' Survey Questionnaire} 
\label{expSurveyQs}
\begin{tabular}{p{6.27in}|p{0.7in}} 
Statement & Answer Options \\ \hline
THIS SURVEY CONSISTS OF 5 QUESTIONS THAT SHOULD TAKE LESS THAN 3   MINUTES TO COMPLETE. In conflict settings, distributing humanitarian aid while supporting the dignity and inclusion of vulnerable populations is challenging, in particular where there is a risk of capture by hostile regimes. The widespread adoption of mobile phones suggests one innovative solution: direct aid via digital   financial platforms. Digital payments have the potential to empower   recipients to meet basic needs using local markets, and to improve   transparency while minimizing opportunities for diversions compared to   physical cash or commodity distribution. We are piloting a direct aid program to Afghan women using a commercial   platform called HesabPay. Local partners have identified 2500 highly   vulnerable women in three major cities (Kabul, Herat, and Mazar), each of   whom will receive four semi-monthly digital payments of 50 USD. All   beneficiaries complete an in-person onboarding that includes identity   verification, registration of a digital wallet linked to a unique mobile phone number, and a test transaction using the digital wallet to purchase   goods from a registered merchant. Aid payments are unconditional   and can be used for purchases at any HesabPay-registered local   merchant. To inform our assessment of this program, we are collecting the views of   experts like yourself about the likelihood of operational success and   anticipated impacts. Your responses will be anonymized. If you opt in, we   will contact you again after the study is complete to provide the final   research findings as well as the aggregate views of those who completed this   survey. Based on the short program description above, please respond to the   following questions - please select ``Not Applicable'' if you are not able   to provide a forecast for a given question. &      \\
& \\ \hline
Please check this box to consent to participate in this survey. Only the researchers at the London School of Economics and Political Science will have access to your personal information.    &    Yes  \\
& \\ \hline
Q1) What percentage of beneficiaries do you expect will be able to use their digital payments to purchase goods directly without resorting to cashing out the aid? &   0-100\%   \\ 
& \\ \hline
Q2) How many meals per week do you expect beneficiaries of direct aid to eat only bread and tea? (At baseline, they reported ~14 meals of only bread and tea out of 21 meals in the last week.)      &  0-21 Bread and Tea Meals \\
& \\ \hline
Q3) What share of beneficiaries do you expect will report efforts to tax or divert their payments?   &   0-100\%   \\
& \\ \hline
Q4) What do you expect the delivery cost to be for our digital direct aid payments - not including costs of beneficiary selection or impact evaluation? (By one recent estimate, the delivery cost of humanitarian aid distributed in physical cash in Afghanistan is approximately 17\%.)    &     0-100 cents \\
\hline
\end{tabular}
\end{table}

\begin{table}
\scriptsize
\begin{center}
\caption{Existence of Necessary Conditions in Fragile Settings} \label{CountryCharacteristics}
\begin{tabular}{l|llll|lll}  
\hline \hline
 & \multicolumn{2}{c}{Food insecurity}  &   &  &  & Mobile  & Mobile   \\ 
Country & Millions & Share of population & Conflict  & Freedom & CDC & phone & money  \\ 
 \hline 
DRC             &  26.4 &  26\% &  Yes  & Not free    & Yes   & 45.50 & Medium \\
Ethiopia        &  23.6 &  21\% &  Yes  & Not free    & Yes   & 49.44 & Medium \\
Afghanistan     &  19.9 &  46\% &  Yes  & Not free    & Yes   & 58.26 & Very low \\
Nigeria         &  19.5 &  12\% &  Yes  & Partly free & Yes   & 99.07 & Medium \\
Yemen           &  17.3 &  55\% &  Yes  & Not free    & Yes   & 60.49 & No data \\
Myanmar         &  15.2 &  27\% &  Yes  & Not free    & Yes   & 95.36 & Medium \\
Syria           &  12.1 &  55\% &  Yes  & Not free    & No data    & 95.20 & No data \\
Sudan           &  11.7 &  24\% &  Yes  & Not free    & Yes   & 80.26 & Very low \\
Ukraine         &  8.9  &  25\% &  Yes  & Partly free & No data    & 100.0 & No data \\
Pakistan        & 8.6   &  43\% &  Yes  & Partly free & Yes   & 79.51 & Medium \\
\hline 
\end{tabular}
\end{center}

\parbox{\linewidth}{\scriptsize \justifying \(Notes \): Food insecurity  \citep{grfc2023}, with the share of population being the share of the population analyzed by FSIN. Active armed conflict \citep{DaviesEtAl2023}. Level of freedom \citep{FreedomHouse2023}. Community development council (CDC) presence coded from various sources (authors' research). Mobile-cellular subscriptions per 100 people  \citep{ITU2022}. Mobile money prevalence index \citep{Andersson-Manjang2021}.}

\end{table}

\begin{landscape}
\begin{center}
\begin{table}
\tiny
\scriptsize
\caption{Cash-based transfer programs with rigorous evaluation in crisis contexts} \label{23ProgramsSummary}
\begin{tabular}{llll|cccc|ccc}  
\hline \hline
Country/ & Target & Cash        & Delivery & Usage & Food & Mental  & Diversion & Cost & Decent- & Trans- \\ 
Study & Beneficiaries & Type    & Mechanism & (Mobile) & Security & Well-Being & Discussed  & (TCTR) & ralized & parent \\ 
  \hline
1. Ecuador \citep{HidroboEtAl2014}       & Refugees            & C/V/F     & ATM/physical & & + &  &  & 1.075 & \checkmark &      \\
2. Philippines  \citep{MercyCorps2022} & Affected HHs    & C & Mobile money & \checkmark &  \O & &  & 1.020 &  &   \\
3. Philippines \citep{KandpalEtAl2016} and \citep{CrostEtAl2016}  & Poor HHs  & C & Physical & & + & & & & &   \\
\hline
4. Lebanon  \citep{SaltiEtAl2022}  and \citep{MoussaEtAl2022}   & Refugees            & C   & ATM & &  + & + & & & \checkmark &    \\
5. Lebanon  \citep{LehmannMasterson2020}  & Refugees             & C          & ATM  & & + &  \O  & \checkmark & & \checkmark &         \\  
6. Lebanon  \citep{deHoopEtAl2019}  & Refugees              & C         & Not stated & & & & & & &      \\
7. Sri Lanka   \citep{SandstromTchatchua2010}  & Affected HHs       & C/F & Physical & &  + &  & & \checkmark & &    \\
\hline
8. Bangladesh  \citep{PopleEtAl2021}  & At-risk HHs   & C  & Mobile money & \checkmark &  + &  + &  & & \checkmark & \checkmark    \\
9. Uganda  \citep{BlattmanEtAl2016}      & Poor women  & C & Physical & &  + & & \checkmark  & \checkmark & &    \\
10. Niger    \citep{AkerEtAl2016}     & Poor women   & C  & Mobile/physical & \checkmark & + & &  & 1.059 & & \checkmark    \\ 
11. Niger    \citep{HoddinottEtAl2018}    & Poor HHs    & C/F  & ATM &  & + &  & &   1.053 & & \\  
12. Niger \citep{BossuroyEtAl2022} and  \citep{PremandBarry2022}  & Poor women  & C & Physical & &  + &  + & &  1.252 & &  \\
13. Niger \citep{LangendorfEtAl2014} & Poor children  & C/F & Physical & & + &  & & & &      \\
14. Iraq \citep{KurtzEtAl2021} & Poor HHs  & C & Physical & & + & & & & &     \\
15. D.R.C.    \citep{Aker2017}    & Displaced          & C/V       & Bank/physical & & \O & & \checkmark &  1.087 & &    \\
16. D.R.C.  \citep{BonillaEtAl2017}      & Displaced   & C/V & Multiple  & & \O & & & & &  \\ 
17. D.R.C. \citep{QuattrochiEtAl2022} & Displaced  & V & Physical & & \O &  + &  &  1.437 & &  \\
18. C.A.R. \citep{AlikLagrangeEtAl2019} & Poor HHs  & C & Physical & & \O &  + & & &  & \\ 
19. Yemen \citep{Schwab2020} and \citep{SchwabEtAl2013}   & Poor HHs  & C/F & Physical & &  + & &  &   1.083 & &  \\
20. Yemen   \citep{Kurdi2021}       & Poor women     & C & Physical & & + & & & & &   \\
21. Afghanistan \citep{BedoyaEtAl2019} & Poor women  & C & Physical & & + &  + & &  2.017 & &  \\ 
22. Afghanistan \citep{LyallEtAl2020} & At-risk youth  & C & Mobile money & \checkmark & &  &  & \checkmark & \checkmark & \checkmark \\ 
23. South Sudan \citep{ChowdhuryEtAl2017} & Poor HHs  & C & Physical & & + & &  & & &   \\
\hline
\end{tabular}
 \parbox{\linewidth}{\scriptsize \justifying \(Notes \): Abbreviations: HHs=households; C=Cash; V=Voucher; F=Food. Symbols apply to cash/voucher findings/discussion: +=analytically examined and positive effect; -=analytically examined and negative effect; \O=analytically examined and null effect; \checkmark=discussed and/or a possible advantage, but either not analyzed or, as with the cost metric, insufficient information to derive the TCTR; Organization: countries are ordered by increasing level of fragility based on \citep{OECD2022}. Ecuador and Philippines are not ranked or discussed; Lebanon and Sri Lanka are not ranked, but discussed as exhibiting early-warning signs.}
\normalsize
\end{table}
\end{center}
\end{landscape}

\begin{landscape} \scriptsize
\begin{center} 
\begin{longtable}{p{1.5in}|p{3.25in}|p{3.25in}} 
\caption{Related food security, mental well-being, and diversion results} 
\label{LiteratureFindings1}

\\
\hline \hline

Program & Relevant study context \& notes & Selected results  \\ 
  \hline
1. 2011 WFP food security \& Food security for refugee integration (Ecuador, \cite{HidroboEtAl2014}).         & RCT evaluation of 240 USD cash / voucher / food / control to 2,357 HHs to address non-acute, persistent displacement (by government request).  & Log calorie intake per capita increased by 12\% (cash), 18\% (vouchers) 21\% (food); food consumption scores (FCS) improved by 11\% (cash), 16\% (vouchers), and 12\% (food). \\
\hline
2. 2014 Mercy Corps TabangKO for typhoon recovery (Philippines, \cite{MercyCorps2022}).      & RCT evaluation of 89 USD lump-sum / varying disbursement cash to 25,480 HHs to assist immediate recovery \& build resilience (with government targeting assistance).  &  None of the treatment arms had differential impacts on food security outcomes (coping strategies index, FCS dietary diversity). \\
\hline
3. 2009--11 Pantawid Pamilya poverty reduction program (Philippines, \cite{KandpalEtAl2016} and \cite{CrostEtAl2016}).        &  RCT evaluation of 176--330 USD cash / control for longer-term poverty reduction in 65 non-acute, high-need villages in 714 study HHs (government implemented).   &   Dairy and egg consumption increased by 6.9pp and 8.2pp, and broader increases in height-for-age, increased antenatal \& prenatal care, decreased severe stunting, \& less violence. \\
\hline
4. 2017--18 WFP/UNHCR multi-purpose cash assistance (Lebanon, \cite{SaltiEtAl2022}  and \cite{MoussaEtAl2022}).         & RD assessment of 2100 USD cash (shorter/longer) / control to 56,000 refugee HHs (6,287 in study) to meet survival needs (contributing to government crisis response).    & Dairy consumption +0.6 days (control: 2.52), borrow food -0.27 days (control: 1.74), eating elsewhere +0.1 days (control: 0.17); happiness increased, stress decreased. \\
\hline
5. 2013--14 UNHCR/IRC Winter Cash Assistance Program (Lebanon, \cite{LehmannMasterson2020}).      & RD assessment of 575 USD cash / control to 87,700 Syrian refugee HHs (1,358 in study) to provide winter shelter (with government targeting assistance).      &  Days w/: meals skipped -0.65 (control: 3.25), less-preferred meals -0.6 days (control: 4.7), restricted consumption -0.3 (control: 2.6), \& reduced portions -0.4 (control: 3.2).         \\ 
\hline
6. 2016--17 No Lost Generation education program (Lebanon, \cite{deHoopEtAl2019}).       &  RD assessment of 441 USD cash / control to 1440 Syrian refugee HHs to offset commuting / foregone income (part of government education initiative).       &  No food security, mental well-being, or diversion outcomes reported.              \\ 
\hline
7. 2005--6 WFP cash transfers for tsunami relief (Sri Lanka, \cite{SandstromTchatchua2010}).    & RCT evaluation of cash / food (amount unclear) to 3,200 HHs (1,360 in study) to address medium-term recovery from tsunami a year earlier (goverernment role unclear).     & Estimates not reported: food spending increased; dietary diversity improved; \& consumption of staples decreased. Greatest impact in poorest areas, with neediest HHs.           \\ 
\hline 
8. 2020 WFP anticipatory cash for flood preparation \& relief (Bangladesh, \cite{PopleEtAl2021}).         & Natural experiment of 53 USD cash / control to 23,434 HHs (8,954 in study) ahead of imminent flooding (with government forecasting/targeting assistance).    & Percentage of HHs reporting: days w/out meals -10pp (control: 28\%) and children eating 3 meals +3pp (control: 80\%); Increased life satisfaction +12.5\% (control: 2.8) on a 10-pt scale. \\ 
\hline
9. 2009--11 AVSI Women's Income Generating Support (Uganda, \cite{BlattmanEtAl2016}).            &  RCT evaluation of 150 USD cash, training, \& supervision / control to 1,800 women to promote microenterprise in post-conflict region (supporting government programming).   &  Number of times in past week going hungry -0.1 (control: 0.2), usual number of meals per day +0.06 (control: 1.76), and diversion less than 1\% of grants. \\  
\hline
10. 2009--10 Concern Worldwide social protection program (Niger, \cite{AkerEtAl2016}).           & RCT evaluation of 225 USD cash (mobile vs. physical vs. physical+phone) to 10,000 women (1152 in study survey) for drought response (government role unclear).   &  Mobile increased dietary diversity +0.28--0.51 (physical: 3.17; physical+phone: 2.94) on 12-pt scale and \# of meals by under-5 children in past day +0.33 (physical: 3.17).        \\ 
\hline
11. 2011 WFP cash/food aid in Zinder region (Niger, \cite{HoddinottEtAl2018}).         & RCT evaluation of 50 USD cash-for-work / food-for-work to 2,209 poor HHs to address lean-season drought among highly poor (by government request).         &   Dietary diversity reduced -0.56 (food mean: unclear) on 25-pt scale, food consumption reduced -3.9 points (food: 41.5) on 112-pt scale, \& increased grain purchases.     \\ 
\hline
12. 2016--19 Supplement to government safety nets program (Niger, \cite{BossuroyEtAl2022}).        & RCT evaluation of 127 USD cash / psychosocial / joint (w/core support) to 4712 poor women (1191 in cash) in HHs in government safety nets program.        & Cash increased food security +0.20sd (joint: +0.25sd), dietary diversity +3.69 (joint +6.11), days eating foods from 8 groups, \& life satisfaction +0.20sd (joint +0.45sd).   \\ 
\hline
13. 2011 Cash for prevention of young child malnutrition (Niger, \cite{LangendorfEtAl2014}).        & RCT evaluation of 260 USD cash / cash+various food supplements to 5,395 HHs with at-risk 6--23 month old children (in collaboration w/government ministries).    & Cash+food reduced moderate acute malnutrition by double (Hazard ratios: 2.07--2.42). Cash+super cereal reduced severe acute malnutrition by triple (Hazard ratio: 3.13).  \\ 
\hline
14. 2019--20 multi-purpose cash assistance for economic recovery (Iraq, \cite{KurtzEtAl2021}).            & RCT evaluation of 1200 USD cash (lump-sum, multiple disbursements) / control to 827 poor HHs in protracted crisis (supporting the government safety nets program).      &  Food consumption +4.515 (control: appx. 44) on 112-pt scale, reliance on coping strategies -12.18 (control: 25.01), and insecurity -0.18 (control: unclear) on 1--5 scale.    \\ 
\hline
15. 2011 Concern Worldwide social protection program (D.R.C., \cite{Aker2017})       & RCT evaluation of 130 USD cash / vouchers to 474 displaced HHs to increase access to food \& non-food items amid ongoing violent conflict (government role unclear).     & No improvements to: diet diversity (3.05 on 12-pt scale; voucher: 3.07), meals / day (1.35; voucher: 1.38), or months of adequate food (1.25; voucher: 1.26).           \\ 
\hline
16. 2013--15 Multi-purpose cash for displaced communities in crisis (D.R.C., \cite{BonillaEtAl2017})                   & 2 RCT evaluations: 120 USD cash (lump sum vs. 3 disbursements) to 196 HHs \& gender of HH beneficiary (male, female, choice) for 157 HHs (government role unclear).    &  No differential impacts of number of transfers or gender of HH beneficiary on food security, children's health, or other key program indicators.   \\ 
\hline
17. 2017--18 RRPM voucher assistance for displaced (D.R.C., \cite{QuattrochiEtAl2022})      & RCT evaluation of 55--90 USD vouchers / control to 488 displaced HHs (488 control) to improve health \& well being (government role unclear).    & Psych. well-being +0.32sd at 6-weeks, +0.18 at 1-year; life satisfaction +0.59 (control: 3.29) on 10-pt scale, \& coping/ meal skipping -0.07 (control: 1.79) on 11-question index.     \\ 
\hline
18. 2016--19 WB Londo cash-for-public works program (C.A.R., \cite{AlikLagrangeEtAl2019})        & Natural experiment of 120 USD cash-for-public-works / control to 3,470 poor study HHs to improve livelihoods amid ongoing violent conflict (government role unclear).    & No change in days with skipped meals (0.30; control: 0.31), happiness increases +0.14sd (control: -0.10), and satisfaction with security increases 0.08sd (control: -0.07).        \\ 
\hline
19. 2011--12 WFP cash \& food transfer program (Yemen, \cite{SchwabEtAl2013} and \cite{Schwab2020}).      & RCT evaluation of 147 USD cash / food to 1,983 poor HHs facing lean-season food insecurity (in collaboration with government).   & Food consumption +2.27 (food: 48.82) on 112-pt scale, days/week reducing meal frequency +0.37 (food: 0.63), and days/week adults reduced intake +0.47 (food: 0.3).  \\ 
\hline
20. 2015--17 Yemen SWF / WB Cash-for-Nutrition program (Yemen, \cite{Kurdi2021}).        & RCT evaluation of 1122 USD cash / control to 1,001 women in high-need HHs facing food insecurity amid violent conflict (in collaboration with government).   & Child dietary diversity +0.616 (control: 1.88) of 7 food groups; likelihood of coping through selling gold -0.104 (control: 0.244) among the highest tercile.          \\ 
\hline
21. 2015--18 Targeting the Ultra Poor program (Afghanistan, \cite{BedoyaEtAl2019}).         & RCT evaluation of 180 USD cash, livestock assets, training, \& coaching / control to 491 ultra-poor female-headed HHs (implemented by government-owned organization).   & HHs in which adults skip/reduce meals -23pp (control: 0.44) \& children skip/reduce -20pp (control: 0.59); life satisfaction +0.44 (control: 0.00) on 10pt-scale. \\ 
\hline
22. 2016 Mercy Corps INVEST vocation/skills program (Afghanistan, \cite{LyallEtAl2020}).     &  RCT evaluation of 75 USD cash / training / combined to 1,841 youth (756 control) at risk of insurgent recruitment (supported by local government).        & No food security, mental well-being, or diversion outcomes reported.          \\  
\hline
23. 2013--14 Cash / Targeting the Ultra Poor (South Sudan, \cite{ChowdhuryEtAl2017}).        & RCT evaluation of 350--410 cash / TUP bundle / control to 125 poor HHs to increase productivity amid ongoing violent conflict (government role unclear).   & Null: \% skip day of meals / week (0.06; control: 0.41), go to bed hungry (0.03; control: 0.40), no food in home (-0.01; control: 0.45), \& limiting portions (-0.04; control: 0.48).  
\tabularnewline \hline
\end{longtable}
\end{center} \normalsize
\end{landscape}

\begin{table}[h!] \centering
\newcolumntype{R}{>{\raggedleft\arraybackslash}X}
\newcolumntype{L}{>{\raggedright\arraybackslash}X}
\newcolumntype{C}{>{\centering\arraybackslash}X}

\caption{Attrition}
\label{attrition}
\begin{adjustbox}{max width=\linewidth, max height=\textwidth}\begin{tabular}{l|cc}

\toprule
& Attrited & By Survey Round \tabularnewline 
{}&{(1)}&{(2)} \tabularnewline
\midrule \addlinespace[\belowrulesep]
\midrule \midrule Treated&--0.001&0.004 \tabularnewline
&(0.004)&(0.004) \tabularnewline
Round 2&&0.003 \tabularnewline
&&(0.003) \tabularnewline
Treated \( \times \) Round 2&&--0.009** \tabularnewline
&&(0.004) \tabularnewline
\midrule Observations&4,818&4,818 \tabularnewline
R Squared&0.017&0.017 \tabularnewline
\bottomrule \addlinespace[\belowrulesep]

\end{tabular} \end{adjustbox}
\\ \parbox{0.6\linewidth}{\footnotesize \justifying \noindent \(Notes \): Controls for strata fixed effects are included. Standard errors are clustered at the individual level. In the first survey round, 29 interviews could not be completed (17 treatment, 12 control). In the second round, 26 interviews could not be completed (10 treatment, 16 control). Households were surveyed once per month, over two months. Each of these months constitutes a survey round. \\   \textit{Levels of significance}: *$ p<0.1$ , **$ p<0.05$ , ***$ p<0.01$.}
\end{table}

\begin{table}[ht] 
\centering \small
\caption{Cost-efficiency calculations} 
\label{Ta:costefficiency}
\begin{tabular}{p{0.5\linewidth} | p{0.25\linewidth}} \hline \hline
     \textbf{Panel A: Raw costs}  &    \\ \hline
    \textit{Costs by category} &  \\ 
    --Distribution: Personnel  & \$4054.18     \\  
    --Distribution: Facilities  & \$914.36    \\ 
    --Distribution: Technology  & \$910.62    \\ 
    --Digital aid transfers  & \$433,620.00    \\ 
    \textit{Total costs without onboarding} & \$439,499.16 \\ 
    --Onboarding costs  & \$23,029    \\ 
    \textit{Total costs with onboarding}  & \$462,528.16   \\ 
        \textit{Total costs treatment group ($N=1,208$)}  & \$231,936   \\
            \textit{Total costs control group ($N=1,201$)}  & \$230,592   \\ \hline
  \textbf{Panel B: Costs per beneficiary (CPB)}  &  \\ \hline 
      \textit{Including the digital aid transfer} &  \\ 
  --With onboarding  & \$192.00    \\  
  --Without onboarding  & \$182.44   \\       \textit{Excluding the digital aid transfer} &  \\ 
  --With onboarding  & \$12   \\  
  --Without onboarding  & \$2.44   \\  \hline
  \textbf{Panel C: Transfer ratios}  &    \\ \hline
        \textit{Total cost-transfer ratios (TCTR)} &  \\ 
  --With onboarding & 1.067  \\ 
  --Without onboarding &  1.014  \\ 
  \textit{Cost-transfer ratios (CTR)} &  \\ 
  --Cost-transfer ratio with onboarding  & 6.7 cents  \\
  --Cost-transfer ratio without onboarding  & 1.4 cents   \\
  \hline \hline
\end{tabular}
\caption*{\footnotesize{\textbf{Note:} Based on 2,409 beneficiary households each receiving 180 USD for a total of 433,620 USD distributed.}}  
\normalsize
\end{table}
  
\begin{table}[ht] 
\centering \small
\caption{Cost-effectiveness ratios: days with skipped meals and corresponding calorie intake } 
\label{Ta:costeffectivenessmealscalories}
\begin{tabular}{ p{0.17\linewidth}  p{0.17\linewidth} p{0.17\linewidth}  p{0.17\linewidth}  p{0.1\linewidth}} \hline \hline
\multicolumn{5}{l}{\textbf{CERs for Reduction in \texttt{Days with Skipped Meals}}} \\ \hline
    \textit{Group}   & \textit{Mean/Week} & \textit{Skipped Days} & \textit{\% Skipped} & \textit{CER}   \\ \hline
\multicolumn{5}{l}{\textit{Panel A: The entire eight-week program (Weeks 1--8)}} \\ \hline
        Control Group      & 2.569 & 24,682.952 & 36.700\% &                    \\
        Treatment Group     & 1.809 & 17,482.176 & 25.843\% &                       \\
        Effect          & -0.760 &  -7,344.640 & -10.857\% &   \$31.58   \\ \hline
 \multicolumn{5}{l}{\textit{Panel B: The second half of the program (Weeks 5--8)}} \\ \hline
        Control Group      & 2.448 & 11,760.670 & 34.973\% &                    \\
        Treatment Group     & 1.401 & 6767.699 & 20.009\% &                       \\
        Effect          & -1.048 &  -5,061.520 & -14.964\% &   \$22.91   \\ 
 \hline
\multicolumn{5}{l}{\textit{Panel C: The final week of the program (Week 8)}} \\ \hline
        Control Group      & 2.508 & 3,012.108 & 35.823\% &                    \\
        Treatment Group     & 1.088 & 1,314.304 & 15.543\% &                       \\
        Effect          & -1.42 &  -1,715.36 & -20.286\% &   \$16.90   \\ \hline \hline
\multicolumn{5}{l}{\textbf{CERs for Reduction in \texttt{Foregone Calories}}} \\ \hline
\textit{Group}   & \textit{Mean/Week} & \textit{Foregone calories} & \textit{\% Foregone} & \textit{CER}   \\ \hline
\multicolumn{5}{l}{\textit{Panel D: The full eight-week program (Weeks 1--8)}} \\ \hline
        Control Group      &  34,041.82 & 327,073,797  &  36.700\% &                    \\
        Treatment Group     &  23,971.06 & 231,656,314  &  25.843\% &                       \\
        Effect          & -10,070.76 & -97,323,825  & -10.857\% &   $\$2.38$  \\ \hline
 \multicolumn{5}{l}{\textit{Panel E: The second half of the program (Weeks 5--8)}} \\ \hline
        Control Group      &  32,439.77 & 155,840,670 & 34.973\% &                    \\
        Treatment Group     &  18,559.35 &  89,678,782 & 20.009\% &                       \\
        Effect          & -13,880.42 & -67,070,202 & -14.964\% &   \$1.73   \\ \hline
\multicolumn{5}{l}{\textit{Panel F: The final week of the program (Week 8)}} \\ \hline
        Control Group      &  33,233.51  &  39,913,443 &  35.829\% &                    \\
        Treatment Group     &  14,417.088 &  17,415,842 &  15.543\% &                       \\
        Effect          & -18,816.42  & -22,730,235 & -20.286\% &   \$1.28   \\ \hline
 \hline
\end{tabular}

\parbox{\linewidth}{\scriptsize \justifying \(Notes \): The cost-effectiveness ratio is computed as the treatment group's costs divided by the reduction in days with skipped meals for the beneficiaries (treatment group costs / (\# beneficiaries * treatment effect * \# of weeks)) and is interpreted as the number of dollars required for a reduction of a day with skipped meals. For aggregate estimates, there are 67,648 and 67,256 possible days for the treatment group and control group respectively ($1208$ or $1201$ participants $*$ $56$ days). The cost-effectiveness ratio for calorie intake is computed similarly and is interpreted as the number of dollars required for a reduction of $1,000$ foregone calories. Results for calorie intake are based on the Afghanistan CVWG estimate of $2,100$ calories/person/day, which translates to $13,251$ calories/household/day and $92,757$ calories/household/week. Across the eight-week program, there are $896,403,648$ and $891,209,256$  possible calories for the treatment group and control group respectively ($1208$ or $1201$ participants $*$ $92,757$ calories per household week $*$ $8$ weeks).} 
\normalsize
\end{table}

\end{document}